%% file: Masterarbeit.tex
\documentclass[12pt,a4paper,twoside,openright]{memoir} 

\usepackage[utf8]{inputenc} 
\usepackage{lmodern}
\usepackage[T1]{fontenc}
\usepackage{microtype} 
\usepackage[english]{babel}
\usepackage[autostyle=true,english=american]{csquotes}
\usepackage{verbatim} 
\usepackage{graphicx} 
\usepackage{url}
\usepackage[style=authoryear,backend=biber]{biblatex}
\usepackage{float}
\usepackage{listings}
\usepackage{algorithm2e}
\usepackage{color}
\usepackage{framed}
\usepackage[fleqn,tbtags]{mathtools} 
\usepackage{MnSymbol} 
\usepackage[ngerman]{todonotes}
\usepackage[hidelinks,pdftitle={Towards the Improvement of Automated Scientific Document Categorization by Deep Learning},pdfauthor={Thomas Krause}]{hyperref} 
\usepackage[acronym,toc,nomain,nonumberlist]{glossaries} 
\usepackage{ellipsis}
\usepackage{multirow}
\usepackage{booktabs}
\usepackage[nott]{inconsolata}
\usepackage{pdfcomment}
\usepackage{numprint}
\usepackage{pgfplots}
\usepackage{pgfkeys}

\usepackage{xparse} 
 
\makeglossaries

\definecolor{dkgreen}{rgb}{0,0.5,0}
\definecolor{dkred}{rgb}{0.6,0.1,0.1}
\definecolor{lightgray}{rgb}{0.9,0.9,0.9}
\definecolor{mauve}{rgb}{0.58,0,0.82}
\definecolor{shadecolor}{rgb}{0.8,0.8,0.8}

\lstdefinelanguage{blank}
{
  sensitive=true, 
  keywords={},
  morestring=[b]", 
  morestring=[b]' 
}

\lstdefinelanguage{rest}
{
  sensitive=true, 
  keywords={POST,GET,PUT,null},
  morestring=[b]", 
  morestring=[b]' 
}

\lstMakeShortInline[breaklines=false]|

\SetKw{Yield}{yield}

\newcommand{\coderef}[1]{{\small\fontfamily{pcr}\selectfont#1}}

\newcommand{\norm}[1]{\lvert #1 \rvert}

\pgfkeys{
 /figgraph/.is family, /figgraph,
 default/.style = 
  {width = \textwidth},
 width/.estore in = \figgraphWidth,
}

\newcommand{\figgraph}[3][width=\textwidth]{%
 \pgfkeys{/figgraph, default, #1}%
\begin{figure}[htbp]
	\centering
	\includegraphics[width=\figgraphWidth]{images/#2}
	\caption{#3}
	\label{fig:#2}
\end{figure}
}

\newlength{\eqnspace}
\bibliography{sources} 

\OnehalfSpacing 

\settrimmedsize{297mm}{210mm}{*} 
\settypeblocksize{9.0in}{30pc}{*}

\setulmargins{*}{*}{1} 
\setlrmargins{*}{*}{1.3}
\setheaderspaces{*}{*}{0.3}

\checkandfixthelayout 
\pretitle{\begin{center}\sffamily\huge\MakeUppercase}
\posttitle{\par\end{center}\vskip 0.5em}

\chapterstyle{hangnum} 

\hangsecnum 
\maxsecnumdepth{subsection} 

\setFloatBlockFor{section} 

\setsecheadstyle{\Large\sffamily\raggedright} 
\setsubsecheadstyle{\large\sffamily\raggedright} 

\makepagestyle{plain}
\makeevenfoot{plain}{\thepage}{}{}
\makeoddfoot{plain}{}{}{\thepage}

\title{Towards the Improvement of Automated Scientific Document Categorization by Deep Learning}
\author{Thomas Krause}

\newcommand{\acr}[1]{\gls{#1}}
\newcommand{\acrpl}[1]{\glspl{#1}}
\newcommand{\Acrpl}[1]{\Glspl{#1}}

\newacronym{gpu}{GPU}{Graphics Processing Unit}
\newacronym{api}{API}{Application Program Interface}
\newacronym{svm}{SVM}{Support Vector Machine}
\newacronym{cnn}{CNN}{Convolutional Neural Network}
\newacronym{bptt}{BPTT}{Backpropagation Through Time}
\newacronym{lstm}{LSTM}{Long short-term memory}
\newacronym{gru}{GRU}{Gated Recurrent Unit}
\newacronym{gpgpu}{GPGPU}{General Purpose Graphics Processing Unit}
\newacronym{dto}{DTO}{Data Transfer Object}
\newacronym{paas}{PaaS}{Platform-as-a-Service}
\newacronym{ddc}{DDC}{Dewey Decimal Classification}
\newacronym{cbow}{CBOW}{Continuous Bag of Words}
\newacronym{sg}{SG}{Skip Gram}
\newacronym{relu}{ReLU}{Rectified Linear Unit}
\newacronym{w3c}{W3C}{World Wide Web Consortium}
\newacronym{rdf}{RDF}{Resource Description Framework}
\newacronym{rest}{REST}{Representational State Transfer}
\newacronym{json}{JSON}{JavaScript Object Notation}

\newcommand{\chapref}[1]{\enquote{\textbf{\nameref{#1}}}}

\begin{document}
\selectlanguage{english}


\include{chapters/titlepage}

\floatstyle{ruled}
\restylefloat*{table}    

\restylefloat*{figure}   

\captionnamefont{\footnotesize\bfseries}
\captiontitlefont{\footnotesize}

\begin{abstract}
This master thesis describes an algorithm for automated categorization of scientific documents using deep learning techniques and compares the results to the results of existing classification algorithms.

As an additional goal a reusable API is to be developed allowing the automation of classification tasks in existing software.

A design will be proposed using a convolutional neural network as a classifier and integrating this into a REST based API. This is then used as the basis for an actual proof of concept implementation presented as well in this thesis.

It will be shown that the deep learning classifier provides very good result in the context of multi-class document categorization and that it is feasible to integrate such classifiers into a larger ecosystem using REST based services.
\end{abstract}

\cleardoublepage

\tableofcontents* 
\cleardoublepage
\renewcommand{\glossarymark}[1]{\markboth{#1}{#1}}

\renewcommand\lstlistlistingname{List of Listings}

\listoffigures
\cleardoublepage
\listoftables
\cleardoublepage
\addcontentsline{toc}{chapter}{\lstlistlistingname}
\lstlistoflistings
\cleardoublepage

\printglossary[type=\acronymtype,style=listdotted]
\cleardoublepage

\chapter*{Notation}
\addcontentsline{toc}{chapter}{Notation}
This list gives an overview about frequently used symbols and their significance in this thesis:

\begin{tabular}{l l}
$D$	&	Corpus consisting of $\norm{D}$ documents\\
$d \in D$	&	An individual document, consisting of $\norm{d}$ terms\\
$n$ & Sample/document index\\
$N$ & Number of samples/documents\\
$t \in d$ & A term within a document (also used as an index)\\
$tf(t,d)$ & Frequency of the term $t$ within a document $d$\\
$df(t,D)$ & Frequency of documents in the corpus $D$ that contain the term $t$\\
$tfidf(t,d)$ & The tf-idf weight of a term $t$ in a document $d$\\
$v$ & The vector representation of a term, with $\norm{v}$ components\\
$i,j$ & Neuron indices\\
$w_{i,j}$ & The weight a neuron $i$ applies to the output of a neuron $j$\\
$netinput_i$ & Weighted sum of a neuron's input\\
$a_i$ & Output of neuron $i$\\
$\sigma(netinput)$ & The activation function of a neuron\\
$\sigma_{tanh}$ & The tanh activation function\\
$\sigma_{sigmoid}$ & The sigmoid/logistic activation function\\
$\sigma_{softmax}$ & The softmax activation function\\
$\sigma_{relu}$ & The ReLU activation function\\
$\sigma_{leaky}$ & The Leaky ReLU activation function\\
$X$ & Input of a classifier (a Tensor)\\
$y$ & Vector representing the expected output of a classifier\\
$c_k \in C$ & The categories/classes to be differentiated\\
$\norm{y} = \norm{C}$ & Number of categories/classes\\
$E$ & Error function used to train a neural network\\
$E'$ & First derivative of the error function\\ 
$E_{sq}$ & Quadratic error function\\
$E_{bc}$ & Binary cross-entropy error function\\
$E_{cc}$ & Categorical cross-entropy error function\\
\coderef{symbol} & Style used for code symbols, such as variables\\
\end{tabular}


\hyphenation{back-end}
\hyphenation{to-ken}
\hyphenation{pre-pro-cessing}

\glsresetall


\include{chapters/01.introduction}

\include{chapters/02.fundamentals}

\include{chapters/03.algorithm}

\include{chapters/04.implementation}

\include{chapters/05.evaluation}

\include{chapters/06.prospect}

\appendix
\include{chapters/A.lessons}
\include{chapters/B.installation}
\include{chapters/C.usage}

\printbibliography\cleardoublepage
\thispagestyle{empty} 
\vspace*{\fill}
I hereby declare that the thesis submitted is my own unaided work. All direct or indirect sources used are acknowledged as references. This paper was not previously presented to another examination board.
\vspace{2cm}

\hspace*{\fill}\begin{tabular}{@{}l@{}}
\hline
\makebox[5cm]{(Thomas Krause)}
\end{tabular}

\vspace{0.5cm}
\hspace*{\fill}Cologne, 26th of June 2016

\end{document}

%% file: chapters/titlepage.tex

\begin{titlingpage}

\begin{center}

\begin{figure}[!ht]
	\centering
		\includegraphics[scale=1.8]{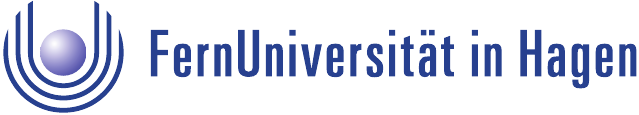}
\end{figure}

\vspace{2.0cm}
{Master Thesis}\\
\vspace{0.2cm}

\large{\textbf{Towards the Improvement of\\Automated Scientific Document Categorization\\by Deep Learning}}\\
\normalsize

%

\vspace{1.5cm}

{presented at the\\}
\vspace{0.2cm}
\begin{large}
\textbf{University of Hagen}
\end{large}

\vspace{1.5cm}

{written by}\\
\vspace{0.2cm}

\begin{large}
\textbf{Thomas Krause}\\
\textbf{\# 9234322}\\
\end{large}

\vspace{1.5cm}

{Examiner and Supervisor}\\
\vspace{0.2cm}

\begin{large}
\textbf{Prof. Dr.-Ing. Matthias Hemmje}\\
\textbf{MSc. Tobias Swoboda}\\
\end{large}

\vspace{1.5cm}

\begin{large}
Cologne, June 2016
\end{large}

\end{center}
\end{titlingpage}

%% file: chapters/01.introduction.tex

\chapter{Introduction}
\label{chap:Introduction}
This chapter will include a small overview of the motivation behind text classification, describes common difficulties and provides a short overview of the content of the thesis and its goals.

\section{Motivation}
Digital document libraries have many advantages compared to traditional  (``paper based'') libraries. To mention only a few:

\begin{itemize}
   \item They can be made available over networks such as the internet to allow access from any location.
   \item Copying digital documents is cheap and fast, so instead of lending documents, they can be simply copied to anyone interested. This means a single document can be made available to any number of users at the same time.
   \item A single state of the art hard drive can hold the equivalent of thousands or even millions of books and documents. 
\end{itemize}

\parencite[][4\psqq]{arms99}\\

Combining these advantages means that it is now very cheap to build large document libraries and making them available to audiences. 

One difficulty however faced by both traditional and digital libraries is how to make the documents discoverable so that someone searching for a particular topic or a particular book can easily find what he is looking for. To achieve this goal in traditional libraries, all documents are usually primarily ordered by topic. An example is the \acr{ddc} \parencite{dewey1876} used by many libraries world wide. The \acr{ddc} uses a hierarchical numbering scheme to differentiate topics. For example books about algebra would be assigned the class 512. The leading number (5) designates the main class (``science''). The second number is a division inside the class referring to ``mathematics'' and the last number is a specific section inside of the division which refers to ``algebra''.

As a physical book can only be in one place at a time, only a single classification system can be used to arrange the books, which is very limited. If someone wants to look for a book written by a specific author, maybe even knowing the specific title, but being unsure about the topic in which the book was classified, he will have a hard time finding it, if the library is organized by topic. And of course someone generally interested in books from the 18th century will have similar difficulty if the books are not distinguished by publication date.

Traditional libraries can use additional indices (also called catalogs) that are ordered by additional criteria and denote, where in the library the book can be found \parencite{arms99}. These catalogs can be in the form of a physical register, book or nowadays stored on computers. They only record specific metadata about the book without replicating the content itself. This is a very powerful concept as it expands the one-dimensional classification system for topics into a multi-dimensional system where the topic is only one of many attributes of a book that can be used to find it.

Using an index the actual classification system used to physically locate the book in the library is no longer that important. In fact books could be located in random places at the library as long as there are indices for the interesting attributes of a book which can be used to locate it. The primary classification system is reduced to a convenience of not having to look through the index first before locating the book (therefore saving time). The same idea is used in relational databases on computer systems where tables employ a primary index used for the storage location and additional indices that point to this record \parencite[][chapter 14]{Ullman2008}.

An important question is though, where does the data used for indexing come from? It was already said that one can use attributes of the document such as the title, year or topic to build indices. But looking at various documents it is not always obvious where this data comes from or even what attributes can and should be used. The title can usually be found at the beginning of a document, maybe the author and the date of publication can as well. The general topic category on the other hand is often not explicitly mentioned in the document itself. In this case it might only be possible to get an idea about the topic by reading the book and making a judgment based on that. Even for the attributes readily available in the document, such as the title, there is no standardized and formal way to get this information from an unstructured document. A librarian in a traditional library needs to inspect every single document and extract the information out of the document manually. \parencite[][chapter 10]{arms99}

This process of extracting information out of a document and assigning various attributes to a document is called annotation. Annotation requires a classification system using one or more attributes which should have a clearly specified syntax and semantics. Attributes can use simple texts, numbers, dates or even complex systems such as taxonomies to define their values. \label{desc:taxonomy}A taxonomy is a hierarchical classification system like the \acrlong{ddc} system described before \parencite{Rees03}.

A description of available resources (such as documents), a set of attributes that the resources can take and relationship between resources is also called an ontology \parencite{Pretorius04,Rees03}. Creating an ontology can be a very hard task as it should take in account many different aspects:

\begin{itemize}
  \item How is the ontology going to be used? What are the primary use cases?
  \item What fundamental attributes are essential to fulfill this use case?
  \begin{itemize} 
	\item Every attribute added to the ontology comes at a cost as it has to be clearly defined and increases the complexity of the ontology. Classification will take more time with every attribute added.
  \end{itemize}
  \item What is the exact semantic meaning of these attributes?
  \begin{itemize} 
	\item E.g. for publication date, does this refer to the first time the document was published or does it refer to when a specific edition was published?
	\item What is the publication date of ``The bible''?
	\item How is the decision made what topics a book belongs to?
  \end{itemize}
  \item What is the syntax used by the attributes?
  \begin{itemize} 
	\item Is the forename and surname of the author stored separately or together?
	\item What format is used for dates?
  \end{itemize}
\end{itemize}

\parencite{Noy01,Antoniou08}\\

After an ontology has been defined it can be used to annotate resources such as documents using the rules of the ontology. After annotation the attributes can be used to build indices as described above to facilitate access to the resources. Of course the annotations can also be used for other tasks such as building statistics.

Manual annotation is a labor intensive task. It is therefore desired to automate or at least facilitate some of these aspects using computer science. Existing examples of this include:

\begin{itemize}
	\item Databases can be used to store attributes for documents and index them automatically for easy retrieval
	\item Ontologies can be created using standardized formats and tools such as RDF Schema\footnote{\acr{rdf} is a conceptual model to describe resources with their properties and relationships to other resources. RDF Schema defines a common vocabulary to create simple ontologies using RDF. Both are maintained by the \acr{w3c}. See: \url{https://www.w3.org/TR/rdf-schema/}, retrieved 20th of June 2016} facilitating automated processing of the data
	\item Structured documents (such as HTML) can be used to ease the extraction of information
	\item Some file formats can store attributes (such as the author of a document) directly as metadata.
	\item User interfaces can make the search for documents easier by allowing rich interactions such as filtering and sorting over multiple attributes which are not available in traditional libraries
	\item Full text indices can be employed to allow searching the entire content of documents
	\item Machine Learning can be used to extract information and annotate documents automatically
\end{itemize}

\parencite{Husain06,arms99,Pretorius04}\\

Arguably the most difficult of these points is the last one on which this thesis will focus on.

\section{Problem Statement}
While computers are very good in storing, querying and displaying structured information, they lack a fundamental understanding of the data being processed. When faced with unstructured data such as plain text, it is difficult for an algorithm to extract new information without being able to understand the meaning of the text like a human would.

The fundamental research question is therefore how these obstacles can be overcome and ultimately how the process of annotation can be automated.

More specifically the following questions can be asked:

\begin{itemize}
  \item How can a computer ``learn'' to classify documents?
  \item What techniques can be used for this?
	\begin{itemize} 
	\item Is it possible to use algorithms and techniques that are successful in other domains and to adapt them?
  \end{itemize}
  \item Assuming that a perfect classification algorithm cannot be built:
  \begin{itemize} 
	\item How can the quality of the predictions be measured?
	\item How can the classifications be improved / the error be reduced?
  \end{itemize}
  \item What input does an algorithm need to start predictions?
    \begin{itemize} 
	\item Does the algorithm need to learn from existing known assignments?
	\item Or can the algorithm even classify documents into categories by itself without having to specify the categories beforehand?
  \end{itemize}
  \item Can the process be generalized enough so it can be applied to various datasets without having to modify the algorithm itself?
  \item Do the resulting algorithms perform fast enough to be of practical use, even when used on large datasets?
  \item Is it possible to automate the process enough so it can be easily integrated into digital libraries?
\end{itemize}

\section{Scope of Work}
Most of the questions raised in the previous section are already addressed by various techniques and algorithms, but there is still a lot of room for improvement.

One approach in machine learning which has been very successful in raising the state of the art in related problems is called ``Deep Learning''. The idea behind deep learning is that complex predictions require high level abstractions. These abstractions are achieved by using architectures with many levels of operation \parencite{Bengio09}. Applying deep learning to the field of document categorization could therefore also result in better predictions. Deep learning will be discussed in detail in \autoref{chap:Fundamentals}.

As will be shown later there is also little research on practical applications that integrate algorithms into reusable components usable in digital libraries.

As an example of a digital library that is used throughout this thesis the RAGE ecosystem portal will be used.

The RAGE project\footnote{\url{http://rageproject.eu/}} acts as a social space to interconnect people interested in applied gaming. The RAGE ecosystem portal is one part of this and provides users access to a variety of documents and other resources related to applied gaming. The project is partly funded by the European Union and the University of Hagen participates in the development of the ecosystem portal.

The idea is that the RAGE portal will benefit from this research ultimately reducing the amount of manual work required to maintain its document collection. A focus and the basis of evaluation will therefore be the classification of scientific documents, but the scope is not limited to these kind of documents.

To limit the scope of work, the thesis will only look at the special case of document categorization using a predefined set of categories instead of annotation in general.

This thesis will therefore focus mainly on how existing classification systems can be improved using deep learning, how they can be applied to the domain of (scientific) document categorization and how they can be implemented in real digital library systems.

\section{Approach and Goals}
To answer these refined question, existing solutions will be researched and discussed to see how they can be utilized or improved. Armed with this knowledge a model will then be designed and implemented in the form of a prototype. Finally the resulting system will be evaluated and conclusions will be made.

The main goal of this thesis will be to find a suitable algorithm that can categorize scientific documents automatically. The aim is to make use of the latest research in this area (specifically Deep Learning) and to see if it provides significant enough advances to warrant its use in this area. As a second goal this algorithm should be implemented as a prototypical component which can be integrated into existing document portals to aid the automated classification of documents.

Specifically the following sub-goals should be achieved:

\begin{itemize}
  \item Preparation of a generic model that can be used for document classification
  \item Design of a suitable classification algorithm based on deep learning
  \item Definition of a generic \acr{api} between digital libraries and document classification systems
  \item Building of a prototypical implementation of the model containing the algorithm, capable of being integrated as a component into existing digital libraries using the \acr{api}
  \item Selection of a suitable performance metric to compare different text classification algorithms
  \item Evaluation of the algorithm and comparison with state of the art systems
  \item \textbf{NOT:} Integration into the RAGE ecosystem portal
\end{itemize}

The system should be generic enough to be used in all kind of applications that require automated document classification. The RAGE ecosystem portal will be used to guide the implementation and evaluation of the system. However the actual integration into the portal will be part of a future project at the university and will therefore not be discussed within this thesis (see also \autoref{sec:related_work}).

\section{Structure of this Thesis}
The rest of the thesis is structured like this:

Chapter \chapref{chap:Fundamentals} will describe both former work in this area and general techniques used in the area of text classification.

Next, the chapter \chapref{chap:Algorithm} will focus on describing the deep learning algorithm used in the thesis and will also propose a solution on how this algorithm can be integrated into existing digital libraries. It will describe the theory independent of the concrete practical implementation.

In the chapter \chapref{chap:Implementation} the prototypical implementation will be described. This includes the general components and libraries used to develop the system.

The chapter \chapref{chap:Evaluation} will have a short overview of metrics that can be used to compare algorithms and possible problems. It will then describe the metric that will be used to measure the developed algorithm and provide the evaluation results for various experiments.

Lastly the chapter \chapref{chap:Conclusions} will provide a summary of the obtained results as well as a short prospect of possible future improvements and further research that could be done in this area.

%% file: chapters/02.fundamentals.tex

\chapter{Fundamentals and State of the Art}
\label{chap:Fundamentals}
\section{Machine Learning and Automated Document Classification}
Machine learning is the art of training an algorithm to generate new information from previous experiences. The goal of machine learning is not to model explicitly how to extract this information, but to let the computer itself learn how to do this. \parencite{Mohri12}

Historically many different methods have been used to achieve this task with various degrees of success. A lot of these methods are not specific to applying categories to documents but are used in a wide area of applications. Some examples are neural networks or support vector machines. While they provide respectable results there is still room for improvement.

In recent years significant advances have been made in many applications by deep learning methods that use multiple processing layers and other smart designs to tackle problems that were previously difficult to solve. Deep learning achieved impressive results (often record breaking) in multiple classification benchmarks; this includes language processing tasks like text categorization. \parencite{LeCun15}

In this chapter some existing machine learning algorithms for text classification will be explained together with the mathematical and theoretical groundwork needed to understand them. It is impossible to mention all possible methods as this would not fit into this thesis. Instead only important and popular methods as well as methods used later as building blocks in the design and implementation will be described.

\section{Text Categorization}
\subsection{Definitions}
Broadly one can refer to text categorization as a way to label documents with a set of categories. As this, text categorization is a special topic within the more general field of text mining which tries to extract useful information from unstructured texts \parencite{Khan10}. An algorithm that can assign labels to some input data is called a classifier. In this document the term ``classifier'' will be used specifically for the classification of documents as described above.

It is important to note that while a category usually will have a specific meaning to the users of the classifier, the classification algorithms will usually treat the categories as purely symbolic and do not interpret them in any way. Also in general is is assumed that the set of categories is fixed, meaning the categories are known before a classifier is created \parencite{Sebastiani02}.

While this is a simple and easy description, further restrictions or additions are often done to this definition when required by a specific applications. Some of these will be discussed in the following section.

\subsection{Modes of Text Categorization}
\subsubsection{Multi-Class vs Multi-Label}
One important distinction for many applications is the question whether a single document can only belong to one category (single-label) or if it can belong to multiple categories at the same time (multi-label). The first case is also called the multi-class case since is describes the problem to classify documents when there is more than one class/category. In contrast to this, in the multi-label case, a classifier has to decide for each category separately whether a document belongs to the category or not -- independent of the other categories. A classifier that decides for a single category whether a document is part of the category is also called a ``binary classifier''.

Some classification algorithms can only be used to create binary classifiers. For the multi-label case this is not a problem as multiple classifiers can be trained independently for each category. For multi-class problems, other techniques can be used.

\parencite{Sebastiani02}\\

\subsubsection{Soft vs Hard-Labeling}
Another important distinction is whether the classification simply determines if a document belongs to a given category or not or if it provides some sort of rank between different categories to order them based on the likelihood that they should be assigned to the document. Often classification algorithms can provide a probability for each of the categories. The latter is suited especially well for semi-interactive classification, where the final decision is done by a user, but the system automatically suggests categories. \parencite{Sebastiani02}

\subsubsection{Hierarchical or Flat Categories}
In some applications, especially when the number of documents and categories is large, the categories will form a hierarchy (also called a taxonomy, see also \autoref{desc:taxonomy}), where membership in a child category also implies membership in the parent category (and so on). One possibility to implement such a classifier is to use a separate classifier for every branch in the category tree which decides to which of the child nodes the document belongs. Starting from the root node the classifiers can then be used to make branching decisions on each level of the tree \parencite{Sebastiani02}. Depending on the application the classifier might decide to assign only the label of a parent category if no suitable child category for the document was found.

\subsection{Structure of Text Classification Algorithms}
According to \textcite{Sebastiani02} the automated text categorization was first described by \textcite{Maron61}. This first attempt of text categorization used a probabilistic model and a so called bag-of-words approach that is still very popular today. The basic idea was that the individual words in a document are clues that can be used to predict the category of a document. By gathering statistics such as the frequency in which a word appears in a document and comparing this to the average values for a given category as well as the average values for all documents, a probability that the document belongs to the category can be calculated.

To understand the exact process better, one needs to look at the various phases that are shared by most text classification algorithms \parencite{Swoboda14}:

\subsubsection{Text Extraction}
The first step for many algorithms is the extraction of the actual natural language text from documents. The documents might be in a binary format with formatting and layout applied and they could be mixed with other forms of media such as images. The text could even come from audio sources, requiring speech recognition, before further processing can be done.

Text data is necessary because most algorithms require a stream of characters, words or sentences. More sophisticated algorithms can however make use of the additional information contained in the original source (e.g. giving higher priority to words that are formatted as a heading). But even in this cases some sort of preprocessing to transform the documents into a common format is almost always required.

\subsubsection{Feature Extraction}
\label{sec:feature_extraction}
One challenge in text categorization is the selection of suitable features that can be used for the classifier. Most classifiers work on numeric vectors, so it is important to transfer the input domain (like characters or words) into a vector space. A reverse transformation might be necessary to map the output of the classifier to a category.

There are many ways to represent textual data as vectors. Many text classifiers use high-dimensional vectors in which each word of a shared dictionary is represented as a unique dimension of the resulting input vector. In the simplest case a binary model is used where the presence of a word is denoted as a 1 in the vector component and all other values in the vector are 0. More sophisticated models use real numbers that also take into account how often a word occurs in the document and how often it occurs in the complete document corpus. An example is the $tf idf$ value (see also \autoref{sec:tfidf}), which is popular in many applications of text processing.

Depending on the application additional factors can be included to further determine the importance of a single term in the document. Also, instead of using words, the same can be done using phrases or smaller units such as $n$-grams.

\subsubsection{Feature Selection}
Some terms provide more information about the category of a text than others. Certain words like ``the'', ``a'' or ``by'' do not give any meaningful information about the document topic and can be removed without negative effects on the classification effectiveness. This is often done by using a black list or white list of words even before feature extraction takes place.

Other techniques reduce the term space by checking which of the features contribute most to classification effectiveness and remove the terms that contribute little to nothing. This term reduction techniques reduces the number of dimensions in the vectors and as a result not only improve resource usage but may also increase the classifier effectiveness and reduces the risk of overfitting. Overfitting will be explained further down.

\parencite{Sebastiani02}

\subsubsection{Classification}
All previous steps can be thought of as preprocessing steps for the input of the actual classification algorithm, which is applied in this final phase. Depending on the classifier used not all steps may be necessary or additional steps have to be added. However in the general case after preprocessing is done, a classifier is presented with the final feature vectors and will output one or more possible output categories, often paired with a number indicating the probability that a document belongs to the category in question.

In order to achieve this, most classifiers have to be trained beforehand with a set of already labeled documents. The training steps will be discussed further later in this chapter when talking about the various types of classifiers.

\subsubsection{Evaluation}
\label{sec:evaluation}
In order to compare the classification result with other classifiers, model validation is used. While this is not strictly part of the classification or training process itself, it is often the first step after training a new model. The details of evaluation methods will be discussed in \autoref{chap:Evaluation}, but the general idea is to present a new classifier with prelabeled data and to compare the prediction of the classifier with the expected output. These results can then be used to calculate various metrics and to compare them with other classifiers.

It is important that the data used to validate a classifier was not used during the training itself. Many classifiers are prone to overfitting to the training data. Overfitting occurs when a classifier adapts to strongly to the training data instead of generalizing from it. The result is a classifier that has good performance when presented with the training data, but bad performance when presented with new data. Using validation data it is easy to detect such problems -- \autoref{chap:Evaluation} will discuss this further.

\section{Classifiers}
There are many types of classifiers. This chapter will focus on two types which are especially popular in text categorization: \acrpl{svm} and neural networks. Within neural networks a special focus will be set on deep neural networks which are used in this thesis. As the name implies, classifiers are mainly concerned with the last step of a text classification algorithm (the actual ``Classification'').

\section{Support Vector Machines}
Support vector machines categorize a set of vectors into exactly two categories (or looking at it the other way two apply a single label to some of the documents). \acrpl{svm} are therefore a binary classifier. Multiple \acrpl{svm} can be combined to form a classifier capable of differentiating multiple categories.

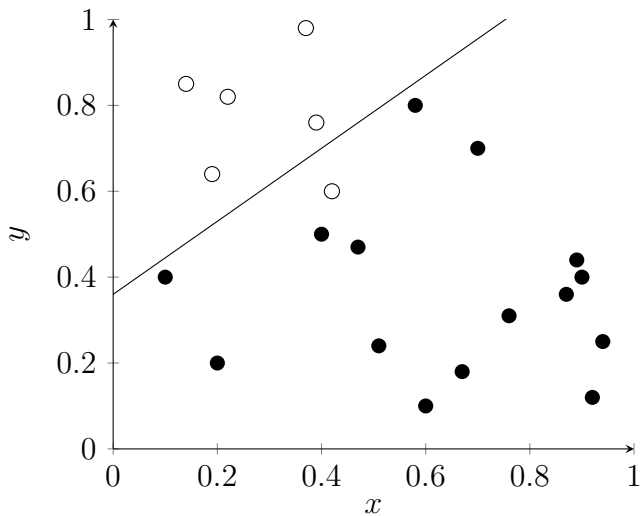
\begin{figure}[htbp]
	
\begin{tikzpicture}
  \begin{axis}[ 
    xlabel=$x$,
    ylabel={$y$},
    axis x line=bottom,
    axis y line=left,
    xmin=0,
    xmax=1,
    ymin=0,
    ymax=1
  ] 
    \addplot[mark=none] {0.85*x+0.36}; 
    \node[circle,fill,inner sep=2pt] at (axis cs:0.1,0.4) {};
    \node[circle,fill,inner sep=2pt] at (axis cs:0.7,0.7) {};
    \node[circle,fill,inner sep=2pt] at (axis cs:0.6,0.1) {};
    \node[circle,fill,inner sep=2pt] at (axis cs:0.4,0.5) {};
    \node[circle,fill,inner sep=2pt] at (axis cs:0.58,0.8) {};
    \node[circle,fill,inner sep=2pt] at (axis cs:0.9,0.4) {};
    \node[circle,fill,inner sep=2pt] at (axis cs:0.2,0.2) {};
    
    \node[circle,fill,inner sep=2pt] at (axis cs:0.51,0.24) {};
    \node[circle,fill,inner sep=2pt] at (axis cs:0.47,0.47) {};
    \node[circle,fill,inner sep=2pt] at (axis cs:0.87,0.36) {};
    \node[circle,fill,inner sep=2pt] at (axis cs:0.76,0.31) {};
    \node[circle,fill,inner sep=2pt] at (axis cs:0.94,0.25) {};
    \node[circle,fill,inner sep=2pt] at (axis cs:0.67,0.18) {};
    \node[circle,fill,inner sep=2pt] at (axis cs:0.89,0.44) {};
    \node[circle,fill,inner sep=2pt] at (axis cs:0.92,0.12) {};

    \node[circle,draw=black,inner sep=2pt] at (axis cs:0.42,0.6) {};
    \node[circle,draw=black,inner sep=2pt] at (axis cs:0.39,0.76) {};
    \node[circle,draw=black,inner sep=2pt] at (axis cs:0.14,0.85) {};
    \node[circle,draw=black,inner sep=2pt] at (axis cs:0.37,0.98) {};
    \node[circle,draw=black,inner sep=2pt] at (axis cs:0.19,0.64) {};
    \node[circle,draw=black,inner sep=2pt] at (axis cs:0.22,0.82) {};
  
      \end{axis}
\end{tikzpicture}
	\caption{Attempt at linear separation of samples in two dimensions}

	\label{fig:linsep}
\end{figure}

In order to separate the vectors, \acrpl{svm} use a simple but brilliant concept. All input data needs to be presented as vectors as described before. These vectors are usually highly multi-dimensional. A \acr{svm} uses a hyperplane to split the vector space into two parts. Vectors on one side of the hyperplane are deemed to belong to the labeled category, while the vectors on the other side are assigned to the unlabeled category. \autoref{fig:linsep} shows an example of a hyperplane in two-dimensional space (a line), that separates vectors (shown as points) into two groups. The vectors that belong to the target category are white, while all other vectors are black. The hyperplane does a pretty good job at separating the vectors into a white group and a black group, but it does not succeed completely. It is easy to see that there is no possible line which would separate the two groups perfectly. The correct terminology for this is that the two sets of vectors are not \emph{linearly separable}. \parencite{Cortes95}

In practice feature vectors have a lot more dimensions and it is easy to see that additional dimensions can help to make vectors linearly separable. If all points in the example would have an additional depth dimension and the hyperplane would also extend in this dimension, it is clear that additional dimensions never hurt separability, because the hyperplane can always ``choose'' to ignore the additional dimension. However, having additional dimensions gives vectors a higher chance to differentiate themselves from members of the opposite set and a suitable hyperplane can exploit this to provide a better classification. Of course this only helps if the additional dimensions do actually represent meaningful features that can be used for differentiation and do not simply add noise. If the natural dimensions of the vector are not sufficient to make the vectors linearly separable, there are ways to compute additional dimensions by projecting the vectors (so called kernel trick). \parencite{Cortes95}

\subsection{Choosing the Hyperplane}
Of course \acrpl{svm} need to be trained beforehand to split the term space in such a way that the predictions of the classifier are most effective. Based on a training set of prelabeled data, a logical choice is to compute a hyperplane that splits the data into the correct set and which also tries to maximize the margin between the hyperplane and the nearest vectors from both sets. This can be formulated as an optimization problem and treated accordingly. The vectors lying on the minimum margin are called the support vectors. \parencite{Cortes95}

\section{Neural Networks}
Neural Networks describe a family of models that can be used for various classification tasks or more general for the approximation of functions that often have many input variables. There are many types of neural networks that can work in very different ways.

Fundamental to all neural networks is the concept of small units (also called neurons or nodes) which are connected with each other. One of the earliest models is the perceptron model introduced by \textcite{Rosenblatt58} as a mathematical abstraction of a biological neuron. This model was later refined and simplified several times and is now the basis for most neural networks \parencite{rojas93}.

The next section will describe a modern perceptron model, similar to the model introduced by \textcite{Minsky69} \parencite[as cited in][]{rojas93}, in contrast to the original perceptron model described by \textcite{Rosenblatt58}. 

\subsection{Perceptron Model}
In the perceptron model a neuron receives one ore more input signals and produces an output signal. The output signal is computed using a simple function (the activation function $\sigma$) dependent on the input signals. The input a neuron receives can either come from external sources or from the output of other nodes. The generated output is forwarded to other neurons or is used as an output of the network itself. The input of neurons are weighted so that the incoming signal is either weakened or strengthened by the weight. Mathematically this is usually achieved by multiplying the real valued weight and input with each other. The weighted inputs of the node are then added together before applying the activation function. In this way some inputs influence the activation function of a neuron more than others. This is similar to the way biological neurons work, where neurons are also connected with each other and some connections are stronger than others. \parencite{rojas93,rey11}

\subsection{Activation Functions}
There are many possible activation functions. Most of them mimic the behavior of biological neurons of mostly ignoring their inputs until a certain threshold is reached at which point the neuron ``fires'' and sends a signal to connected neurons \parencite{rojas93}. There are other desirable properties of activation functions such as the restriction of the activation level, biological plausibility and differentiability. Often it is required to restrict the activation level to avoid that numbers in the network ``explode'' into infinity. Many popular activation functions map an infinite input domain of $\left[-\infty,+\infty\right]$ into a fixed range of $\left[0,1\right]$ or $\left[-1,1\right]$ to achieve this. Such a restriction is also biological plausible because biological neurons also have a limited activation potential. Differentiability on the other hand is required for many training algorithms of neural networks, as will be explained later.

\begin{figure}[htbp]
	
\begin{tikzpicture}
  \begin{axis}[ 
    xlabel=$x$,
    ylabel={$\sigma_{tanh}(x) = tanh(x)$},
    axis x line=bottom,
    axis y line=left,
    xmin=-3,
    xmax=3
  ] 
    \addplot {tanh(x)}; 
  \end{axis}
\end{tikzpicture}
	\caption{Plot of the hyperbolic tangent activation function}

	\label{fig:tanhplot}
\end{figure}
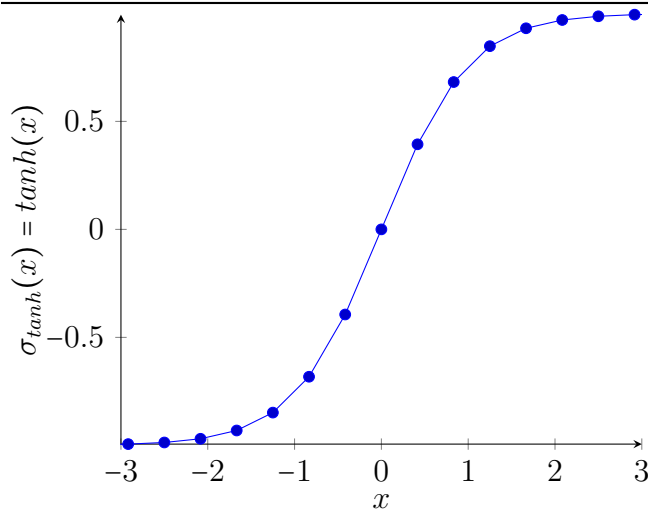

Mathematically these properties can be achieved by sigmoidal functions such as the hyperbolic tangent seen in \autoref{fig:tanhplot} or the logistic function (described later). There are however some popular activation functions that do not posses all these properties. One example are the linear activation functions often used within deep networks \parencite{LeCun15}. The concept behind these special activation functions will be discussed further in \autoref{sec:relu}.

\subsection{Layered Networks}
Most neural networks are organized in strict layers where a neuron only has inputs from neurons of a previous layer and presents its results to the next layer. Neurons are not connected within the same layer (some neural network architectures make exceptions to this rule). Since the first perceptron networks did not feature layers or connections between nodes, the basic layered variant is also called a multilayer perceptron network.

In a multilayer perceptron network, all nodes nodes in one layer are connected to the outputs of all nodes in the previous layer. For this reason, yet another name for these layers are ``Dense Layers''. In other network architecture the layers are only partially connected. Some of these will be discussed later. In general, network architectures where layers only depend on the output of a previous layer are called feed-forward networks.

\parencite{Krose1996}\\

\subsection{Vector Space Models}
Since Neural Networks work on numbers, the input and output domains need to be transformed to and from a numeric representation (see also \autoref{sec:feature_extraction}). The complete input and the complete output of a neural network is often represented as an input and an output vector. One of the challenges in the design of a neural network is therefore how to find a suitable vector model to present the input data to the network. For many input domains it is easy to find a possible representation, however as there are usually many ways to represent the data the challenge remains. One possible model to represent textual data as vectors is the td-idf model mentioned before.

\subsubsection{The tf-idf Measure}
\label{sec:tfidf}
One of the first successful vector space models for text was described by \textcite{Salton75} who used a combination of term frequency (how often does a term occur in the text) and (inverse) document frequency (how often does the term occur in a larger corpora) to determine how important a specific term within a document is. This is known as the tf-idf measure.

Given a corpus $D$ with $\norm{D}$ being the number of documents, the tf-idf weight $tfidf(t,d)$ for a term $t$ within a document $d$, is defined as:
\begin{align}
tfidf(t,d) = {tf}(t,d) \cdot log \frac{\norm{D}}{df(t,D)}
\end{align}

With $tf(t,d)$ being the number of times the term $t$ occurs in the document $d$ (often normalized by document length) and $df(t,D)$ being the number of documents (in the complete corpus) in which the term occurs.

When calculating this weight for all words of a defined vocabulary, it is possible to compute a vector for each document. The vector components are simply the individual weights $tfidf(t,d)$ for each term in the vocabulary.

\subsubsection{Sparse and Dense Vector Representations}
Using the tf-idf model with a large vocabulary will lead to vectors with many zero components, since a single document will only use a small part of the complete vocabulary.  

Taking this even further, one can imagine scenarios where an input text has to be analyzed word by word by a neural network. If a single word is represented as a unique dimension, then the input vector will only encode a single non-zero value in the vector. Such encodings where only one component contains a value and all other components are zero, are called one-hot encodings. Vectors where most of the components are zero are called sparse vectors.

Such sparse vectors are shown to deliver good results in many applications due to the inherent linear separability. The high dimensionality of sparse vectors does have disadvantages however, so that in some cases dense vector representations are preferred. Some of these disadvantages and applications will be discussed later.

\parencite{Joachims98}

\subsection{Training of Neural Networks}
\label{sec:training}
\subsubsection{Definitions}
Another similarity between almost all types of neural networks is that the variables defining it (especially the weights between neurons) are not completely set by humans or by a set of strict formulas but are most often learned automatically by training the network in some way. Training can be \emph{supervised} (e.g. with a set of already labeled test samples) or \emph{unsupervised} (test data is not labeled). Combinations of these are possible as well \parencite[e.g.][]{Erhan10}.

Is is important to understand how training of most neural networks works as this will lead us to the problem of vanishing gradients and the solution that deep learning offers. This requires some formal definitions first. The most common way to train neural networks is using backpropagation and gradient descent, which will be described now.

First one should remember that each neuron receives input from other neurons or from the network input itself. The input that a neuron $i$ receives from a neuron $j$ will be called $input_{i,j}$ (note that there is no differentiation made between input from other neurons or input from the network itself. The input of the network can in fact be modeled as being special neurons with no input that output the corresponding input value, so the differentiation is not important). This input is calculated by:
\begin{gather}
input_{i,j} = a_j w_{i,j}
\end{gather}
where $w_{i,j}$ refers to the weight applied to the connection between neuron $i$ and $j$ and $a_j$ is the output of the neuron $j$.

The weighted inputs of the various connections that enter a neuron are usually summed together to form the net input of the neuron. More formally one can call this the propagation function:
\begin{gather}
netinput_i = \sum_{j}{input_{i,j}} = \sum_{j}{a_j w_{i,j}}
\end{gather}

This input is then passed to an activation function $\sigma(netinput_i)$ which computes the activation level $a_i$ of the neuron. The actual neuron output is sometimes determined by a separate output function. However in most cases the output function is simply the identity function and the output is therefore the same as the activation level $a_i$ of the neuron \parencite{rey11}. In this thesis no differentiation will be made and the activation level of the neuron is assumed to be identical to the neuron output.

\parencite{rojas93,rey11}

\subsubsection{The Error Function}
The basic (supervised) learning principle for a neural network is to apply a known training sample to the neural network and check if the outputs are the same as the desired outputs (the known correct outputs for the training sample). If the values are not the same, the parameters of the network can be adjusted to correct the error. However the goal is not to optimize the network for a single training sample but rather optimize it for all training samples (and beyond). To compute the total error for all training samples in all output neurons an error function is defined:
\begin{gather}
E_{sq} = \frac{1}{2}\sum_{n}\sum_{i}{(y_{n,i} - a_{n,i})^2}
\end{gather}
$y_{n,i}$ is the desired output for neuron $i$ in the $n$-th training sample and $a_{n,i}$ is the actual output for the current network weights. By squaring the difference between these values and summing it one gets the overall squared error of the network. The factor $\frac{1}{2}$ is used to make differentiation easier later on. This error function is called the quadratic error function.\footnote{There are other alternative error functions. The choice of error function will be discussed later again.}

The goal of training a network is to minimize this error function. It is not possible to directly adjust the $a_{n,i}$ as they depend on the activation levels of the neurons in the previous layer and the weights connecting the neurons together. Of course the activation level of the neurons in the previous layer will again depend on the layer before that and so on. So the only variables that can be adjusted are the weights between the neurons. The result is an optimization problem where one tries to minimize the error $E$ in regards to all weights $w_{i,j}$ in the network.

\parencite{rojas93,rey11}

\subsubsection{Gradient Descent}
One possible way to tackle optimization problems is to do a gradient descent. The number of variables (weights) is too high to systematically search the minimum of the function. Instead of this some randomly initialized weights will be chosen as a starting point and then a stepwise descent in the direction where the function values are getting smaller will be done until a minimum is reached.

It is easy to know in which direction to descent if the derivative of the function and therefore its slope (or gradient) is known. Of course there is no guarantee that the function minimum reached is a global minimum of the error function or just a local one. In fact with many variables it is almost certain that one will only reach a local minimum \parencite{Choromanska14}. In practice this does not matter as long as the local minimum is ``good enough''. In addition, there are techniques that can be applied to avoid that the descent will get stuck in a ``bad'' minimum.

The only remaining problem is now how to calculate the derivative of the error function. This is where the ``back propagation'' part becomes relevant. The whole neural network can be viewed as a simple concatenation and composition of functions in which one starts with an input vector $X$ and then repeatedly applies functions (propagation function, activation function, ...) until the output of the network is reached (forward propagation). To calculate the derivative of the error function ($E'$) the reverse has to be done by starting at the output neurons and using the chain rule of differentiation repeatedly to get to the input of the network (backward propagation). Once the value for $E'$ is computed, it can be used for the descent.

\parencite{rojas93,rey11}

\subsection{Advanced Topics}
\subsubsection{Output Layer Activation and Softmax}
\label{sec:softmax}
For classification problems the output of a neural network is often expected to represent a probability in the range $\left[0,1\right]$. In the case of multi-label problems, every single output neuron can take a value between 0 and 1, describing the probability that a class should be assigned to an item. In the case of multi-class problems, where one item can only be assigned a single label, the probabilities of the output neurons are not independent however and should sum to 1.

To achieve this, the output layer of a neural network often uses a special activation function independent of what the rest of the network uses. For the multi-label case, the sigmoid or logistic activation function can be used as it ``squashes'' all input values into the range $\left[0,1\right]$:

\begin{gather}
\sigma_{sigmoid}(netinput_i) = \frac{1}{1+e^{-netinput_i}}
\end{gather}

The multi-class case it a bit more complicated since the various neurons are not independent of each other. A popular choice is the softmax activation function defined as:
\begin{gather}
\sigma_{softmax}(netinput_i) = \frac{e^{netinput_i}}{\sum_{j} e^{netinput_j}}
\end{gather}

In which the $netinput_j$ are the internal activation levels (before applying the activation function) of all the neurons $j$ in the output layer. The fact that the output of the function depends on the internal state of the other neurons ($j$) in the same layer, makes the softmax function special, as this was not the case for all other activation functions discussed so far.

The softmax function satisfies all given constraints: Since the exponential function can not be negative by definition, the fraction and therefore the softmax function itself can also not be negative. It is also easy to see that summing all outputs together will result in $1$. Therefore the range of each individual item is also in the range $\left[0,1\right]$. 

\parencite[][chapter 3]{Nielsen2015}

\subsubsection{Cross-Entropy Error Function}
\label{cross_entropy}
In \autoref{sec:training} the quadratic error function was introduced as an error function to minimize during neural network training. When using sigmoidal functions the plateaus near $y = 1$ and $y = 0$ causes the gradient at these points to be very small. This affects also the gradient of the quadratic error function in a way that when the expected output of the network is very far from the desired output, the gradient for the error function will be small as well. Since the weight update in training is directly proportional to the gradient of the error function, the training itself will be slow as well. This is of course not desired as a big difference between the actual and the expected output should also result in a big training step.

An error function that is specifically designed for this purpose is the log loss or binary cross-entropy error function:

\begin{gather}
  E_{bc} = -\frac{1}{N} \sum_n \sum_i \left[y_i \ln a_i + (1-y_i) \ln (1-a_i) \right].
\end{gather}

Here $y$ is the desired neural network output, $a$ is the actual output, $n$ is summing over all training samples and $i$ is summing over all output neurons.

For the sigmoid function, the derivative of this function can be simplified to:

\begin{gather} 
  E'_{bc} =  \frac{1}{N} \sum_n \sum_i x_i(a_i-y_i).
\end{gather}

As can be seen, the derivative is directly proportional to the difference between the expected output and the actual output, so that there is no training slowdown.

The complete proof is omitted in this thesis for brevity, but can be found in \textcite{Nielsen2015}.

The above error function is valid, when every output neuron is an independent binary classifier, like in the multi-label case. For the multi-class case, where only a single label is assigned to each sample, the output neurons are not independent of each other. The softmax activation function (see \autoref{sec:softmax}) is usually used for such cases and instead of the binary cross-entropy error function, the categorical cross-entropy should be used instead:

\begin{gather}
  E_{cc} = -\frac{1}{n} \sum_n \sum_j \left[y_j \ln a_j \right].
\end{gather}

In case there are only two classes, the result will be the same as the binary cross entropy.

To summarize, it is recommended to pair softmax with the categorical cross entropy error function for multi-class problems, and the logistic function with the binary cross entropy error function for multi-label problems.

\parencite{Nielsen2015}.

\subsubsection{Gradient Descent Optimization}
\label{sec:adam}
In a previous section the training using gradient descent was described. To recapitulate, on every training step the derivative (with regards to the weights) of the error function is calculated to get the current gradient. The gradient descent will then take one step in the direction of this gradient (descending). Since the error function is dependent on the weights of the network, descending means adjusting these weights accordingly. After setting the new weights, the process is repeated until the algorithm reaches a (local) minima of the error function or the process it interrupted otherwise. \parencite{rey11}

One important consideration that was not discussed so far is the step size (or learning rate) that is used in each iteration. Since the error function can be quite complex, setting a step size which is too high can easily lead to problems such as skipping over the minima completely or not being able to reach the minima because of oscillating effects. Setting the step size too small on the other hand can make training slow. Another problem is that the gradient descent might get stuck in a bad local minima, where the performance of the network is far from optimal and the gradient descent is unable to ``escape''\footnote{There is evidence however, that the problem of bad local minima is actually quite low for practical applications \parencite{Choromanska14}}. Several optimizations have been proposed to optimize the basic gradient descent. \parencite{rey11}

\figgraph{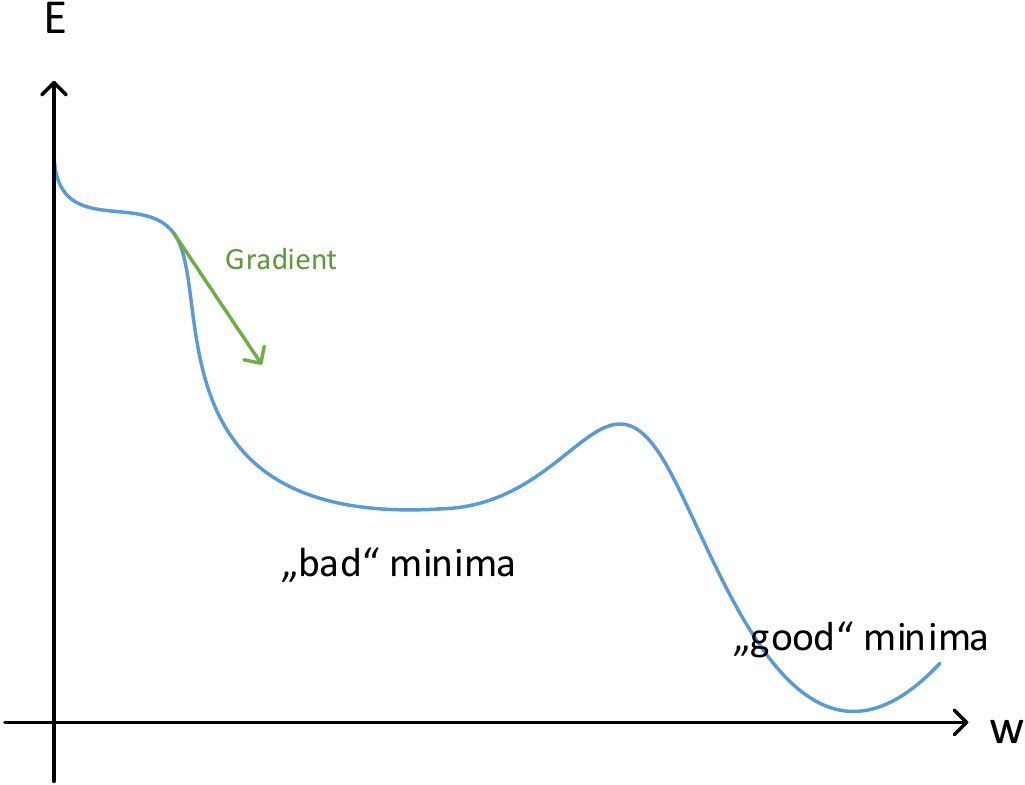}{Schematic gradient descent for a single weight $w$ showing error function $E$ and a gradient}

One of the easiest modifications is to set a variable learning rate which is high at the beginning and then slowly decreases. The idea is that initially one wants to rapidly adjust the weights to find regions where the classifier performs good. Then with time the learning rate is decreased so that local minima can be reached better (fine tuning). \parencite{rey11}

This simple modification can help reduce the training time but it still requires a manual determination of the best (starting) learning rate and it doesn't solve the problem of local minima.

An additional modification is therefore to add a momentum term to the update equation. With a momentum term, the learning rate starts slow and then slowly increases as long as the gradient points to a similar direction -- similar to a ball catching momentum when running down a hill. The effect of this is that small local minima can be escaped from when enough momentum has been reached. Another advantage is that certain oscillations are canceled out since the algorithm prefers directions in which the error decrease is stable (even if low) and penalizes directions in which the gradient is fluctuating (these are canceled out by the momentum term). Also the training process is faster and since the learning rate adapts to the current situation, it is easier to set the initial parameters. \parencite{rey11}

It is possible to further optimize this by making the step size even more adaptive. On example of this is the Adam algorithm, which usually performs very well without the need to adjust any parameters of the algorithm. \parencite{Kingma14}

\section{Deep Learning}
Neural networks made a big step forward in recent years by employing so called deep learning methods. Neural networks with deep learning uses multiple layers of simple modules. Starting from the raw data, each layer provides a slightly more abstract version of its input data to the subsequent layer. For image processing applications the first layer could receive the raw color values of an image and detect simple edges or other basic features in the image. The next layer could detect more abstract shapes as a combination of various edges. Subsequent layers could then combine these shapes to detect more and more complex objects. These layers do not need to be designed explicitly, but are learned automatically using the deep-learning techniques. \parencite{LeCun15}

Deep learning has been very successful with this approach and can solve many difficult tasks. One extreme example of a neural network with over 150 layers recently won several image recognition competitions and the authors experiment with even larger networks well above 1000 layers \parencite{HeZRS15}. Other examples where deep learning outperforms existing technologies are speech recognition \parencite{Hinton12} and many scientific applications \parencite{LeCun15}.

Even more interesting for this thesis however are the advances made by deep learning in various natural language processing tasks, such as sentiment analysis or text categorization. Similar to the way that deep learning can start with simple pixel values and then use layers to build abstractions, \textcite{ZhangL15} show that deep neural networks perform well in these tasks even if they are presented only with character level data. So without any prior understanding of syntactic or grammatical structures such as words or sentences deep neural networks can be trained to extract semantic information from these texts.

It should be noted that while the focus of this section was mainly on neural networks, deep learning is not specific to neural networks as it only describes the general technique of using multiple processing layers to achieve high level abstractions.

\subsection{The Vanishing Gradient Problem}
Unfortunately it is not quite as easy as adding additional layers to a regular network. Up until some years ago the gradient descent back propagation algorithm used to train neural networks did not scale good with additional layers due to the ``vanishing gradient'' problem. The effect of repeatedly applying the chain rule of differentiation while back-propagating the error through the network is that the initial error term is repeatedly multiplied with small values resulting from the gradients of the activation functions. This means that the front layers of the neural network train much slower than the back layers. Even for a small number of layers the cost quickly becomes prohibitive. Changing the activation function so that the derivatives have larger numbers on the other hand can lead to an ``exploding gradient'' problem. \parencite{Hochreiter01}

Since the first beginnings of neural networks, computing power has increased immensely. Computers are now thousands of times faster than they were 20 or 30 years ago. Massively parallel processing units have become mainstream thanks to \acrpl{gpu}. This has certainly helped in some part to overcome the vanishing gradient problem \parencite[e.g.][]{Ciresan2010}. Equally important however has been the research of new techniques on the structure and functionality of the various layers of deep networks \parencite{LeCun15}.

\subsection{Multi-Level Hierarchy and Auto Encoder}
One way to overcome the problem is to initialize the network with unsupervised training sessions and only use supervised training with back propagation in the final stages of training to fine tune the network. Auto encoders do this by training the network to reproduce the original input and at the same time forcing the network to abstract from the original representation and learning higher level features.
\figgraph{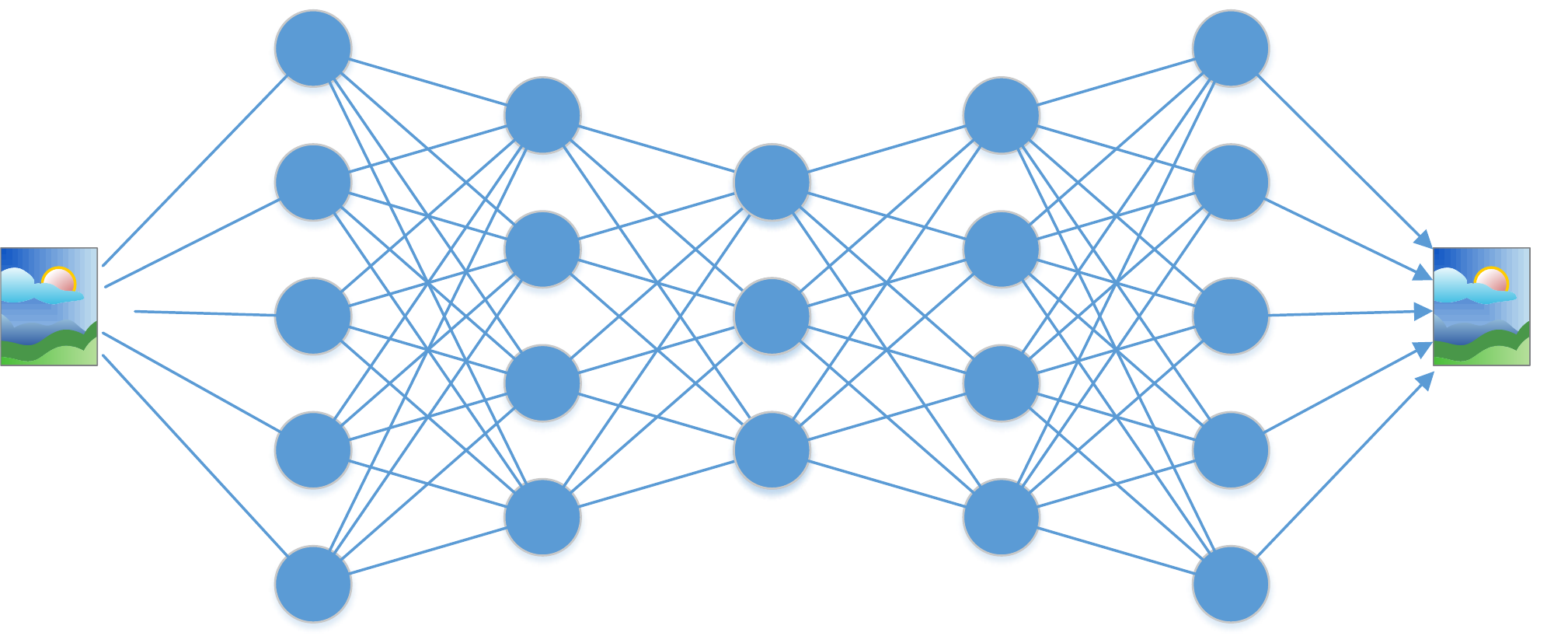}{Auto-Encoder}
The layers of the auto encoder are mirrored and the weights are the same for the corresponding layers of both sides. \autoref{fig:autoencoder} shows an example of such a network. To force the network to abstract the input and not simply use the original image and passing it on to the end of the network, the number of nodes can be decreased in the inner layers. Also often the original input is distorted (noise added) and the network is trained to restore the original input. This forces it to learn higher level concepts (e.g. lines instead of pixel in the case of an image) in order to be able to remove the noise. \parencite{Erhan10,Bengio09}

\subsection{Long Short-Term-Memory}
An alternative approach is the \acr{lstm} model \parencite{Hochreiter97}. In such networks the neurons of a network are partially or completely replaced by smarter \acr{lstm} blocks. These blocks have the ability to ``remember'' values for arbitrary amounts of time.

While all recurrent neural networks are generally capable of remembering recent events, they suffer even stronger from vanishing or exploding gradients \parencite{Hochreiter01}. Most of these networks use the \acr{bptt} algorithm that works like the normal backpropagation algorithm but also propagates backward in time. The network is basically unfolded various time so that it behaves like a network with many layers. The practical result of this is that after few time steps the error gradient either explodes or vanishes so that only very recent events can be used in training.

\acr{lstm} solves this problem by having units with explicit ``gates'' that control when a value is stored, cleared or reproduced. Coupled with an adjusted learning algorithm, the error signal remains constant during training so that long running dependencies can be learned. \parencite{Hochreiter97}

\subsection{Gated Recurrent Units}
Recently another variant of recurrent units have been proposed by \textcite{Cho14}. The new \acr{gru} is similar to a \acr{lstm} unit, but simpler. Most importantly a \acr{gru} will always output its current hidden state while the output of a \acr{lstm} node depends on a special output gate. Empirical evidence suggests that even though \acrpl{gru} are simpler and therefore more limited, they perform comparable or even slightly better than \acr{lstm} nodes in many tasks \parencite{Chung14}. Having a simpler architecture also means the nodes can be computed faster which reduces training time.

\subsection{Convolutional Neural Networks}
Yet another popular deep learning method are \acrpl{cnn} \parencite{Lecun95}. Instead of using fully connected perceptron layers, \acrlongpl{cnn} use filters which are slid over the input range to produce output values.

For example, to train a neural network using fully connected perceptron layers to detect rectangles in an image of 1000 by 1000 pixels, every neuron in the first layer would have 1000000 different input weights to train. Not only does this lead to a high probability of overfitting, but the resulting network is also not invariant with respect to translation, rotation, scaling or distortions. If the training data did not include rectangles in the upper left part of the image, then the resulting network will not be able to recognize them.

In a \acrlong{cnn} the same weights are applied to all parts of an image. A single filter will receive only a part of the input data - for example a region of 32 by 32 pixels. This filter is then applied step by step over the whole image usually with some overlap. The resulting network will be able to recognize features such as rectangles independent of their position in the image.

By using several filters with different sizes and by rotating the filter, the network can also be made invariant to other image transformations. Since there are a lot less weights to train, the training will also proceed faster and with less probability of overfitting.

This on the other hand opens the possibility to add additional layers to improve network performance further. Combining this technique with the auto-encoder approach discussed previously is a popular choice.

\subsubsection{Pooling Layers}
\label{sec:pooling}
\acrpl{cnn} are often used in conjunction with pooling layers (also called subsampling layers). Pooling layers reduce the number of outputs from a previous layer while trying to preserve the essential data that the layer provides for the classification task. A single node in a pooling layer takes the output from a range of neurons in the previous layer and combines them into a single output \parencite{Scherer10}.

For image classification tasks, the nodes typically reduce the output of a rectangular region in the input data. For example a 4x4 rectangle could be used with an overlap of 2 pixels between nodes in each dimension. Like this, the pooling layer would have only a quarter (half in every dimension) of the outputs than the original layer.

To preserve the important information of the previous layer, a common approach is to use the $max$ function as to only preserve the input with the highest signal \parencite{Scherer10}. This variant is also called max-pooling. For data that can be interpreted as representing a sequence, like video or text, this is also called max-over-time pooling, emphasizing that the ``compression'' is done along the time axis. \parencite{Kim14}

\subsection{Rectified Linear Activation Functions}
\label{sec:relu}
\textcite{glorot11} demonstrated that neural networks and especially deep neural networks can benefit from linear activation functions and proposed \acrfullpl{relu}. \Acrpl{relu} are special in that they do not saturate like sigmoidal functions that reach a plateau when the internal activation is high enough. They are also special in that they force the output to be zero if the internal activation becomes negative.

\begin{figure}[htbp]
	
\begin{tikzpicture}
  \begin{axis}[ 
    xlabel=$netinput$,
    ylabel={$\sigma_{relu}(netinput) = max(0,netinput)$},
    axis x line=bottom,
    axis y line=left,
    xmin=-3,
    xmax=3
  ] 
    \addplot {max(0,x)}; 
  \end{axis}
\end{tikzpicture}
	\caption{Plot of the ReLU activation function}

	\label{fig:reluplot}
\end{figure}
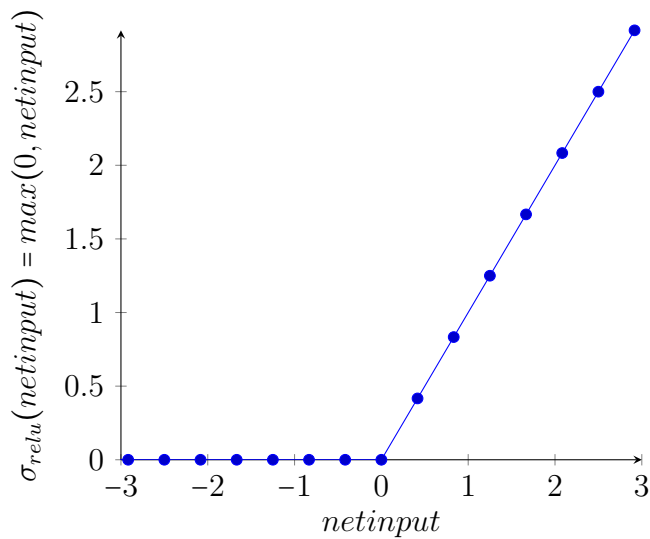

Several variants of the basic \acr{relu} function exist. The Leaky \acr{relu} function also has a linear activation for negative $netinput$, reduced however by a constant factor (e.g. $netinput \cdot c$, for $netinput < 0$). For Parametric \acrpl{relu} the factor is variable and is trained using backpropagation along with the weights of the network. \parencite{Xu2015}

\textcite{glorot11} mentions increased sparsity as the main factor why \acrpl{relu} perform so well, this is however disputed \parencite{Xu2015}. In addition, the fact that the functions do not saturate helps to avoid vanishing gradients. These advantages enable faster training speeds and makes \acr{relu} a popular choice in recent years \parencite{LeCun15}.

\subsection{Dense Vector Representations and Word Embeddings}
As was discussed before, regular neural networks often perform better with sparse vector presentations because they promote linear separability. Deep neural networks are not as dependent on linear separability because they are capable of learning more complex relationships due to the higher number of layers. Using dense vectors in this case provide a number of advantages. First of all sparse vector representations for text models often have tens of thousands of dimensions (every dimension representing a single word). This means that a classifier needs to learn a lot of weights. This in turn can easily lead to overfitting. With dense vectors the number of dimensions can be a lot smaller (often in the range of 50 to 1000) and the risk of overfitting is reduced. In addition less parameters to train means less computations to perform which results in higher training speed. \parencite{Bengio03,Goldberg15}

Another advantage of the dense encoding is that words with similar semantic meanings can be closer in the vector space than words that do not share similarity. There are various techniques to compute such vectors. The most common ways are the \acr{cbow} technique, the \acr{sg} technique and dimensionality reduction on word co-occurrence matrices. The first two are implemented by the word2vec project \parencite{Mikolov13} while the last is implemented in the GloVe project \parencite{pennington14}. Together these techniques are called word embeddings.

Regardless of which technique is used, the resulting word vectors $v$ show remarkable semantic properties when trained on a large document corpus. For example the distance between the ``queen'' and the ``king'' vector is similar to the distance of the ``woman'' and ``man'' vectors.\figgraph{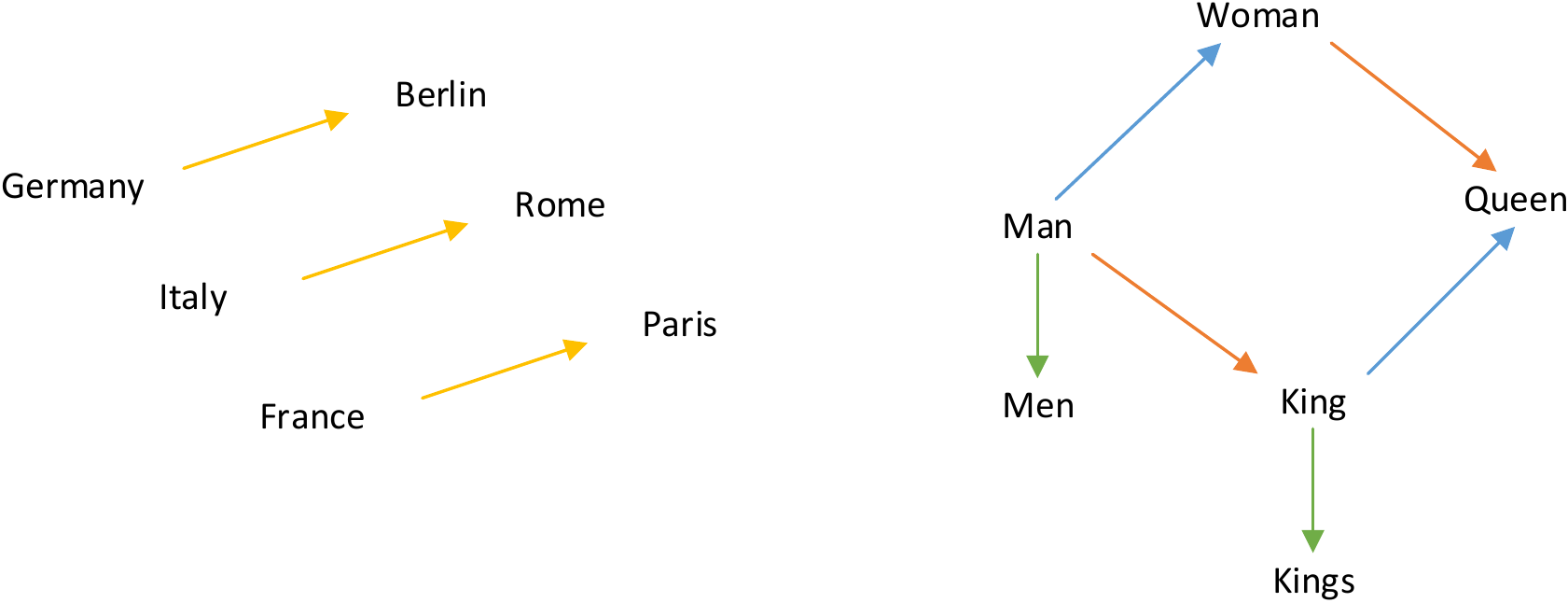}{Word analogies and relationships encoding gender, plural and royalness} Knowing this the vectors be used for calculations (e.g. $v_{queen} - v_{king} + v_{man} \approx v_{woman}$) or for answering analogy questions (e.g. Rome is to Italy as Berlin is to \ldots ?). \parencite{mikolov13b}

\figgraph{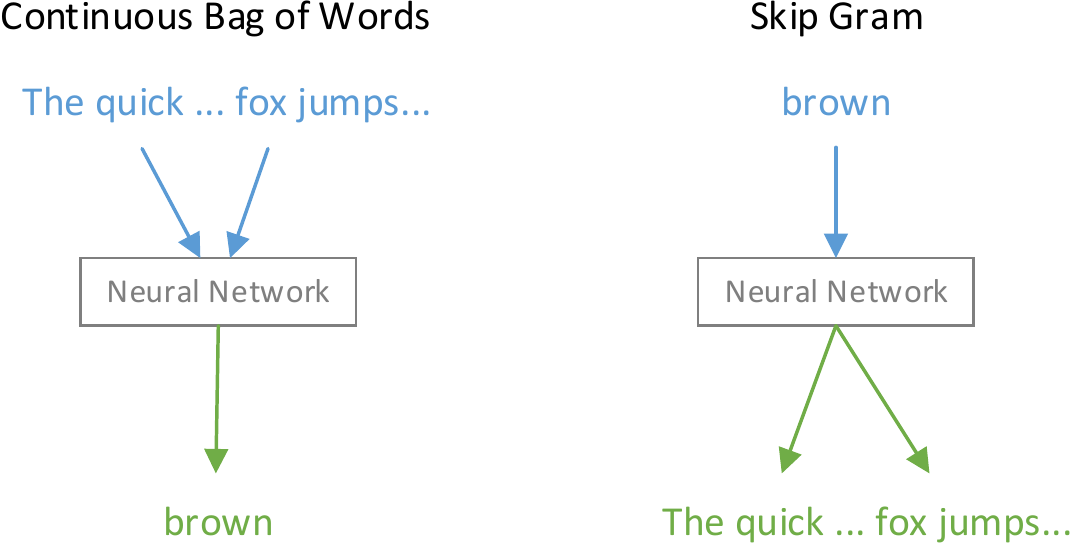}{Continuous Bag of Words and Skip Gram compared}

\subsubsection{Continuous Bag of Words}
The \acr{cbow} technique uses a neural network to learn word embeddings. For learning, a large text corpus is scanned word by word. The current word is called the \emph{target word} and the words that precede or succeed the target word in the text are called the \emph{context words}. The number of context words is limited by a window size. The neural network is now trained to predict the target word (using a one-hot sparse vector representation) given only the context words (also as sparse vectors). The neural network has a hidden layer with a fixed size that it will use for its predictions. When the training is finished the hidden layer is used to compute the final dense representation of words. A simple way would be to use the neuron output of the hidden layer for each word as the word vector. The actual process is a bit more complex and combines both an input vector and an output vector derived from the weight matrix to and from the hidden layer. Additional optimizations are needed to make this process fast enough for large corpora. \textcite{rong14} does a good job explaining all details of the process.

\subsubsection{Skip Gram}
The \acr{sg} technique is very similar to the \acr{cbow} technique, but reverses the role of target word and context words. The target word is the input of a neural network which is then trained to predict the context words. The dense vector representation is again obtained from hidden layer in the network. \parencite{rong14} 

\subsubsection{Dimensionality Reduction on the Word Co-Occurrence Matrix (GloVe)}
The algorithm used by GloVe \parencite{pennington14} is very different from \acr{cbow} and \acr{sg} in the sense that GloVe does not use a neural network to compute word embeddings. The starting point for GloVe is a word word co-occurrence matrix in which both the rows and columns represent the words in the target vocabulary. The cell value denotes with what probability the word in the row occurs together with the word in the column.

In theory, one could already use the rows of this matrix as word vectors. Similar words could be expected to have similar co-occurrence probabilities. However the resulting word vectors would have a large number of dimensions containing mostly noise that does not add much value to discriminate them from other word vectors. For example when trying to differentiate between ``ice'' and ``steam'' it is not helpful to find that both words frequently occur together with the word ``water''. The relationship with words like ``solid'' or ``gas'' are much more discriminative in nature.

The goal is therefore to reduce the number of dimensions but still retaining most of the information expressed in the vector. There are several ways to do this. GloVe tries to compute a factorization of the logarithmic probabilities in the matrix. To do this the factorization along with some optimization parameters is expressed in terms of an error function which is then trained using gradient descent. The details of the this training process are quite complex to explain here. The interested reader is encouraged to read the original GloVe paper \parencite{pennington14}.

\subsection{Dropout}
\textcite{LeCun15} mention Dropout \parencite{Srivastava14} as yet another fundamental enhancement that helped deep learning to become successful. Dropout aids in avoiding network overfitting by temporarily dropping random neurons from the network. Because of this, neurons in higher layers can no longer rely on individual neurons and tend to generalize better.

In addition to reducing overfitting, Dropout can also lead to better performance of the network. Due to Dropout every training step will only affect a subnet of the complete network. The effect is similar to that of a Committee classifier where multiple networks are trained separately and then combined to form a stronger classifier. \parencite{Srivastava14}

\section{Integrated Systems for Document Categorization}
Having a powerful classifier is important, but not sufficient, to do classifications in a production environment. The classifier has to be integrated into a system capable of classifying documents on demand and possibly allowing the creation of new classifiers based on training data.

Most of existing systems that can be found by researching fall within two categories. There are some tools that provide content classification as a part of another product (such as document scanning and archiving solutions). On the other side are products that offer machine learning and classification in general without being specifically tied to document categorization.

The following sections will list two products which are examples of these two sides.

\subsection{ABBYY FlexiCapture and Smart Classifier}
ABBYY FlexiCapture\footnote{\url{https://www.abbyy.com/flexicapture/}, last retrieved on June 6th, 2016} advertizes an ``intelligent self-learning classification''. For developers the same company also offers the ABBYY Smart Classifier which seems to offer similar features as a standalone library that can be integrated into other products.

Unfortunately there is little information on how the software works, how the performance is and how the classification can be integrated into existing solutions. There is however some evidence that suggests that the offered library is based on the .NET Framework, which would make integration into non-windows system or using languages other than those based on the .NET runtime difficult without developing suitable wrappers.

When the website was last retrieved there was also no mention on what the cost of the library is.

\subsection{Azure Machine Learning Web Service}
Azure is Microsoft's cloud computing offering. With the Azure Machine Learning Web Service\footnote{\url{http://azure.microsoft.com/en-us/documentation/services/machine-learning/}, last retrieved on June 6th, 2016} it is possible to define and train classifiers which run on the cloud platform. Similar offerings exist from competitors such as Amazon AWS\footnote{Amazon Machine Learning API, \url{http://aws.amazon.com/machine-learning/}, retrieved 20th of June 2016} and Google Cloud Platform\footnote{Google Prediction API, \url{http://cloud.google.com/prediction/}, retrieved 20th of June 2016}.

With Azure Machine Learning one has to first define a model using one of several algorithms provided such as \acrpl{svm} or Neural Networks. Some of these algorithms allow customization. For neural networks it is possible for example to specify the exact architecture using some basic building blocks (e.g. layers).

After the model is defined and the algorithm was chosen, training can commence. For document classification this would mean providing a prelabeled set of documents as a training set. When the user is satisfied with the classifier performance, the training model can be converted into a predictive model which is accessible using standard web service methods for integration into custom solutions.

The big advantage of this model is that it is highly customizable and can be used not only for document classification but for many other tasks. There are graphical tools available to quickly define models and to train them in the cloud. This also means that there is no special hardware required as everything is provided in the cloud. Using web services means that the resulting classifiers can be integrated into other products using almost any programming language.

The disadvantages are however that the flexibility causes increased complexity. Building models for the specific task of document categorization requires specific knowledge of machine learning and a lot of practice to find a suitable model. Integration of the web service into existing solutions requires additional code to transform the conceptional model into the specific input format required by the web service. It is therefore similar in complexity to using a machine learning library directly.

Using cloud resources, while convenient, can also lead to additional problems. Data protection laws could for example prohibit the transfer of personal documents to a cloud service. Access to the cloud service requires permanent internet access which can be problematic in some scenarios. Lastly the use of the Azure is not free of charge, so pricing also has to be considered.

All in all however the Azure Machine Learning Web Service is an interesting offering and could be used as the backend for a custom web service that is specialized for document classification. Using cloud services for document categorization is also a topic of ongoing work at the University of Hagen as can be seen in \autoref{sec:related_work}.

\subsection{Using REST Services for Interoperability}
Common to all three cloud offerings by Microsoft, Amazon and Google is the use of \acr{rest} services as an \acr{api} to train or publish classifiers. \acr{rest} is an architectural style for distributed systems that is used on the world wide web \parencite{Fielding2000}.

RESTful services (or simply REST services) are web services that are based on the REST principles. Since \acr{rest} uses HTTP as a protocol and adds little complexity, it is easy to consume such services from all modern programming languages. The support from all major cloud vendors indicates that it is the current protocol of choice for interoperable and distributed web based \acrpl{api}.

\section{Related Work}
\label{sec:related_work}
This thesis is one of various works at the University of Hagen that are related to the RAGE project. RAGE (Realizing an Applied Gaming Eco-system) is a project funded by the European Union that \enquote{aims to develop, transform and enrich advanced technologies from the leisure games industry into self-contained gaming assets that support game studios at developing applied games easier, faster and more cost-effectively.}\footnote{Quoted from \url{http://rageproject.eu/}, last retrieved on June 25th 2016}.

One part of RAGE is the creation of an ecosystem portal for the exchange of gaming related resources and documents. It is envisioned to integrate the work presented in this thesis into the portal so that documents can be automatically categorized after upload into the system. The integration of this and other algorithms into the portal is planned as part of a thesis by Oleksiy Lebsack \parencite{Lebsack16}. This work will include a modernization of the portal to allow the usage of external classifier services.

One algorithm that already exists and is integrated into the portal is a classifier based on support vector machines and a bag-of-words approach done by Tobias Swoboda \parencite{Swoboda14}.

Another alternative currently evaluated by Michael Hoffmann will use Named Entity Recognition which could be combined with \acrpl{svm} to form yet another classifier \parencite{Hoffmann16}. Comparing all these different approaches or possibly combining them would be an interesting topic for future work in this area.

Also looking into the future is the work from Wolfang Müller \parencite{Mueller16} that looks at the role of cloud services for document categorization, more specifically on how they can be evaluated and what the decision criteria for one or the other cloud provider could be.

\section{Remaining Challenges}
The field of text classification is big and there is a broad range of technologies to choose from when implementing a text classification algorithm. As has been described, deep learning recently brought new innovations and there is evidence that it can often outperform existing classifier technology such as \acr{svm}. There are however some unique challenges in the application of scientific document categorization which are not thoroughly addressed by existing papers.

Most of the research on modern classification algorithms is quite theoretical and they do not mention if the algorithms are employed in any production level system. The presented results indicate on the contrary that they are used for one-time classification of known datasets in order to compare the pure classification performance to other algorithms. While this is a legitimate approach, it is also important to see how these approaches work in practice and how they can be integrated into a complete system to aid users in maintaining large document corpora.

More specifically it needs to be investigated how these modern algorithms and especially deep learning could be integrated into the RAGE ecosystem portal.

The documents that will be uploaded to the RAGE portal will be mostly of scientific nature, such as research papers or articles on applied gaming and related sciences. This brings us to another problem. One of the advantages of machine learning is that most of the algorithms are general enough to apply them to a broad area of different tasks. Some algorithms may however perform better on one task or the other. For example sentiment analysis is the popular task of subtracting subjective information (the sentiment) out of text or other media. An example could be trying to determine if a movie review is positive or negative, only by looking at the text of the review. Many of the algorithms used in sentiment analysis can also be applied to text categorization in general possibly by changing only a few parameters. But of course to compare the performance of algorithms directly they have to be applied to the specific task.

Even within the same task it is important that all algorithms are tested on the same data set (document collection). Algorithms can perform differently on different data sets. More specifically, many popular datasets (such as the Reuters news corpus or the 20newsgroup data set to be discussed in \autoref{chap:Evaluation}) have texts which are much shorter than a typical research paper \parencite{Lewis04,Lang07}. This difference in size could affect the runtime or performance of algorithms. For neural networks this could mean requiring a higher amount of neurons in some architectures while classic classifier such as a \acr{svm} using only tf-idf values of the document might be affected less. One challenge will therefore be to find a suitable dataset to test the solution in a realistic scenario and to see how deep learning performs under these circumstances. This will be addressed further in the evaluation chapter.

Finding the right architecture and a good set of hyper parameters for a specific task is a challenge of its own. The architecture and hyper parameters should be generic enough so that they can support all required use cases, but still be optimized in order to achieve the highest performance.

Some authors provide good performance with their models but use highly specialized components or algorithms which are not easy to replicate and will be hard to maintain without special knowledge when integrated into a system.

The goal for the next chapters is therefore to address the following questions:

\begin{itemize}
  \item Can the algorithms (specifically deep learning) be used in practical systems and what are the remaining challenges?
  \item Are the algorithms suitable for the specific task of document categorization? Do they provide an advantage over other more traditional algorithms?
  \item Are they easy to implement using standard libraries and programming languages or do they require deep knowledge and are therefore hard to maintain?
  \item Can the results obtained in theoretical research be reproduced on a document collection consisting of scientific documents?
  \item How can the different approaches be compared to pick the one that provides the best performance?
\end{itemize}

In the next chapters a system will be described and implemented that tries to answer these questions.

%% file: chapters/03.algorithm.tex

\chapter{Solution Concept}
\label{chap:Algorithm}
To address the problems mentioned in the previous chapters some of the existing deep learning technologies will be adapted to the specific needs of scientific document categorization. Then the resulting classifier will be taken and a general software design be proposed that can be used to easily embed document categorization into other software. In the following chapters the prototypical implementation will then be described in detail and later evaluated by comparing it to another classifier. During the evaluation the general question on how to compare different classifiers will be discussed as well. 

Various different network architectures such as recurrent neural networks and \acrpl{cnn} with many different parameters have been tested to find out which one achieves the best performance. At the end a \acr{cnn} similar to that proposed in \textcite{Kim14} provided the best empirical results (see \autoref{fig:cnn}).

The original solution is enhanced with advice taken from \textcite{ZhangWallace15}, who conducted multiple experiments to determine the effectiveness of various convolutional network configurations. Also considered in the design are new research, personal experiences and the specific needs outlined in \autoref{chap:Introduction}.

One example of these specific needs is multi-class classification as the original paper deals mostly with sentiment analysis and only two possible outcomes. The texts (such as reviews) tested in the original papers are also significantly shorter than the usual scientific paper and the vocabulary is different.

Taking this into account, experiments were performed to find the best network configuration. Regardless of this even with the original configuration used in \textcite{Kim14} with minimum modifications for multi-class output good results were achieved, indicating that \acrpl{cnn} are generally capable of adapting easily to different problems. Earlier experiments with recurrent neural networks (e.g. using \acrpl{gru} or \acr{lstm}) proved to be much more difficult and required more fine tuning.

The following section will describe the final network being used in more detail.

\section{Basic Structure of the Classifier} 
\figgraph{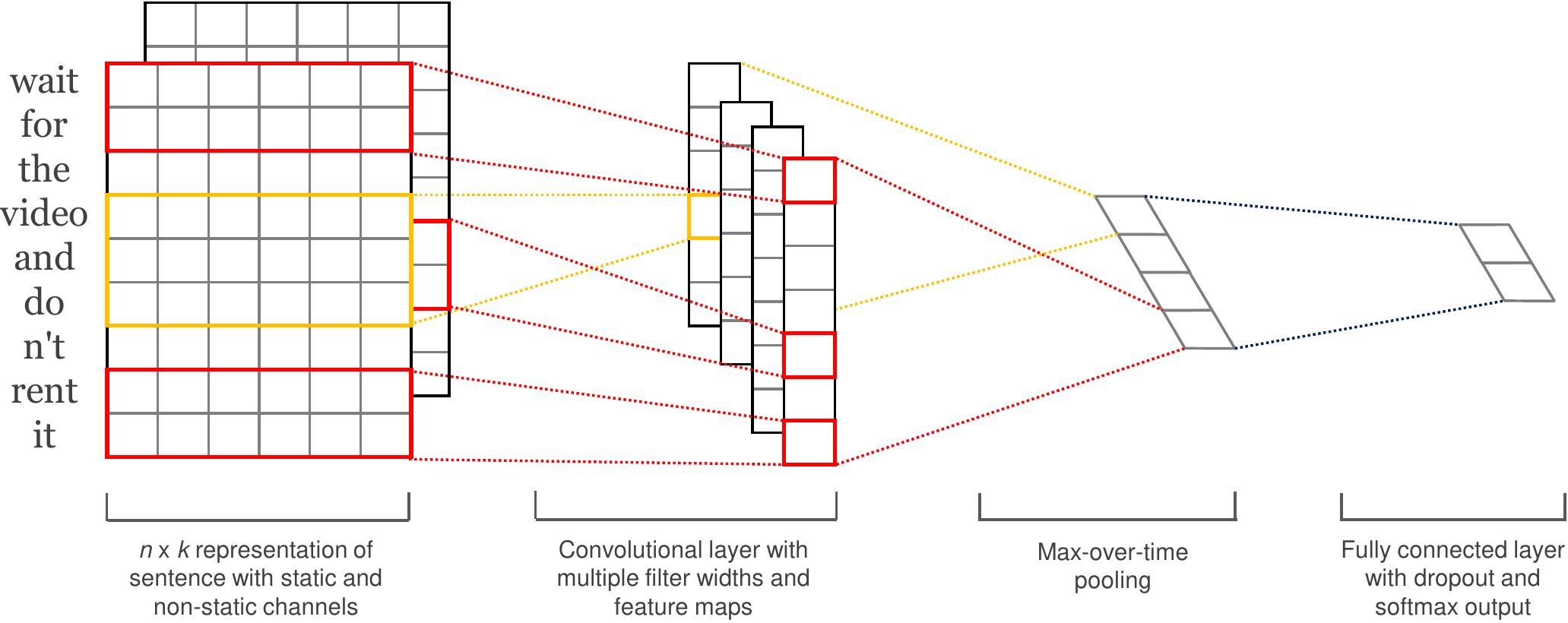}{\acrlong{cnn} for text classification. Graphic adapted from \textcite{Kim14} with permission of author}
\subsection{Input}
The input of the neural network is a list of word vectors based on word embeddings like they are trained by word2vec. The list is simply extracted from the first $t$ words in the documents and then transformed using a word embedding model that is given to the classifier. The model can either be a pretrained model as provided on various websites or a custom trained one (the difference will be evaluated in \autoref{chap:Evaluation}).

\subsection{Convolutional Layer}
As the second general layer, multiple convolutional layers with different filter lengths are used in parallel. For every filter length there are multiple filters to detect various features in the input stream. The filter length determines the number of word vectors that are fed into a single feature filter at the same time. A step size of one word is used, so that the number of output vectors for a single feature vector is the same as the number of input (word) vectors except for a small difference for padding reasons. For example with a filter length of 3, the filters would first be fed the word vectors $v_0$, $v_1$ and $v_2$ ($v_t$ being all word vectors in a single document). In the second step, the word vectors $v_1$, $v_2$ and $v_3$ would be fed to the same filters yielding a new vector.

It should be noted that unlike convolutional filters typically used in 2d images, where the filter moves over the images in two dimensions using a filter of e.g. 3x3 pixels, the filter ``height'' used by the filters in this case always encompasses the whole word vector. So for example with a word vector of 300 dimensions and a filter width of 3, the actual filter will get an input of size 3x300. Unlike images where the position of pixels in both x and y dimension are important, the order of the different dimensions in a word vector do not have any significance, so applying a moving filter over this direction doesn't make sense.

The result of this layer is that every convolutional layer produces vectors with a length equivalent to the number of filters used. The number of vectors produced in each convolutional layer is the same as the number of input word vectors (again, except for some minor difference due to the alignment of the window at the beginning and end).

\subsection{Max-Over-Time Pooling}
Every convolutional layer (for every filter length) is followed by a max (over time) pooling layer (see \autoref{sec:pooling}). Similarly to convolutional layers, pooling layers also have a width which is used as a window to slide over their input. However instead of calculating an output using neural nodes, the output is simply calculated by applying an element wise maximum of its input vectors to compute a new output vector. In this project the width of the pooling layer is set to the total number of input vectors so that the result is a single output vector. Effectively in combination with the previous convolutional layer the maximum value for each single filter over the whole input is calculated and used as an output.

\subsection{Merging and Dense Layers}
For every filter length / convolutional layer there will be a single result vector due to the max-over-time pooling. All these result vectors will be subsequently concatenated into a single large vector, which is used as the input vector for a regular dense (fully connected) layer.

\subsection{Output Layer}
The final output of the network is another fully connected layer with the number of nodes equal to the number of output classes. Depending on whether single label or multi label classification is needed either softmax or sigmoid activation functions are used. Softmax activation has the property of normalizing the result of all outputs in the layer together so that they sum to 1, effectively converting the output to class probabilities. In case of multi-label classification, more than one class should be able to receive a high probability. The sigmoid function will crush its input into an output range of $\left[0,1\right]$, which can be taken as probabilities for each individual class (see \autoref{sec:softmax}).

\subsection{Dropout}
While the layers described above are the fundamental functional layers of the network, additional dropout layers are employed that help to reduce overfitting. In contrast to \textcite{Kim14} who used dropout only between the output layer and the previous dense layer, an additional dropout layer was added right between the input layer and the convolutional layers. Intuitively this seems logical as the convolutional layers should not depend too much on certain inputs.

Overfitting in convolutional layers is often seen as less of a problem since they have less parameters than dense layers. For a convolutional layer trained on an image, each filter could for example take an input of 8 by 8 gray scale pixels resulting in only 64 input weights to be trained.

Using multiple word vectors there are often significantly more inputs however. With a filter size of 5 and word vectors with 300 dimensions, every filter needs to train 1500 input weights. The original dropout paper \parencite{Srivastava14} uses dropout between all layers and they mention using higher dropout rates at the beginning of the network. In experiments done during the project adding dropout at the beginning did in fact seem to improve performance, although further experiments on the best combination of dropout weights are necessary to validate this finding.

\subsection{Other Network and Training Parameters}
\subsubsection{Activation Function}
The activation function used in the inner layers of the network is the Leaky ReLU function (see \autoref{sec:relu}), which seemed to perform better than hyperbolic tangent or the logistic function in initial tests and also had some slight advantage compared to the plain ReLU activation function.

\subsubsection{Loss Function}
As an error function to minimize during training the cross-entropy (log loss) is used as is generally recommended for classification tasks since it is better suited for probability distributions. In the case of multi-label classification, binary cross-entropy is used, for multi-class cases the categorical cross-entropy is used (see \autoref{cross_entropy})

\subsubsection{Optimizer}
Adam \parencite{Kingma14} is used as the optimizer, since it worked well in initial experiments and no problems were encountered during usage. The results of multiple iterations were always very similar indicating that the optimizer does not get stuck in high-error local minima easily (see \autoref{sec:adam}).

\section{Software Design}
\subsection{Requirements}
The classifier described above needs to be embedded into a general software architecture so that it can be easily used from other software.

The design should fulfill the following requirements:

\begin{itemize}
\item The algorithm described in the previous chapter should be implemented.
\item The documents on which to run the algorithm should be selectable by the user (text-only).
\item The categories used by the algorithm should be selectable by the user.
\item It should be possible to train classifiers using the algorithm.
\item After training a classifier it should be possible to classify newly added documents using this classifier.
\item The functionality should be accessible using a documented \acr{api} accessible with standard web technologies (e.g. REST).
\item The \acr{api} should be general enough to support other classifiers, even though only one classifier might be implemented.
\item There should be rudimentary support to authenticate users in some way in order to protect the functions from unauthorized access.
\end{itemize}

The term ``user'' in above requirements refer to users of the \acr{api}, which will usually be other applications.

\subsection{Non-Requirements}
The implementation does not need to fulfill the following items:

\begin{itemize}
\item Enhanced authentication and authorization. There will be no fine grained access control. It is assumed that when a user or application is authenticated, he has access to the whole system.
\item Support for documents that do not have a text-only representation (e.g. there will be no document filters that automatically extract text from various document types). At most, some predefined document types (like PDF) might be supported, but this should not be seen as a requirement.
\item User interface. It is assumed that access to the system will be programmatically from other applications, so there will be no need for a general user interface.
\end{itemize}

\subsection{Architecture}
\figgraph{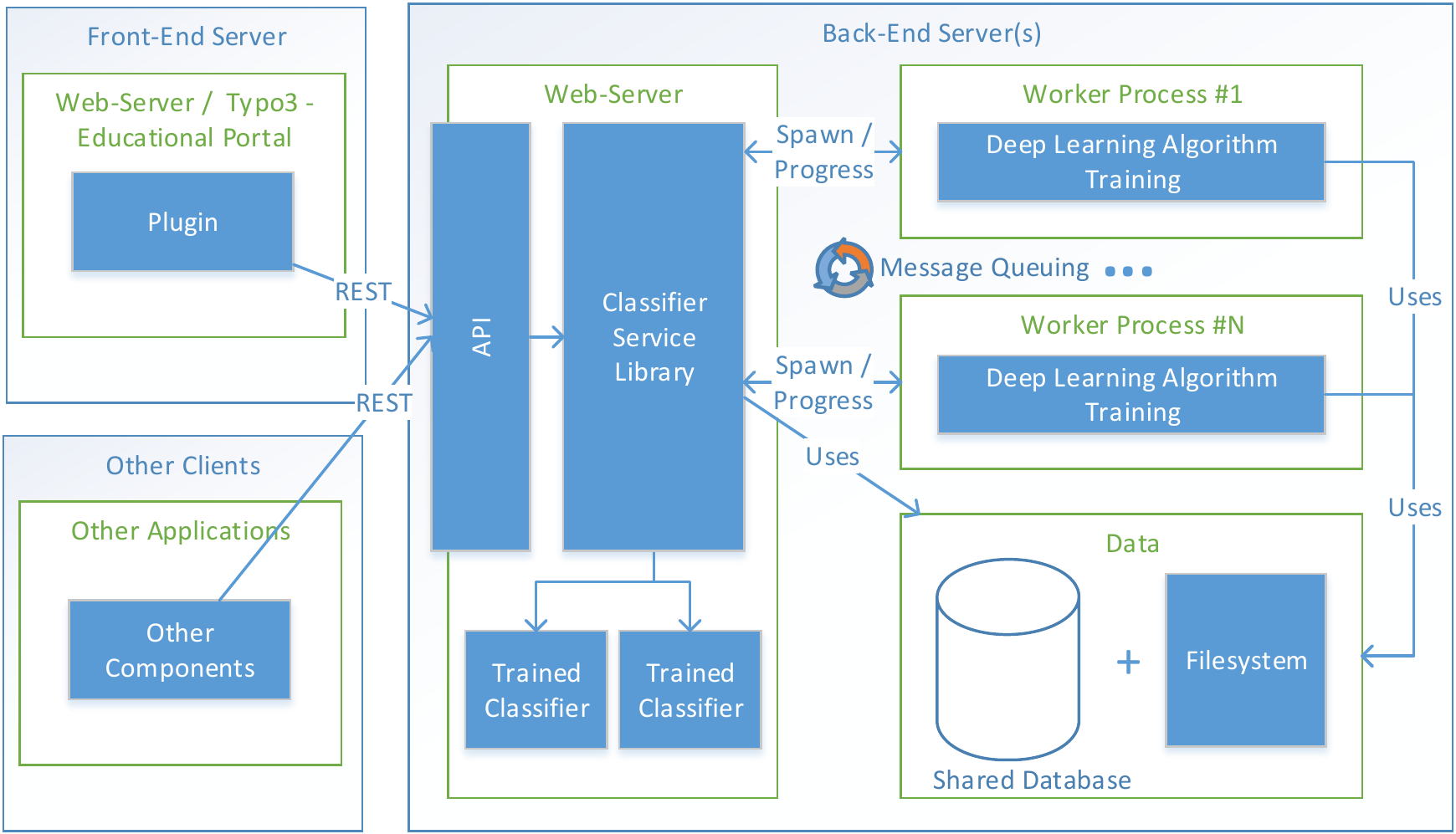}{Architecture of the proposed solution}
To comply with the requirements stated above, the following solution is proposed:

The general architecture of the implementation should follow a classic three tier architecture. There can be multiple clients or applications, one of them being the ecosystem portal that will use the classifiers for automatic categorization. Each of the clients has only access to a public API using a standard protocol such as HTTP/REST. This allows easy access to the system and provides standard mechanisms for things such as authentication. All applications together that call the API will be called the client tier in this thesis. As stated above, the client tier is not within the scope of this thesis (see \autoref{sec:related_work} for related work).

The public \acr{api} can be hosted on a web server and provides all methods necessary to upload documents for classification or to train new classifiers. The implementation of these methods should be done in a service library in order to provide further abstractions and to make is feasible to use the methods without the network and protocol overhead of the API. The service library should have access to the pretrained classifiers and can use them to classify new documents on demand.

In order to train new classifiers separate worker processes that are controlled by the service library and which will report their current progress in regular intervals should be used. Worker processes are useful since the training operation is a time consuming operation and will usually take much longer than a typical web request on the REST \acr{api} will take. Long running HTTP requests may time-out at some point and there is no good way to provide feedback to the \acr{api} user during the operation. With worker processes the API is able to only return a general acknowledgment for the training request and then run the worker process in the background. The \acr{api} can then be used to get the current state and progress of the training task. The communication between the workers and the service library can be realized by message passing. Multiple workers can be run at the same time easily. The service library and all associated worker processes will form the middle tier of the implementation.

Lastly there should be a data tier consisting of a shared database and a file system storage used for documents uploaded into the system.

%% file: chapters/04.implementation.tex

\chapter{Prototypical Implementation}
\label{chap:Implementation}
\section{Overview}
In the previous chapter a design was proposed on how to structure the software in order to fulfill the stated requirements. This chapter will describe how this design is implemented in detail.

\subsection{Platform and Libraries}
The project itself is cross platform. The development and testing was done mainly on a Windows machine while the final production environment is a Linux system.

The programming language of choice in this project is Python for various reasons. First of all, Python is cross platform so that it is easy to use the same sources for both Windows and Linux. While this is true for many languages, Python also has a big user base\footnote{The TIOBE index from June 2016 lists Python as the 4th most popular programming language, after Java, C and C++, \url{http://www.tiobe.com/tiobe_index}, retrieved 17th of June 2016} and many libraries available to rapidly develop applications of almost any type\footnote{The Python Package Index lists a total of 82671 packages as of June 17th 2016}.

More specifically Python offers a lot of libraries for computer science, mathematics and machine learning\footnote{A good overview is available at \url{https://wiki.python.org/moin/NumericAndScientific}, retrieved on the 17th of June 2016}. Projects like Theano \parencite{theano16} offer advanced mathematics while at the same time being very fast as computations are automatically compiled to machine code and can even be offloaded to a graphics card. This makes Python popular in scientific applications and makes it a good choice for this project.

Python is also a supported platform on various \acr{paas} providers such as Google App Engine\footnote{\url{https://cloud.google.com/appengine/}, retrieved 17th of June 2016}. This could make it feasible in the future to host the project as a cloud service for easy scale-out (see also \autoref{sec:related_work}). 

As for libraries Theano was already mentioned. Theano is the computational base of the project. It provides fast methods for things such as matrix multiplication and has many features such as computing function derivatives which can be useful for machine learning algorithms such as neural networks. Theano itself is not tailored specifically for neural networks. Building neural networks directly on top of Theano is possible and there are projects which do this. There are however libraries which make it even easier by providing many ready to use neural network techniques and abstracting the underlying Theano framework. In this project the Keras library \parencite{chollet15} is used for this aspect.

For the hosting of the REST \acr{api} Flask\footnote{\url{http://flask.pocoo.org/}, retrieved on 17th of June 2016} is used as a lightweight web framework. There are several alternative web frameworks in the Python environment, but Flask seemed like a good choice by being small and at the same time very easy to use.

To distribute work load among worker processes and to handle communication with them (progress reports, etc.) Celery is used which is a task queue based on message passing. Celery can be used with various different message queues such as RabbitMQ. While not an explicit goal of this thesis, Celery supports worker processes running on multiple machines, so it should be easy to make the solution more scalable by putting the worker processes on separate machines. These machines could be especially equipped to produce fast results such as having a \acr{gpu}.

Lastly as a database SQLite and the SQLAlchemy library are being used. SQLite is very easy to set up (file based, no installation necessary) and still has many features making it an ideal choice. SQLAlchemy makes it easy to switch to another database system though if desired.

\section {Static Structure}
In this section a description of all relevant packages and their respective modules will be given.

\figgraph{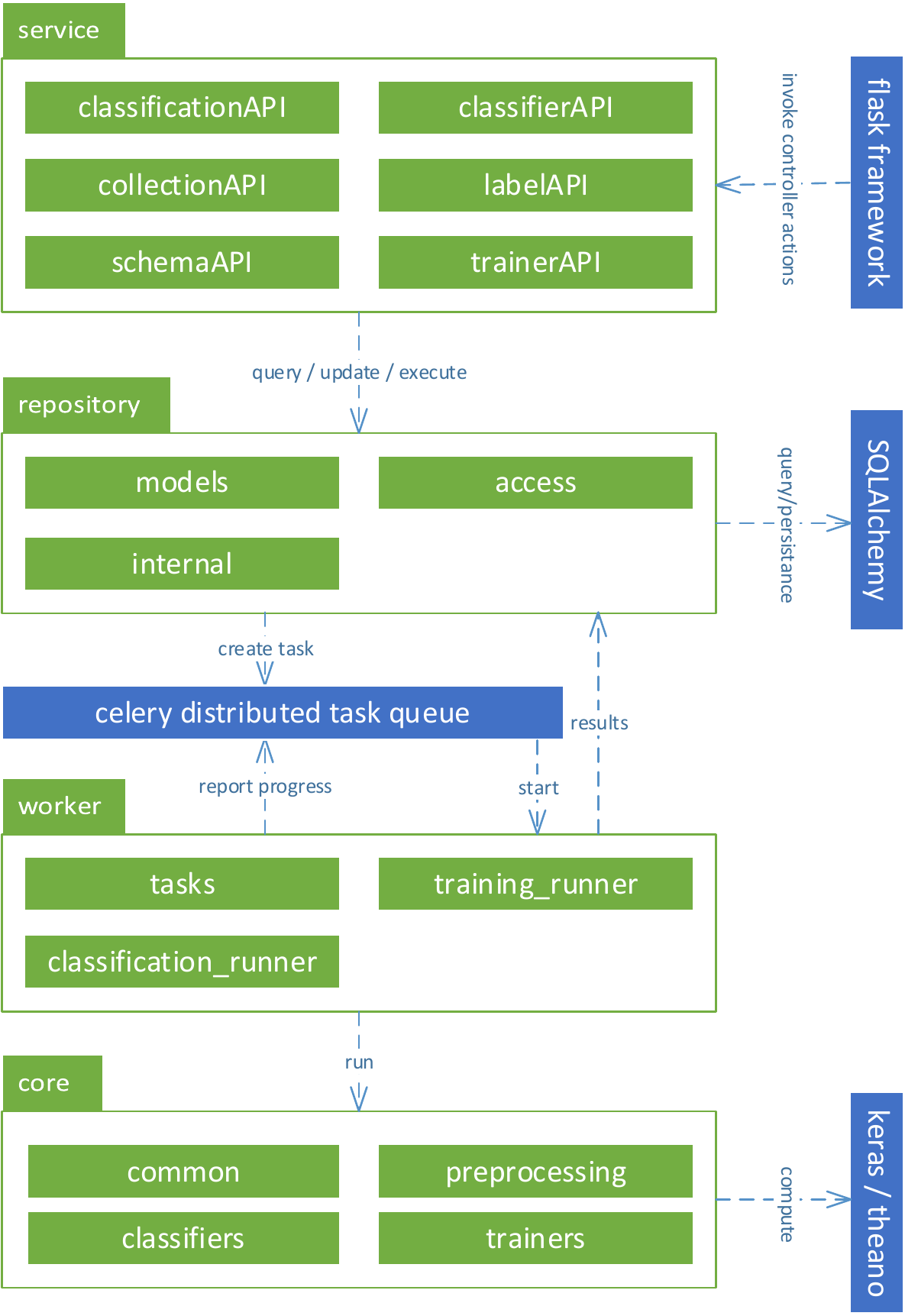}{Relevant packages and modules of the project and their dependencies}

\subsection{The core Package}
The \coderef{core} package is the heart of the implementation. It provides the actual implementation for the deep learning classifier and the general infrastructure used to implement new classifiers.

There are four main modules within the package. The \coderef{common} and \coderef{preprocessing} modules are mostly used internally by the other modules in the package and contain utility types and functions. One example are the tokenization classes in the \coderef{preprocessing} module, which are used to separate the words in a text. The \coderef{classifiers} module and the \coderef{trainers} module contain the actual implementations for the various classifiers supported out of the box by this project.

\subsubsection{Preprocessing}
\figgraph{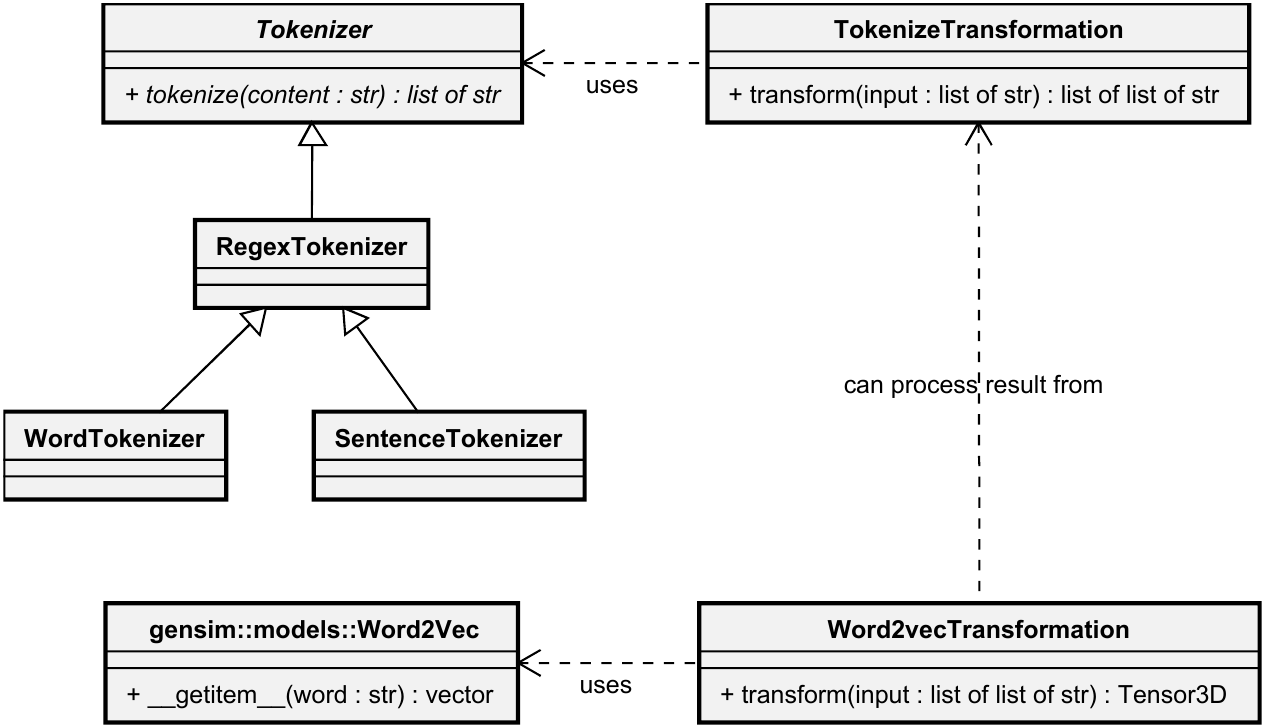}{Preprocessing classes and their correlation}
The main classes within the \coderef{preprocessing} module are the tokenization classes and the word2vec transformation. \autoref{fig:classes_preprocessing} shows their correlation. The \coderef{WordTokenizer} is a simple tokenizer that uses regular expressions to find words within a text. It splits words by whitespaces and has some extra features optimized for texts extracted from PDF files such as trying to de-hyphenate words at line breaks. The result returned by the tokenizer is a sequence of words. There is also a \coderef{SentenceTokenizer} which can be used to extract individual sentences from a text corpus. This was sometimes used in this project to train the word2vec model.

The \coderef{TokenizationTransformation} class uses a tokenizer and applies it to a list of texts given to it in the \coderef{transform} method. It then returns a list of strings for each input text (a list of a list of a string). The result of this method can directly be used as the input for the \coderef{Word2vecTransformation} class.

As learned in previous chapters, word2vec provides a model that assigns vectors to individual words. These vectors preserve the semantic meaning of the words to some point. For example, words of the same category (like capital cities) are grouped together in the vector space. These models need to be trained separately beforehand. There are however models available online which were trained on big corpora like Wikipedia or Google News Articles. An additional model was trained for this project using the arXiv corpus (see \autoref{chap:Evaluation}).
 
The \coderef{Word2vecTransformation} class applies the word2vec transformation for each word in the input. It uses the \coderef{Word2Vec} class of the gensim class to look up individual words from a pretrained model and then returns the list of all word vectors in a format directly understood by the neural network model that will be discussed later.

\subsubsection{Trainer and Classifiers}
The \coderef{classifiers} module contains the implementation for various classifiers such as the deep convolutional neural network classifier presented in this thesis or the \acr{svm} classifier used for comparison.

Strictly separated from the classifiers are ``trainer'' classes within the \coderef{trainers} module. The general idea is that before a classifier can be used, it needs to be trained. For each type of classifier, there is an associated trainer class which will create such a classifier. The algorithms used for training can be quite complex and require a lot of resources while the algorithms for prediction are usually more simple. Separating these two concerns makes the code more readable and ensures that only resources and dependencies required for the current use case are being used. \autoref{fig:classes_core} shows the relationship between \coderef{Trainer} and \coderef{Classifier} classes as well as the concept of checkpoints. Checkpoints will be explained in detail in \autoref{sec:checkpoints}

There is not necessarily a one to one relationship between trainers and classifiers. For example multiple neural network architectures could be trained using various trainers, but the classifier implementation only needs to know how to load an existing architecture and can use it as long as the input and output formats for the different architectures stay the same.

\figgraph{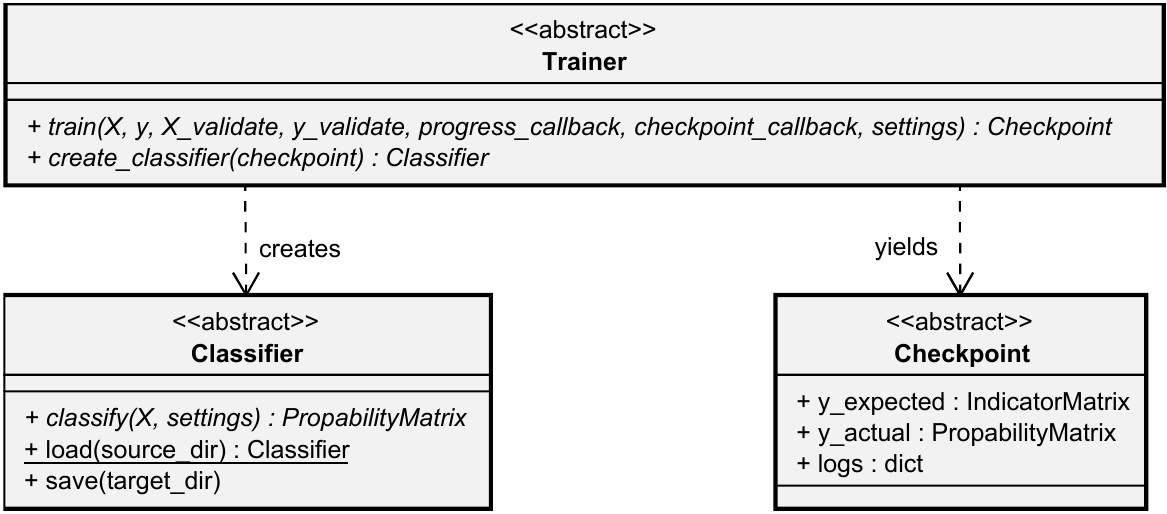}{Core classes}

Classes derriving from the \coderef{Trainer} base class have to implement the \coderef{train} and \coderef{create\_classifier} methods. The \coderef{train} method is called whenever a new classifier should be trained and receives the following parameters:

\begin{description}
\item[\coderef{X}] is a list of documents that should be used during training. The only requirement is that the objects passed provide a \coderef{read} method that can be used to retrieve the document content as unicode text.
\item[\coderef{y}] is a binary indicator matrix that contains a row for every document in \coderef{X} and has as many columns as there are categories to classify. If a specific document belongs to a specific category, the corresponding cell in the matrix will contain a $1$ otherwise it will contain a $0$.
\item[\coderef{X\_validate}] is a list of documents that should be used during training validation. The classifier should not use these documents directly in the training process, but it can use them to determine the current performance of the classifier and to detect overfitting. In this case it may also be used for early stopping in case the performance on the validation documents is no longer improving.
\item[\coderef{y\_validate}] is the binary indicator matrix for the \coderef{X\_validate} parameter.
\item[\coderef{progress\_callback}] is a callback function which is called regularly during the training process to indicate the current progress. The progress includes data such as a message indicating the current operation, the progress percentage, a timestamp and other useful information which can be used by client programs to report the current progress to users.
\item[\coderef{checkpoint\_callback}] is a callback function which can be used by a trainer to report an intermediate result - a so called checkpoint.
\item[\coderef{settings}] is a dictionary containing additional parameters which are implementation specific.
\end{description}

Important trainer classes are the \coderef{CnnTrainer} class, which implements the \acr{cnn} based classifier described earlier and the \coderef{SvmTrainer} class for the \acr{svm} classifier that will be used for comparison in \autoref{chap:Evaluation}.

\subsubsection{Checkpoints}
\label{sec:checkpoints}
Training often involves an iterative process in which the classifier should improve its performance with every iteration until a peak is reached or a maximum time or iteration count has been reached. Especially for long running training sessions it can be very helpful to get a sense of the current classifier performance or even to be able to run a classifier at a specified checkpoint while the training continues. This concept is realized in this project using checkpoints. A \coderef{Checkpoint} object encapsulates the state of a running training process at a specific point in time as well as statistical information about the checkpoint. Among this information are the prediction results for the \coderef{X\_validate} documents (\coderef{y\_actual}). This prediction results can be used by the caller to calculate metrics for the current checkpoint. By using the \coderef{create\_classifier} method of the trainer, the caller can use the checkpoint object to create an actual classifier. This means that for every training session, multiple classifiers can be created and used independently of each other. This is another reason why the implementations for trainer and classifier are strictly separated.

Implementers are not required to provide intermediate checkpoints during the training. But they have to return at least one checkpoint as the final result of the training.

\subsubsection{Classification}
A \coderef{Classifier} instance has a \coderef{classify} method which can be called to evaluate a list of documents. The return value is a probability matrix which indicates for every document and category what the probability is that the document belongs to the category.

The class also methods to save and load classifiers to or from a specified directory. This is essential to be able to persist and restore classifiers and being able to use them on demand.

\subsection{The worker Package}
\label{sec:worker_package}
The worker package contains the execution logic to run training sessions or to evaluate a specific classifier. These tasks are implemented as celery tasks. Celery is an distributed, asynchronous task queue. It allows scheduling of various tasks and automatically handles their execution. The execution is performed by worker instances which can run of different computers. Long running tasks can provide notifications about their current progress and this progress can easily be queried using the celery framework. Celery provides many more features which could be useful in the future of this project, but at the moment only basic features are used.

The \coderef{tasks} module provides small stubs that are called by the celery framework. These stubs forward the calls to the actual implementations provided in the other modules. The \coderef{training\_runner} is executed to start the training of a new classifier. It receives the ids for the classifier that should be trained and the training session with which the task is associated. Using this id, the runner loads all necessary data using the repository package and transforms it to a format understood by the trainer classes. For example it loads the document metadata from the database and transforms the target attribute values into a vector that will be passed in later as the \coderef{y} argument of the training. The module is also responsible for splitting the documents into a validation set and a training set, by randomly selecting up to 10\%\footnote{The number of documents is additionally limited by $100 * \norm{y}$ with $\norm{y}$ being the number of categories. This is to avoid overly large validation sets.} of the documents as validation data. It then proceeds to ensure that common dependencies such as the default tokenizer and the word2vec model are instantiated. After all dependencies are present it instantiates the appropriate trainer class and starts the training process. 

During the training process, the trainer can report progress which is forwarded using a custom state in the \coderef{update\_state} mechanism of celery. The trainer can also provide intermediate checkpoints. These checkpoints are evaluated by the task and a score is calculated. The classifier resulting from the checkpoint is then saved to disk and the score together with other metadata is persisted in the database. These actions are performed by the \coderef{repository} package.

The \coderef{classification\_runner} receives the id of a classifier that should be used and a list of document ids that should be evaluated. Like the trainer it then proceeds to load the associated data from the database and ensures all dependencies for the classifier. It will then instantiate the correct classifier and call it. The result will be a prediction vector for each document that was passed in. The classification runner will use this result to decide which of the classes should be assigned to each document and it will return a dictionary mapping each one of the input documents ids to a list of attribute value ids.

It should be noted that the implementation of classification as a separate worker process is a small deviation from the design proposed in \autoref{chap:Algorithm}, which only used workers for the actual training. For technical reasons, the initialization of classifiers can be time consuming (e.g. requiring to load word embedding models of several gigabytes) and the execution of the classification can be resource intensive. For these reasons the decision was made to use worker processes for the classification as well.

\subsection{The repository Package}
The repository package provides access to the the data store of the project and tries to abstract from concrete implementation details. For example it contains methods that read or write both to the database and the file system or even merges results from the database with the current state of the celery worker queue.

The implementation uses the SQLAlchemy package, which provides object relational mapping for databases in python. Although the project uses SQLite as a backing store, there is no hard dependency on a specific database engine and it should be easy to use alternative databases.

The various database tables used in the project are presented in the \autoref{sec:database}.

\subsection{The service Package}
As has been mentioned, one goal of the project is to provide an easy interface to allow integration into other projects. This interface is realized using a REST API. REST is based on HTTP and can therefore be used from all programming languages that can make HTTP requests.

To process incoming requests the flask framework is used. It provides many features such as the mapping of specific URL patterns to methods out of the box, so that the implementation can focus on providing the actual functionality without having to concern itself too much about the on-wire representation.

The following is an example of an \acr{api} method that returns a specific collection to the user. The example has been slightly modified from the original to improve readability:

\begin{lstlisting}[language=python,caption=Example of service method]
@api.route("/collections/<cid>/", methods=['GET'])
def get_collection(cid):
    """Returns the collection with the given id"""

    model = repo.get_collection(int(cid))
    dto = Collection.from_model(model)

    return dto.to_json()
\end{lstlisting}

All \acr{api} methods have a \coderef{@route} annotation that specifies the URLs that they handle and the HTTP verbs that they support. The URL can contain placeholders, which are automatically passed in as parameters to the method when the URL is requested. The method above would be called for example when someone requests the ``/collections/2/'' URL.

Like many methods, \coderef{get\_collection} forwards the request to the repository package which will load the correct instance from the database and which returns a model object. It is bad practice to return such model objects directly from web services as this couples the internal representation to the external representation and therefore makes it hard or impossible to change one of them without changing the other. It can also lead to involuntary information leakage when internal data (such as passwords) is not properly removed before returning the object to the user.

For these reasons there is a strict separation in the project between the internal models located in the \coderef{repository} package and so called \acrfullpl{dto} located in the \coderef{dtos} module of the \coderef{service} package. The \coderef{from\_model} methods are implemented on the DTO classes as convenience methods that fill the \acrpl{dto} from corresponding model classes. \acr{dto} instances also have a \coderef{to\_json} method which returns a \acr{json} \parencite{rfc7159} string representation of the instance, which can be returned directly from the API methods.

The various \coderef{*API} modules implement all the different API methods which are described in \autoref{sec:rest}.

There is an additional \coderef{util} module contained in the package, with various functions that helps with processing of requests. For example they contain an extension that automatically detects and parses paging parameters in the request (e.g. \coderef{``?offset=20\&limit=100''}) as well as filters  (e.g. \coderef{``?code=mydocument''}) and fills a data structure which can be passed to the repository without having to manually parse the query string in each of the API methods that support paging or filters.

The service has basic access control using basic authentication. If activated, only users in a configurable list with the correct password can access the service. The configuration is described in \autoref{sec:configuration}. The authentication is realized with the \coderef{flask.ext.basicauth} extension.

\section{The Convolutional Neural Network Classifier}
The topic of this project and therefore the main implementation of a classifier is the \acrlong{cnn} classifier and trainer, whose implementation will be described a bit more detailed in this section.

\figgraph{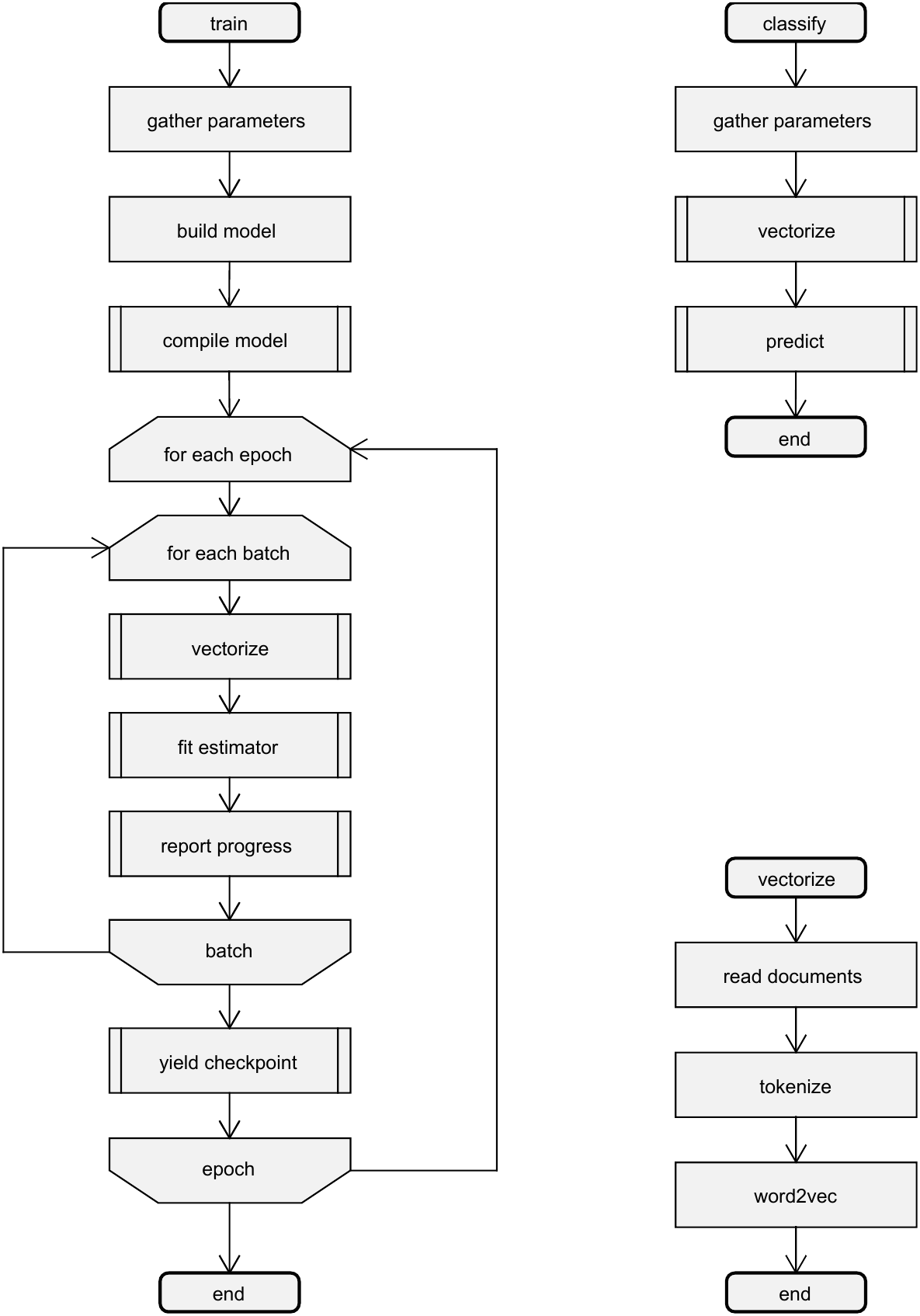}{Program flow for the train function in the \acrshort{cnn} trainer}

\subsection{Operational Settings}
The general program flow for the \acr{cnn} training and prediction is shown in \autoref{fig:flow_train}. The training starts by gathering various settings that can be used to influence the neural network created by the trainer or that influence how the training process itself will work. The settings are passed in the \coderef{settings} parameter of the \coderef{train} function

Some examples of settings which are supported are:
\begin{description}
\item[\coderef{w2v}]: The word2vec instance that should be used to convert words into feature vectors. This is a required parameter.
\item[\coderef{tokenizer}]: The tokenizer instance that should be used to extract words from a text. This is a required parameter.
\item[\coderef{max\_timesteps}]: The maximum number of words for each document that should be considered. Optional, defaults to 1000.
\item[\coderef{batch\_size}]: The maximum number of documents that should be processed at the same time. Increasing the number will provide better performance but needs more memory. Optional, defaults to 200.
\item[\coderef{filter\_count}]: The number of filters to add in the convolution for each filter length. This influences the neural network architecture. Optional, defaults to 200.
\item[\coderef{filter\_lens}]: A tuple specifying the filter lengths that should be added as convolutions. Every entry creates a separate convolution with the specified filter length. This influences the neural network architecture. Optional, defaults to (1,2,3).
\item[\coderef{dense\_size}]: The number of hidden neurons in the fully connected layer present after the convolutions. Optional, defaults to 100.
\item[\coderef{dense\_size2}]: If set, adds an additional fully connected layer after the previous one with the specified number of neurons. Optional, defaults to none.
\item[\coderef{activation}]: The activation function to use in all inner layers. Optional, defaults to `leakyrelu'.
\end{description}

\subsection{Model Generation}
Based on these parameters, the trainer builds a conceptual model of the neural network using the keras library. Keras supports many different layer types which can be connected to each other in a graph like fashion. For simple architectures, Keras also allows the usage of sequential containers which connect all contained layers in a sequence. Such containers can also be connected to other layers or used within other containers (composite pattern).

Individual layers are created and connected like this:
\begin{lstlisting}[language=python,caption=Defining a Keras model]
# Define an 1D input with 100 components
entry = Input(shape=(100,))

# Apply dropout with probability 50%
dropout = Dropout(0.5)(entry)

# Add a dense layer with 50 nodes
dense = Dense(50)(dropout)

# Define a model
model = Model(input=entry, output=dense)       
\end{lstlisting}

The example defines an input of shape $(100)$ (each sample has 100 numerical components), applies a dropout layer which drops inputs with a probability of 50\% and then adds dense layer with 100 neurons. At the end a model is created by specifying an input and an output for the network. The syntax might seem a bit strange. Every layer constructor (\coderef{Dense}, \coderef{Dropout}, etc.) creates a layer object, which also acts like a function. By calling this function, the layer can be connected to a previous layer. There are helper methods to merge the output of multiple layers to allow complex architectures with this pattern.

\figgraph{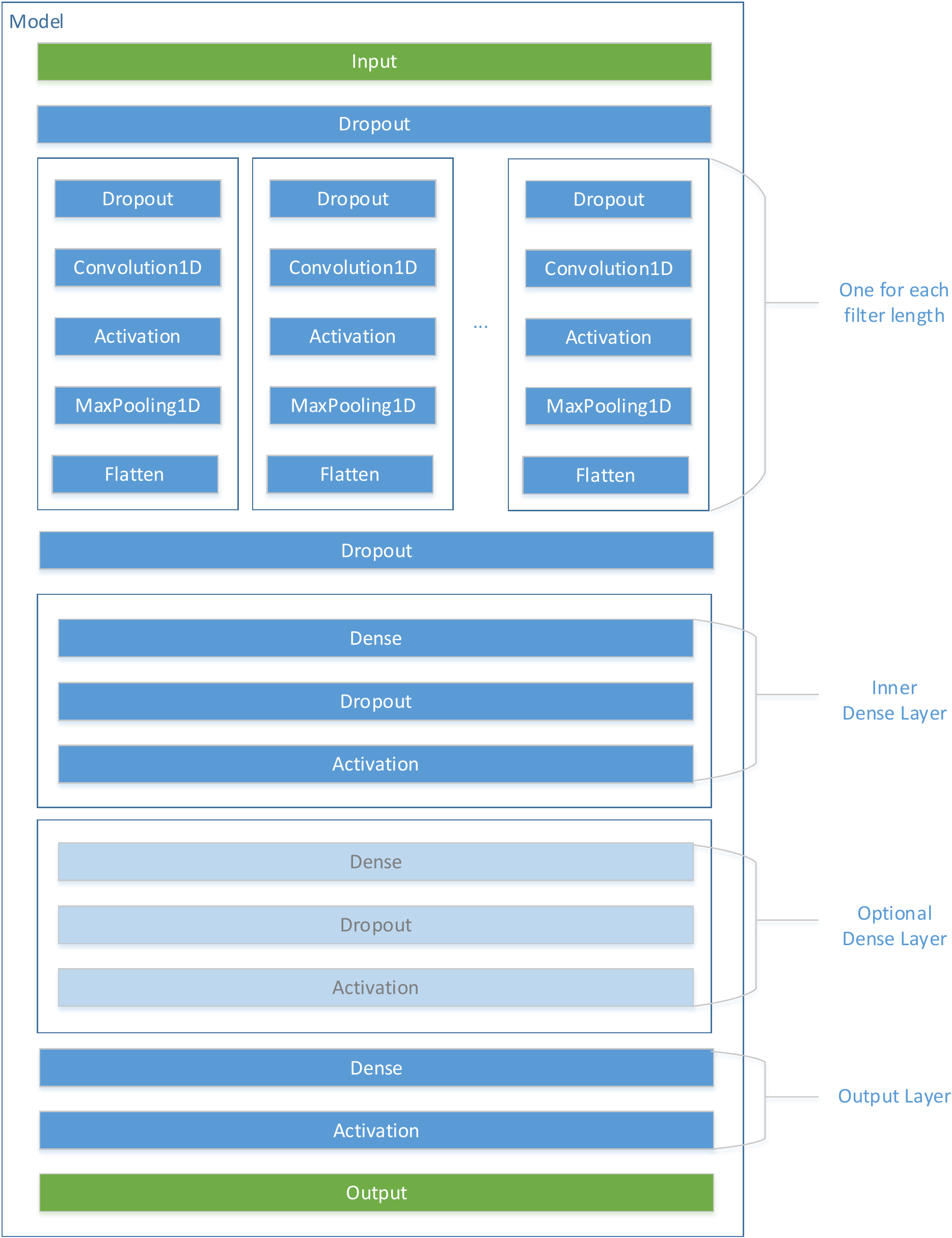}{Keras layers produced by the train function}
\autoref{fig:keras_layers} shows the different layers that are constructed by the trainer. The layers are connected from top to bottom. As can be seen, many concepts and techniques in neural networks are implemented as layers in keras. Activation functions for example are implemented as a separate layer which provides more flexibility. The same applies to the Dropout technique.

\subsection{Compilation}
After the model has been constructed, it needs to be compiled. This is done using the \coderef{compile} function of the Keras framework, which expects an optimizer and a loss function to use as input parameters. While in the implementation this call is only a single line of source code, it is very interesting to take a quick look at what happens behind the scenes.

\figgraph{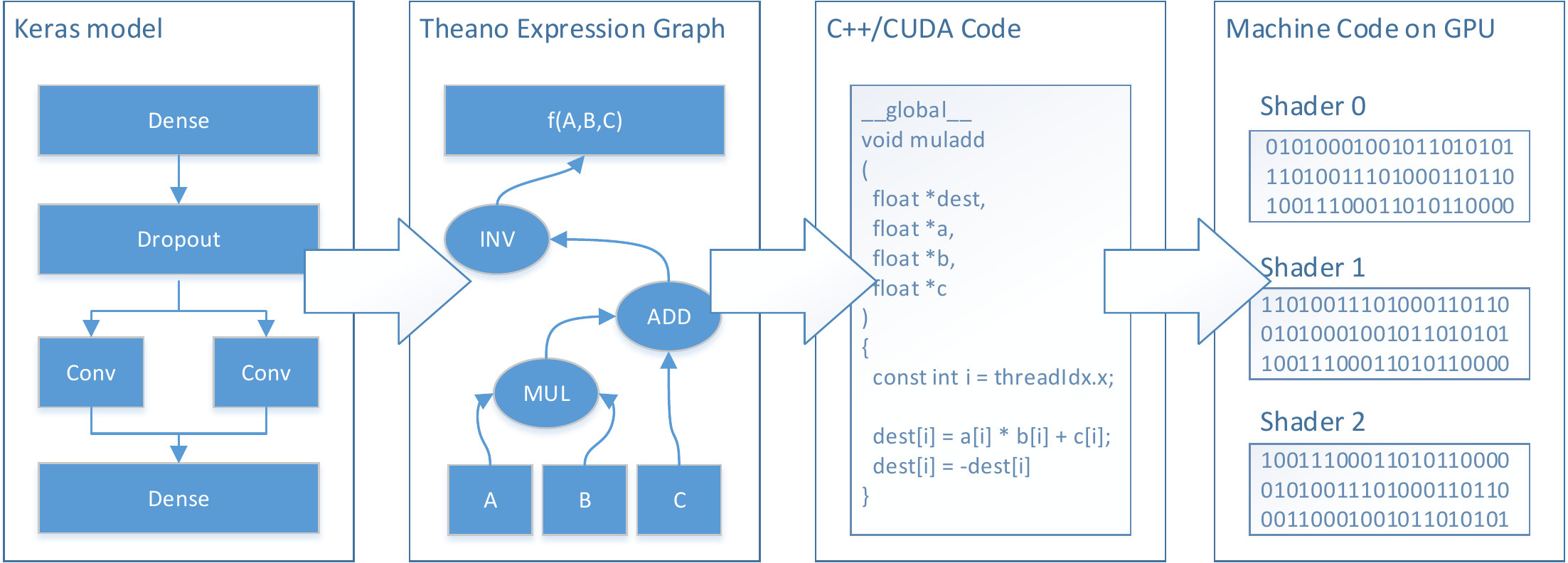}{Compilation steps of Keras model from conceptional model to \acrshort{gpu}}

Keras first transforms the neural network into a computational graph using the Theano framework. It uses the fact that many operations in a neural network can be computed as linear algebra expressions using matrices and vectors. Theano provides the necessary functions to build such expressions and to compute them efficiently. It also has features such as automatic differentiation for functions which can be useful for backpropagation training and many built-in optimizations. Most important is however that Theano supports various backends to perform the actual computation.

Python implementations usually do not offer the same performance as optimized C code let alone hand crafted assembler instructions. On the other hand python allows very high level programming and is usually considered easier to use than these low level languages. Theano bridges this gap by providing a high level symbolic expression language and then employing highly optimized backends that do the actual compilation.

One of these backends generates CUDA compatible C++ code which is then compiled by the Nvidia CUDA compiler. CUDA is a platform created by the graphics card manufacturer Nvidia to facilitate \acr{gpgpu} applications \parencite{Nickolls2008}. After compilation the generated machine code runs directly on CUDA compatible graphics cards.

Since \acrpl{gpu} are very effective at doing linear algebra operations, using many cores at the same time, a great speed up can be achieved in comparison to CPU only solutions.

\subsection{Preprocessing}
After compilation the created model is ready to use and can be fed with training data. The training data consists of a set of inputs ($X$) and their corresponding desired outputs ($y$). The actual input is not the content of documents itself, but rather a transformed view of it. This transformation converts the individual words in a document into vectors. \autoref{fig:flow_train} shows this transformation as the \coderef{vectorize} process.

The process of transforming the document text into individual words is called tokenization. The tokenizer that should be used is configurable. The default tokenizer available in this project was presented earlier in this chapter. The result is a sequence of words for each document. The results of tokenization is then passed on to the word2vec transformation which retrieves the word vectors for each word.

By concatenating all these word2vec vectors that compromise a document, a final representation of the document in vector space is obtained. The result is called a two dimensional tensor with shape $(\norm{d},\norm{v})$ (with $\norm{d}$ being the number of tokens in a document and $\norm{v}$ being the length of a word vector). One could also call this a $\norm{d} \times \norm{v}$ matrix, but Keras and other frameworks use the concept of tensors since matrices lay out components in only two dimensions (rows and columns) and higher dimensional constructs are often necessary to represent a domain.

The tensor created for each document contains the $\norm{d}$ words in the form of $\norm{v}$ values corresponding to the word vector $v$ of the word. The length of the word vector depends on the model used. Most models use between 100 and 1000 components.

All the resulting two-dimensional tensors for the documents are combined into a three-dimensional tensor which is the input for the \coderef{fit} function in Keras. By convention, most frameworks call this input tensor simply $X$.

Unfortunately combining the tensors requires that they all have the same shape. This means that the tensor for shorter documents with fewer than \coderef{max\_timesteps} words will be padded with zeroes where necessary. Longer documents are cut accordingly. There are some techniques to avoid this limitation with keras, but the network architecture already limits the maximum number of processable words due to the fixed number of input neurons. The only disadvantage is the increased memory footprint for documents which are below this limit. For most use cases this should not be a problem.

In addition to the input given as $X$ the training also requires the expected outputs as another tensor $y$. The last layer of the network has as many neurons as there are attribute values for the attribute covered by the classifier. Each neuron stands for a value and will output the probability that the attribute value should be assigned to the document. Combined, these probabilities form an output vector. By concatenating the expected outputs for all documents the 2-dimensional tensor $y$ is obtained. This vector is already passed in as an argument to the train function and therefore needs to be prepared accordingly by the caller. This will be described later.

\subsection{Fitting the Classifier}
Given all this data, the Keras model will automatically start the training and will try to optimize the loss function specified before. It is common for neural network training to train the same pair of expected input and output several times. Usually each pair is presented one time to the network until all training samples have been processed. After this a new iteration starts. Such an iteration is called an epoch. The number of epochs can be specified as an input parameter to Keras.

Within an iteration several training pairs will be processed at the same time. This is called batch processing. The network error is computed/averaged over these batches and a single weight adjustment is done for each batch. This increases performance and therefore leads to faster convergence. Keras allows the adjustment of the batch size as another parameter.

\subsubsection{Batch Processing}
Passing in the complete $X$ and $y$ tensors is convenient and sufficient for many applications. A problem arises however when dealing with large amounts of data as in this project. Assuming one wants to train on 100 categories, each with 1000 prelabeled documents, considering only the first 1000 words and using a 300 dimensional word vectors, the input tensor will have a shape of $(100000, 1000, 300)$. This is a total of 30 billion input values. Using 32 bit floating point values, this would require about 111 gigabytes of data to be passed in a single variable. Within the neural network the memory would increase even further as the samples pass through the different layers.

Fortunately it is possible to train a Keras model batch by batch, in which case only the 3d tensor for a single batch at a time is required. So instead of using a single tensor with shape $(100000, 1000, 300)$ one can pass in 2000 individual batches with input tensor of shape $(50, 1000, 300)$, reducing the memory footprint to about 57 megabyte. In this case the vectorization described above is only performed for one batch at a time.

This batch wise operation is what is shown in the initial flow diagram (\autoref{fig:flow_train}) as it is easy to understand and visualize. There is however a third option which combines the advantage of reduced memory footprint with the advantages of letting Keras handle the details of epochs and batches. This option uses an iterator that is passed to a special \coderef{fit\_generator} method on the Keras model. An iterator in python is an object with a special \coderef{next} function that is called every time a new value is required. Keras uses this iterator to ask for a new batch every time it needs one. Another advantage with this model is that Keras will use a separate thread to ask for a new batch while the training for the previous batch is still running. Since batch preparation is often I/O bound (data needs to be loaded from hard disk), it makes sense to start loading the next batch data already. When offloading to GPU it reduces the time the \acr{gpu} is running idle because it has to wait for new data.

Because of this advantages, this project uses this option in the actual implementation. Separate utility classes and functions are used to split up the initial document list into batches and to run the preprocessing transformations on them.

\subsubsection{Caching of Generated Data}
One remaining disadvantage of batch wise processing is that the preprocessing has to be run for each batch again after each epoch, since the data is only stored temporarily and then discarded. This can be a problem when the preprocessing is slow and the training is run on a fast \acr{gpu}. In this case the \acr{gpu} will be slowed down, decreasing the overall performance. To further increase performance a caching mechanism was implemented which will run the preprocessing only during the first epoch of training and then stores the resulting tensors in temporary files on the hard disk. In the next epochs the tensors will be read directly from hard disk. Of course this requires the hard disk saving and loading to be faster than the actual preprocessing. For modern solid state disks this if often the case, but this needs to be evaluated on a case by case basis.

\subsubsection{Progress Reporting and Checkpoints}
In order to report the current training progress to the user and also to evaluate the current classifier performance after each epoch, the system hooks into the Keras processing pipeline. Special callbacks are passed to Keras which will be called at specific points like when a batch finished or when a new epoch starts. The trainer uses these callbacks to determine the current training state and informing its caller using callbacks of its own (the \coderef{progress\_callback} and \coderef{checkpoint\_callback} parameters). A progress update will be raised every time a batch ends, while the checkpoint callback is called whenever a training epoch ends.

A checkpoint is more than just a hint. In this project a checkpoint represents the complete state of a classifier at the current training progress. A checkpoint is used to create the actual classifier instance. While it can make sense to use the final checkpoint created by the trainer to create the classifier, often the performance of a classifier does not improve after some number of epochs. In some cases the performance might even decrease because of overfitting or other effects. It it therefore a good idea to pick the classifier to be used from the point in time when the performance was best. To support this idea the \acr{cnn} classifier will store the state of the neural network (the architecture and all weights) inside the checkpoint and it also computes the current prediction values for all documents in the validation set that was passed to the function. This allows the caller to calculate a score for this checkpoint which can later be used to select the best one.

The creation of checkpoints is done in the \coderef{create\_checkpoint} method.

\section{REST API}
\label{sec:rest}
The application programming interface (API) used by the implementation is based on Representational State Transfer (REST). The complete functionality of the system is represented as various resources some of which can be read, created or altered using standard HTTP verbs such as GET or POST.

The main resources managed by the API are collections, schemas, training-sets and classifiers. Collections represent a set of documents, schemas represent a collection of attributes that can be applied to a document and training-sets are a manual mapping from a subset of documents to their corresponding attributes. Lastly a classifier uses the aforementioned resources to automatically assign one or more attribute values from a specific schema to a document.

The complete API documentation is too extensive to include in this thesis and will instead be maintained together with the source code as automatically generated HTML pages. The following will only list examples of valid resource URI and possible HTTP verbs with a short description.

\subsection{The /collections Endpoint}

The collection endpoint manages documents and collection of documents. It has methods to create, update, query or delete collections as well as individual documents. The following methods are supported:

\begin{description}
\item[GET /collections/] \hfill 

Returns all collections in the system

\item[POST /collections/] \hfill 

Creates a collection

\item[GET /collections/\textnormal{(colid)}/] \hfill 

Returns the collection with the specified id

\item[DELETE /collections/\textnormal{(colid)}/] \hfill 

Removes a collection and all its documents from the system

\item[GET /collections/\textnormal{(colid)}/documents/] \hfill 

Returns all documents belonging to the collection

\item[POST /collections/\textnormal{(colid)}/documents/] \hfill 

Creates a new document within a collection

\item[GET /collections/\textnormal{(colid)}/documents/\textnormal{(docid)}/] \hfill 

Returns relevant information about a single document in a collection

\item[DELETE /collections/\textnormal{(colid)}/documents/\textnormal{(docid)}/] \hfill 

Removes a document from a collection

\item[GET /collections/\textnormal{(colid)}/documents/\textnormal{(docid)}/content] \hfill 

Retrieves the current document data associated with the document

\item[POST /collections/\textnormal{(colid)}/documents/\textnormal{(docid)}/content] \hfill 

Updates the (text) data associated with a document

\item[PUT /collections/\textnormal{(colid)}/documents/\textnormal{(docid)}/content] \hfill 

Updates the (text) data associated with a document

These functions are implemented in the \lstinline{collectionAPI} module

\end{description}

\subsection{The /schemas Endpoint}

The schemas endpoint allows the creation, modification and query of schemas, attributes and attribute values in the system. It supports these methods:

\begin{description}
\item[GET /schemas/] \hfill 

Returns a list of all available schemas (without details)

\item[POST /schemas/] \hfill 

Creates a new schema

\item[GET /schemas/\textnormal{(sid)}/] \hfill 

Returns the schema with the specified id

\item[DELETE /schemas/\textnormal{(sid)}/] \hfill 

Deletes the schema with the specified id

These functions are implemented in the \lstinline{schemaAPI} module.

\end{description}

\subsection{The /classificationsets Endpoint}

This endpoint allows the creation of classification sets which connect documents with attribute values. Classification sets are used to prelabel documents so that they can be used for training or validation of a classifier. A ``label'' is a single assignment of an attribute value to a document.
The following methods allow the management of classification sets and the labels contained in them:

\begin{description}
\item[GET /classificationsets/] \hfill 

Returns all classification sets in the system

\item[POST /classificationsets/] \hfill 

Creates a classification set

\item[GET /classificationsets/\textnormal{(clsid)}/] \hfill 

Returns the classification set with the specified id

\item[DELETE /classificationsets/\textnormal{(clsid)}/] \hfill 

Removes a classification set and all its labels from the system

\item[GET /classificationsets/\textnormal{(clsid)}/labels/] \hfill 

Returns all labels belonging to the classification set

\item[POST /classificationsets/\textnormal{(clsid)}/labels/] \hfill 

Creates new labels within a classification set

\item[GET /classificationsets/\textnormal{(clsid)}/labels/\textnormal{(docid)}/] \hfill 

Returns all labels for a single document

\item[DELETE /classificationsets/\textnormal{(clsid)}/labels/\textnormal{(docid)}/] \hfill 

Removes a document from the classification set

These functions are implemented in the \lstinline{labelAPI} module.

\end{description}

\subsection{The /trainers Endpoint}

The system is prepared to support multiple types of classifier algorithms. A classifier needs to be trained on a set of examples (given as a classification set) before it can
be used. The trainers endpoints returns the list of training algorithms which are available in the system. Trainers cannot be updated, added or removed using the REST API.
To add a new trainer, one has to add it manually to the table of trainers.

\begin{description}
\item[GET /trainers/] \hfill 

Returns all trainers in the system

These functions are implemented in the \lstinline{trainerAPI} module.

\end{description}

\subsection{The /classifiers Endpoint}

Classifiers are the heart of the system. A classifier is first trained and can then be used to compute possible attribute values for documents in the system. Using this endpoint
classifiers can be created, updated, queried or deleted.

After a classifier is created, it can be trained by creating a training session for it. Each classifier can go through various training iterations. During training the current progress can be
monitored easily.

\begin{description}
\item[GET /classifiers/] \hfill 

Returns a list of all available classifiers

\item[POST /classifiers/] \hfill 

Creates a new classifier

\item[GET /classifiers/\textnormal{(clsid)}/] \hfill 

Returns the classifier with the specified id

\item[DELETE /classifiers/\textnormal{(clsid)}/] \hfill 

Deletes the classifier with the specified id

\item[POST /classifiers/\textnormal{(clsid)}/trainings/] \hfill 

Starts the training for a classifier using a specified test set and parameters

\item[GET /classifiers/\textnormal{(clsid)}/trainings/\textnormal{(trnid)}] \hfill 

Returns the current training progress

These functions are implemented in the \lstinline{classifierAPI} module.

\end{description}

\subsection{The /classification\_requests Endpoint}

Using this endpoint, actual classifications can be requested using one of the previously trained classifiers.

\begin{description}
\item[POST /classification\_requests/] \hfill 

Creates a new classification request and immediately returns the results of the classification.

These functions are implemented in the \lstinline{classificationAPI} module.

\end{description}

\section{Database}
\label{sec:database}
The following is an overview of all database tables in the system. There are some conventions that apply to all tables in the system. In order to reduce duplication, these will be mentioned only once in this section.

\begin{enumerate}
\item Every table has a primary key attribute called \coderef{id} of type \coderef{integer}.
\item Foreign key references to the table will usually contain the name of the target table and the suffix ``\_id''. Exceptions are made only if the role of the foreign key is not clear otherwise.
\item \coderef{code} columns are used to indicate alternative identifiers for the item and can be set by user of the system. This is useful for synchronization with other systems where the code can contain the identifier used in the other system. \coderef{code} columns are usually unique within a specific parent scope. For example a document code must be unique within a collection. The content can be alphanumeric (text).
\item \coderef{name} columns on the other hand are used to give a (human) readable name to an object which can be used directly in user interfaces. They don't have to be unique, but they often are.
\item \coderef{created} columns provide a timestamp given by the server when the object was created.
\end{enumerate}

Not all of the columns mentioned above are present in all tables. If the column is present in a table it can be seen in the database diagram. They will not be mentioned explicitly in the following descriptions unless necessary.

\figgraph{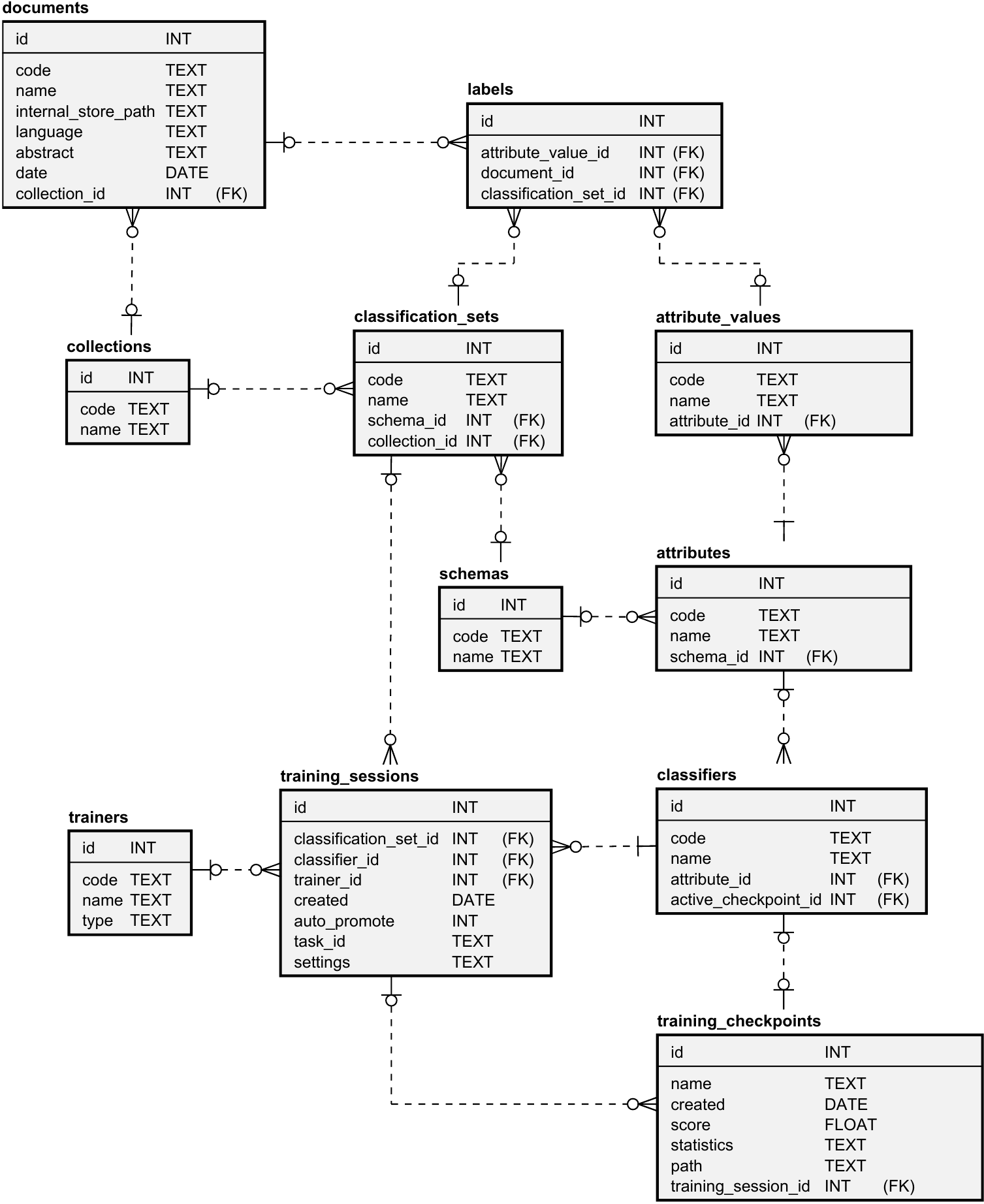}{Database tables}

\subsection{schemas Table}
This table saves schemas when they are created through the REST service. A schema in this project is a collection of attributes. The attributes are saved in the \coderef{attributes} table.

\subsection{attributes Table}
The \coderef{attributes} table stores the attributes that a schema has. An attribute can only belong to one schema, but of course an attribute with the same name and values can be generated within another schema if desired. Every attribute in the system has multiple associated values that the attribute can take on every document.

\subsection{attribute\_values Table}
The values for an attribute are stored in the \coderef{attribute\_values} table. Each value has a \coderef{code} that can be used to distinguish the attribute.

\subsection{collections Table}
This table stores collections of documents. This allows the system to maintain several separate document corpora at the same time. The actual document data for each collection is stored in the \coderef{documents} table.

\subsection{documents Table}
For every document, the \coderef{documents} table stores a name (title), a code (foreign identifier), the (private) physical path on the server where the document data was stored and some metadata such as the language, publication date or a short abstract. While this metadata is not strictly necessary for classification, it allows for some bookkeeping of documents in the system and the ability to query for documents by specific criteria.

\subsection{classification\_sets Table}
A classification set is a labeled subset of documents within a specific collection which is used for training of classifiers.

\subsection{labels Table}
The \coderef{labels} table stores the labels of a classification set. It connects a specific document with one or more attribute values.

\subsection{trainers Table}
A trainer is used to create classifiers. This table stores the available trainers in the system. The table is read only during normal operation and is only edited if new trainer types are added to the system. In this case the \coderef{type} column describes the python package, module and class name separated by dots of the trainer class that is loaded when the trainer type is requested.

\subsection{training\_sessions Table}
Whenever a trainer is invoked, a new training session is created. An important column of this table is the \coderef{task\_id} column that specifies the id of the task in the celery task queue. With this id the current state and progress of a training session can be queried. This information is then merged and presented to the user in a unified way.

\subsection{training\_checkpoints Table}
During the training of a classifier, one or more checkpoints are created (at least one when the training was successful). These checkpoints represent fixed points within the training from which a classifier can be created. The table stores only metadata about this checkpoint such as the standard score achieved on the validation documents and serialized statistics data provided by the trainer. The actual data needed to instantiate a classifier is stored at the path mentioned in the \coderef{path} column.

\subsection{classifiers Table}
The table \coderef{classifiers} stores the classifiers that can be used in the system. From an implementation standpoint a classifier is little more than a symbolic link to a specific training checkpoint that is used for classification. This link is stored in the \coderef{active\_checkpoint\_id} foreign key. This allows users to train the ``same'' classifier multiple times. This is especially powerful in combination with the \coderef{code} column. Other systems using this project can create a classifier once and identify it with a specific code. The classifier can be retrained at will without having to adjust the other systems or pointing them to another classifier.

%% file: chapters/05.evaluation.tex


\chapter{Evaluation}

\label{chap:Evaluation}

\section{Datasets}
\subsection{Standard Datasets}
In literature there are many different datasets used for classification tasks. For the specific case of categorization, there are two datasets which are used very often. These are the Reuters news corpora \parencite{Reuters,Lewis04} and the 20newsgroups dataset \parencite{Lang07}.

The Reuters news corpora consist of the RCV1, the RCV2 and the TRC2 dataset. RCV1 is a collection of 810000 English news articles in 103 categories between 1996 and 1997. RCV2 contains 487000 news articles in 13 different languages covering the same time period as RCV1. Lastly TRC2 contains 1800370 news stories between 2008 and 2009.

The 20newsgroup dataset on the other hand contains 18000 newsgroup posts from 20 different newsgroups.

Both datasets face the problem that they contain relative short texts and that the content is not necessarily of scientific content. For the purposes of this thesis they are therefore not suitable. Regardless of this it would be interesting to get a comparison on these datasets as well to see if the classifier can be generalized to other datasets with ease. This prospect will be discussed later on.

As a perfect dataset for the task on hand. The arXiv database was chosen instead as a dataset.

\subsection{The arXiv Dataset}
arXiv\footnote{\url{http://arxiv.org/}, last retrieved 25th of June 2016} is an e-print service provided by the Cornell University and provides open access to over 1.1 million academic papers. Authors are required to specify a primary category when submitting a paper and can optionally assign additional categories to the paper. The categories of archive are grouped into a hierarchy of three layers which in this thesis will be called subject area, collection and sub-category. For the evaluation of the classifiers, only the collections will be considered.

The following is a list of all subject areas and their collections currently present on arXiv (as of March 2016).

\begin{itemize}
\item Physics
\begin{itemize} 
    \item Astrophysics (astro-ph)
    \item Condensed Matter (cond-mat)
    \item General Relativity and Quantum Cosmology (gr-qc)
    \item High Energy Physics - Experiment (hep-ex)
    \item High Energy Physics - Lattice (hep-lat)
    \item High Energy Physics - Phenomenology (hep-ph)
    \item High Energy Physics - Theory (hep-th)
    \item Mathematical Physics (math-ph)
    \item Nonlinear Sciences (nlin)
    \item Nuclear Experiment (nucl-ex)
    \item Nuclear Theory (nucl-th)
    \item Physics (physics)
    \item Quantum Physics (quant-ph)
\end{itemize}
\item Mathematics
\begin{itemize} 
    \item Mathematics (math)
\end{itemize}
\item Computer Science
\begin{itemize} 
    \item Computing Research Repository (CoRR)
\end{itemize}
\item Quantitative Biology
\begin{itemize} 
    \item Quantitative Biology (q-bio)
\end{itemize}
\item Quantitative Finance
\begin{itemize} 
    \item Quantitative Finance (q-fin)
\end{itemize}
\item Statistics
\begin{itemize} 
    \item Statistics (stat)
\end{itemize}
\end{itemize}

arXiv explicitly allows the bulk download of all its data using standard interfaces. Metadata can be downloaded using the Open Archives Initiative Protocol for Metadata Harvesting (OAI-PMH), using a special REST API (arXiv API) or using an RSS feed. The actual document data can be downloaded as a collection of tar archives from Amazon S3. The tar archives are split by year, month as well as file size (max. ~500MB). There are separate archives for PDF files and Latex source files. Most papers are available in both versions. Downloading requires special tooling capable of communicating with the Amazon S3 cloud. In this project the free and open source S3cmd tool available online was used. The PDF archives can be downloaded using the following command line:

\begin{lstlisting}[language=bash,caption=Downloading arXiv files]
$ s3cmd get s3://arxiv/pdf/arXiv_pdf_* --skip-existing --requester-pays
\end{lstlisting}

This requires the user to have an Amazon S3 account himself, which in itself is free. It should be noted however that arXiv uses a so-called ``requester pays'' bucket, which means that Amazon WILL charge for the download itself. In June 2016 the cost of transfer for one gigabyte was \$0.09\footnote{\url{https://aws.amazon.com/s3/pricing/}}. The complete archive is currently about 270 GB in size. Of course, arbitrary months or years can be downloaded separately if not all data is required. During the development of this thesis initially only a small subset of the archive was downloaded which was then incrementally increased as required.

\subsubsection{Preprocessing}
The following preprocessing steps have been done on the archives downloaded from arXiv.org.

\begin{enumerate}
\item All .tar archives have been extracted into folders which only contain the PDF files.
\item For all PDF files the full texts were extracted using the pdfminer library for python. Few files were not processable by the library, some other files took too long to process with the library. These files were ignored for further processing and classification.
\item Files were also ignored if the resulting full text was smaller than 2 Kilobytes or if the used tokenizer did not detect any word in the resulting file. This step removes some PDF files which do not have extractable text data and also removes a lot of placeholder PDFs for retracted papers.
\item Lastly the python langdetect package was used to detect the language from the full text data. All non-English files were also ignored.
\end{enumerate}

The goal of the processing steps is too produce a high quality dataset where the documents represent real scientific papers. The preprocessing is the same for all experiments and algorithm results shown in this document.

\section{Metrics}
\subsection{Introduction}
To compare the performance of different classifiers, some form of measurement (or metric) is needed which gives us objective values for the various classifiers. Looking at the simple case where only one category exists and a classifier has to decide whether an item is part of this category or not (binary classifier), four possible outcomes can be identified for each classified item:

\begin{enumerate}
\item The item belongs to the category and the classifier correctly predicted this. This is called a True Positive ($TP$)
\item The item item does not belong to the category and the classifier correctly predicted that it does not. This is called a True Negative ($TN$)
\item The item item does not belong to the category, but the classifier predicted that it does. This is called a False Positive ($FP$)
\item The item belongs to the category, but the classifier failed to predict this. This is called a False Negative ($FN$)
\end{enumerate}

\parencite{Sokolova09}\\

It is easy to see that a good classifier should have as many True Positives and True Negatives as possible while at the same time reducing the False Positives and False Negatives as much as possible.

Using absolute numbers makes it difficult to compare various classifiers if the sample size was not the same and having 4 different numbers to compare makes it also hard. This is why usually metrics are used that combine various of these numbers into a single score.

One interesting measure is for example what percentage of the items belonging to a category were correctly identified as such by the classifier. That is called the ``Recall'' defined by $Recall = TP / (TP+FN)$. Another interesting measure is how correct the classifier is with its predictions, meaning what percentage of the items that the classifier predicts as belonging to the category ($TP+FP$), do actually belong to this category ($TP$). This is called the ``Precision'' defined as $Precision = TP / (TP+FP)$. \parencite{Sokolova09}

Unfortunately neither recall nor precision alone are sufficient to measure the quality of a classifier. As an example a classifier that simply predicts the category for all items regardless of whether the item actually belongs to the category\footnote{Such a classifier is also called a ``trivial acceptor'' in comparison to a ``trivial rejector'' which does the opposite \parencite{Sebastiani02}}, will have a recall value of 100\% since all items that had the category were also identified as such. On the other hand a classifier which is overly careful and assigns a category to the item only in ``easy'' to predict cases, might reach a precision of 100\% because it did not falsely assign a category where it did not belong. This ignores however that it might have missed 99\% percent of the samples that actually belong to the category.

Is is important to notice that in the first example the classifier with a recall rate of 100\% would have a bad precision as it just assigned the category to all items. In the second example, the classifier with a precision of 100\% on the other hand would have a bad recall rate since it fails to identify many samples. This shows that there is an interdependency between precision and recall and it is usually not possible to achieve 100\% in both precision and recall. Many classifiers have parameters (such as thresholds) with which the classifier can be tuned to favor either precision or recall. Often this interdependence is plotted in a diagram with one axis being recall and the other one being the precision.

\subsubsection{F1 Measure}
Since the goal is to have a measure to directly compare two classifiers and both precision and recall should be accounted for, one can combine them into a single metric. There are several ways to do this. Arguably the most popular metric in Information Retrieval that combines precision and recall is the $F_1$-measure which is defined as the harmonic mean of both precision and recall \parencite{Sokolova09}:

\begin{align}
F_1 = 2 \cdot \frac{precision \cdot recall}{precision + recall}
\end{align}

In some cases one of the factors, precision or recall, is seen as more important than the other. The measure can be adapted to this by weighting one of the factors higher.

\subsubsection{Accuracy}
Another very popular measure which however is not directly based on precision and recall is the ``accuracy''. Its definition is quite straight-forward and seems very intuitive:

\begin{align}
N &= TP+TN+FP+FN\\
accuracy &= \frac{TP+TN}{N}
\end{align}

Accuracy is the percentage of items that were correctly identified ($TP + TN$) given the total number of items ($N$) \parencite{Sokolova09}. At first glance it seems to solve the problems that were identified with using precision or recall alone, as both true positives and true negatives need to be optimized in order to get a high accuracy value. However the big problem with accuracy are imbalanced groups. If the positive or negative group is much bigger than the other one, it is again easy to build a trivial acceptor or rejector that will reach a high accuracy value without having any predictive power.

As an example it could be desired to build a binary classifier that is able to recognize if a given document is about ``string theory'' or not. It is assumed for this example that only 1\% of the documents given to the classifier are about ``string theory'' (positives) and the other documents are about other topics (negatives). By constructing a simple classifier that always predicts ``NOT string theory'' (negative) for any item regardless of its content, it will reach an accuracy of 99\%! The classifier was ``optimized'' to get the maximum $TN$ value possible (since it predicted negative in all items, it got all the actual negatives right at least) and ignored the $TP$ value which would have been very small anyway (maximum of 1\% of items).

Even worse, a ``better'' classifier, which identifies all positive items correctly and only miss-classifies a few negatives as positive can easily get worse accuracy than the ``stupid'' one.

Since unbalanced groups are very common in practice, it often makes more sense to use the `F1' score or even to report various metrics. Even more important is to report the number of positive and negative samples so that the reader can interpret the results accordingly.

\subsection{Metrics in Multi-Class Classification}
Up to now only the case of binary classification has been discussed where the classifier can only differentiate between two classes (``positive'', ``negative''). Since in this project the goal is to differentiate between many classes, the definitions need to be extended to this case. Looking at how $TP$, $TN$, $FP$ and $FN$ were defined above, one can group them into a two-by-two table according to the predicted vs actual class.

\begin{table}[H]
  \centering
\begin{tabular}{l|l|c|c|}
\multicolumn{2}{c}{}&\multicolumn{2}{c}{Actual}\\
\cline{3-4}
\multicolumn{2}{c|}{}&Positive&Negative\\
\cline{2-4}
\multirow{2}{*}{Predicted}& Positive & $TP$ & $FP$\\
\cline{2-4}
& Negative & $FN$ & $TN$\\
\cline{2-4}
\end{tabular}
  \caption{Binary confusion matrix}
  \label{tab:binary_confusion_matrix}
\end{table}

Normally one would fill the cells of the table with the actual values and it would be called the ``confusion'' matrix since it shows quite well what classes the classifier ``confuses'' with other classes.

Extending such a confusion matrix to multiple categories is simple:

\begin{table}[H]
  \centering
\begin{tabular}{l|l|c|c|c|c|}

\multicolumn{2}{c}{}&\multicolumn{4}{c}{Actual}\\
\cline{3-6}
\multicolumn{2}{c|}{}&$c_1$&$c_2$&\ldots & $c_k$\\
\cline{2-6}
\multirow{4}{*}{Predicted}& $c_1$ & & & &\\
\cline{2-6}
& $c_2$ & & & &\\
\cline{2-6}
& \ldots & & & &\\
\cline{2-6}
& $c_k$ & & & &\\
\cline{2-6}
\end{tabular}
  \caption{Multi-class confusion matrix}
  \label{tab:multi_confusion_matrix}
\end{table}

Now the $TP$, $TN$, $FP$ and $FN$ can be interpreted by looking at a single category ($c_2$):

\begin{table}[H]
  \centering
{
\newcommand{\tn}[0]{\textcolor{gray}{$TN$}}
\begin{tabular}{l|l|c|c|c|c|}
\multicolumn{2}{c}{}&\multicolumn{4}{c}{Actual}\\
\cline{3-6}
\multicolumn{2}{c|}{}&$c_1$&\textbf{\boldmath$c_2$}&\ldots & $c_k$\\
\cline{2-6}
\multirow{4}{*}{Predicted}& $c_1$ & \tn & $FN$ & \tn & \tn\\
\cline{2-6}
& \textbf{\boldmath$c_2$} & $FP$ & $TP$ & $FP$ & $FP$ \\
\cline{2-6}
& \ldots & \tn & $FN$ & \tn & \tn\\
\cline{2-6}
& $c_k$ & \tn & $FN$ & \tn & \tn\\
\cline{2-6}
\end{tabular}
}
  \caption{Multi-class confusion matrix (2)}
  \label{tab:multi_confusion_matrix_2}
\end{table}

\begin{enumerate}
\item All items that belong to the category and were correctly predicted as such are the true positives $TP$
\item All items that do not belong to the category and were predicted as not belonging are the true negatives ($TN$) (regardless of whether they were correctly classified!)
\item All items that do not belong to the category, but for which the classifier predicted that they do are the false positives ($FP$)
\item All items belonging to the category for which the classifier failed to predict this are the false negative ($FN$)
\end{enumerate}

Using this, the precision, recall and the other metrics discussed above can be calculated for a single category. By averaging the values for the metric over all classes, a score for the whole classifier is obtained. This is called a macro average. Example: $F_{1, macro} = avg(F_{1,c_1} \ldots F_{1,c_k})$. Another possibility is calculating a micro average by first summing the individual $TP$, $TN$, $FP$ and $FN$ for all classes and then calculating the metric based on this. Macro averaging treats all classes the same regardless of their size, while micro averaging will favor bigger classes.

\parencite{Sokolova09}\\

It is important to notice that the imbalance problem of the accuracy metric described before is even worse for the multi-class case. Every single class is evaluated separately as if it were a binary classifier that decides if an item belongs to this class or one of the other classes. This means that for multiple classes there are usually a lot more items that do not belong to the category than items belonging to the category. So even if the classes themselves are balanced, the comparison will be imbalanced. This will lead to a high accuracy for classifiers that simply reject most of the items (achieving a high number of true negatives).

\subsection{Validation Methods}
From a practical standpoint one has to decide what data should be used to compute the metrics above when validating the performance of a classifier. In \autoref{sec:evaluation} it was already discussed that reusing the training data for validation is a bad idea, because most classifiers perform a lot better on training data than on data which the classifier did not see before.

The usual way to evaluate a classifier is therefore to separate the labeled data into a validation set and a training set. The training set is used exclusively for the training and the validation set is used for evaluation purposes\footnote{Sometimes a third test set is created so that the validation set can be used for hyperparameter tuning and the test set is used for the final evaluation.}.

A disadvantage of this method is that only a part of the labeled data is used for validation. When the dataset is small, this can lead to imprecise results. One way to solve this is to use \emph{n-fold cross-validation}. With this method, the dataset is first split into $n$ groups (or folds) and the classifier is trained and validated $n$ times. For every training session a different group is used as the validation set and the rest of the groups are used as the training data. At the end the validation results for all runs are averaged to get a final score. \parencite{Mohri12}

Another variant is the \emph{Monte Carlo cross-validation} method, in which the training and validation is also repeated several times. The validation set is however randomly chosen each time. The advantage of this method is that the number of iterations does not depend on the number of folds. A disadvantage is that because of the random selection some of the data might never be used for validation whereas other data is used multiple times. When enough iterations are used however, these imbalances are not a problem as they are averaged out and the model can even provide superior results to n-fold cross-validation due to the higher number of possible configurations. \parencite{Shao93}

\section{Comparison of CNN Classifier to Linear SVM}
To evaluate the performance of the developed classifier, a baseline is needed. The baseline is given by a standard linear \acrlong{svm}. \acrpl{svm} are known as reliable classifiers with very good performance. Using kernel functions they can even solve classification problems that are not linearly separable. However in the case of document categorization a standard linear \acr{svm} seemed to provide the best performance.

Unless otherwise mentioned the following parameters have been used for all tests:

\begin{itemize}
\item The arXiv dataset was used as a dataset. The main collection in which a document is contained was taken as the target category (single-label).
\item For every category a maximum of 1000 documents was considered.
\item 10\% of the documents were randomly chosen as a validation set which is the basis of the reported scores.\footnote{This is done automatically by the service. See remarks in \autoref{sec:worker_package} for details}
\item Each test was run at least three times and the average was taken, so that small variations between test runs do not influence the comparison results. With each run the training/validation data split was different (Monte Carlo cross-validation).
\item The number of categories was limited to 10. In general, for any number of categories, the categories were chosen in this order: cond-mat, hep-ex, hep-th, math-ph, physics, astro-ph, hep-ph, math, quant-ph, gr-qc, nucl-th, cs, nlin, hep-lat, nucl-ex. This order was chosen partly based on the number of documents in each category and partly based on the perceived difficulty in differentiating between the categories.
\item As a performance measure the F1 score is used as described earlier. Since all categories have the same number of documents, the micro and macro average are the same and are therefore not reported separately.
\end{itemize}

Specifically for the \acr{cnn} classifier, the following settings were applied:

\begin{itemize}
\item The classifier was trained during 50 epochs. The highest score obtained within these 50 epochs was chosen.
\item Only the first 500 words in every document were evaluated by the classifier.
\item The batch size was set to 500. This is a quite high number and was made only possible by by the fact that the machine had a high end \acr{gpu} with lots of memory and that some of the other training parameters above are optimized for performance and memory consumption.
\item The neural network itself uses filter lengths of one, two and three words with 200 filters each, a hidden layer of 200 neurons and a dropout of 0.3 after each layer. LeakyReLU was used as the activation function.
\end{itemize}

The parameters described above have been chosen as a compromise between good classification performance and fast training times. In the following sections, some variations to this parameters will be shown and their effects on the overall classification performance.

The \acr{svm} classifier uses tf-idf values computed from the complete documents without restricting it to the first 500 words, unless otherwise mentioned.

\subsection{Overall Performance}
To get a precise estimation on the overall performance, this test was run 10 times instead of the usual 3 times. Also the standard deviation was determined.

The overall (F1) score for the parameters described above was 84.2\% ($\pm$ 0.50\%)  for the \acrlong{cnn} classifier vs. 70.7\% ($\pm$ 2.04\%) for the \acr{svm} classifier. As can be seen, the \acr{cnn} classifier performs significantly better than the \acr{svm} baseline.

\subsection{Number of Categories}
\figgraph{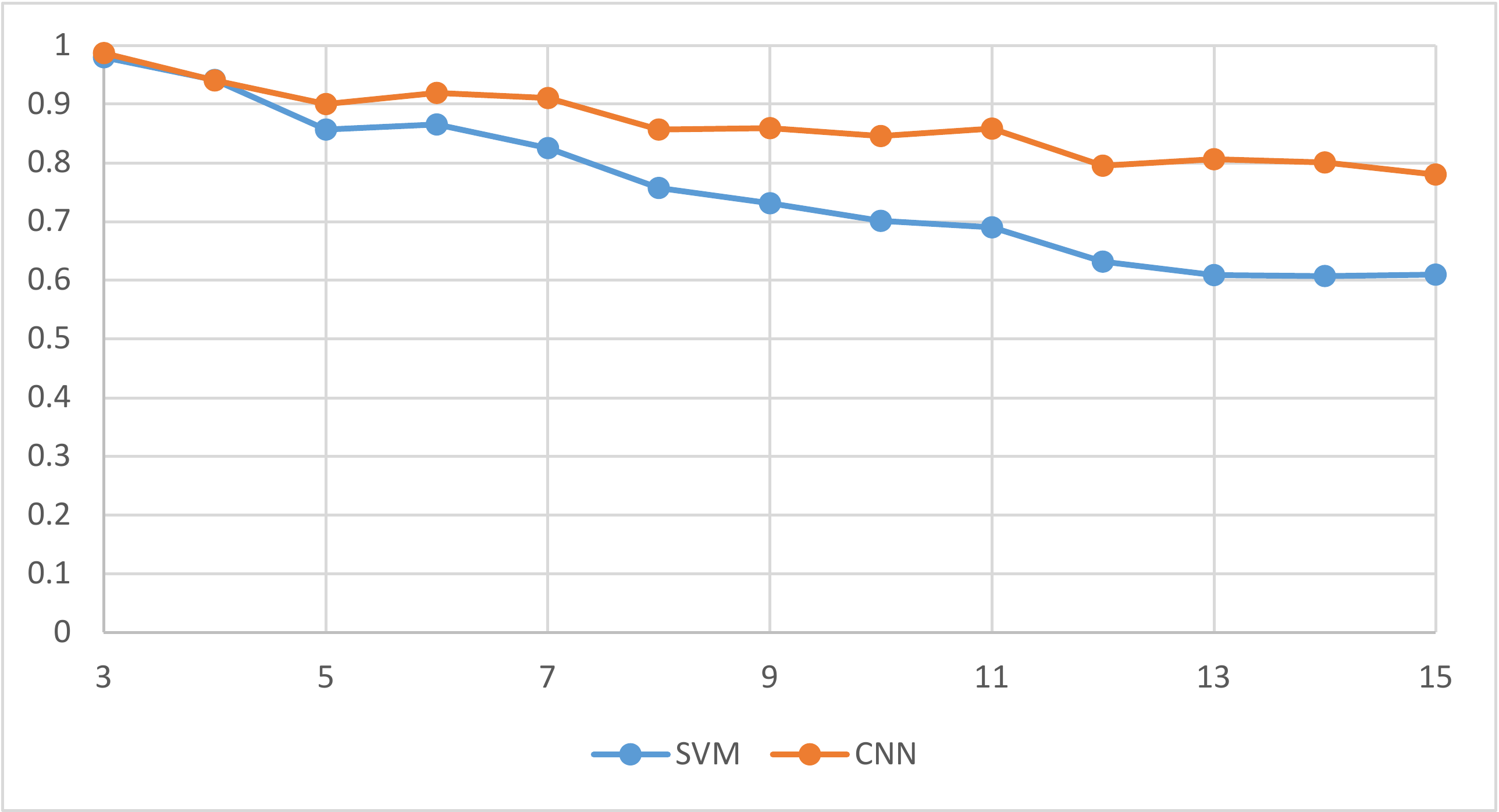}{Performance dependent on number of categories}
When varying the number of categories, it can be seen that the \acr{cnn} performance stays superior, even if the number of categories increases. In fact the distance between the two classifiers seems to be increasing with more categories.

\subsection{Number of Documents per Category}
\figgraph{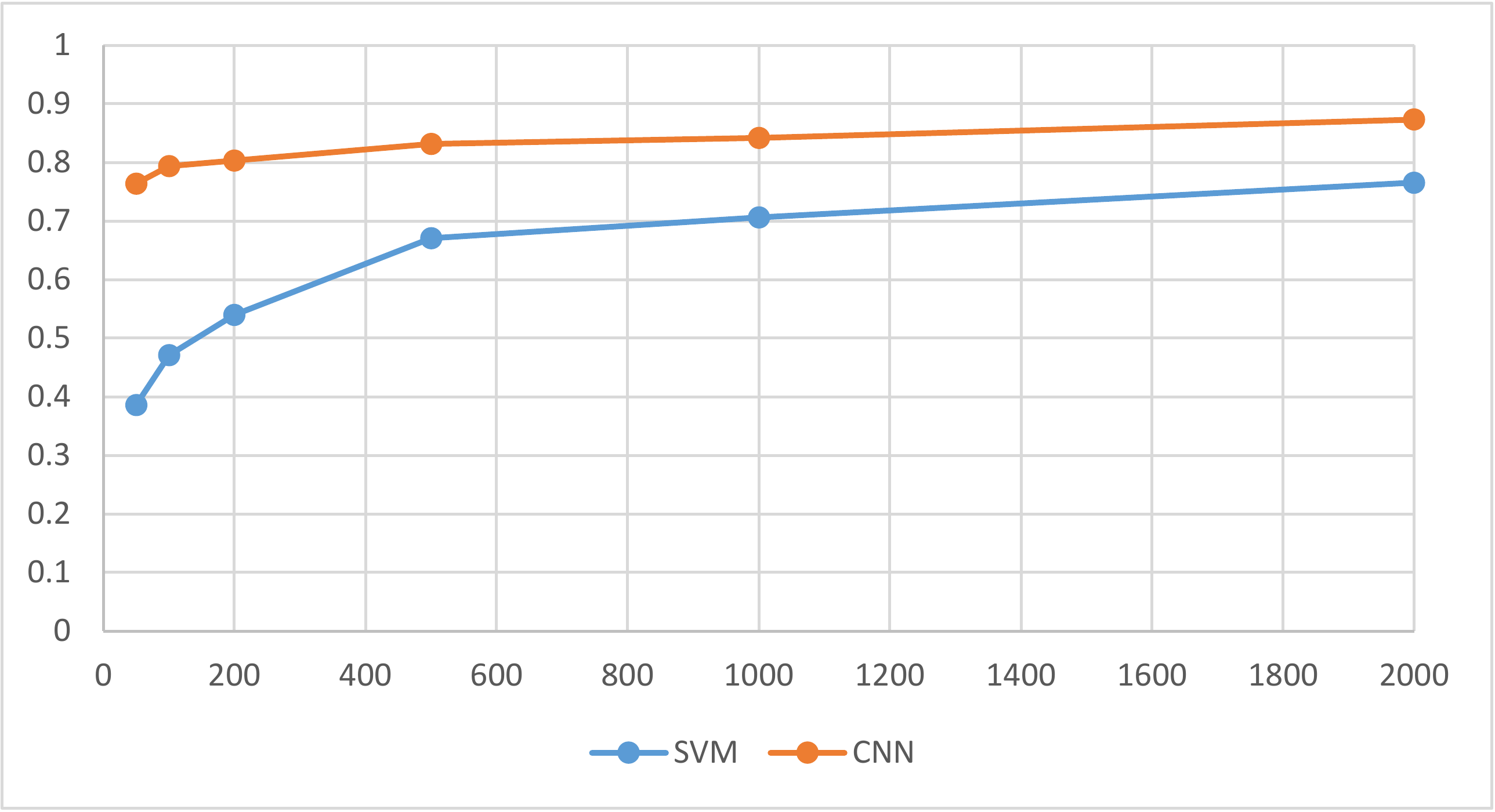}{Performance dependent on number of samples per category}
As another experiment, the number of documents for each category was varied from 50 up to 2000. Interestingly it appears that the \acr{cnn} seems to work a lot better in generalizing from even a small number of samples that the \acr{svm}. This is an important finding because neural networks are often thought of requiring a lot of training samples to perform efficiently. The network parameters were not adjusted in any way, so it seems that the default parameters described above work for small number of documents as well as large number of documents.

\subsection{Number of Tokens Evaluated}
\figgraph{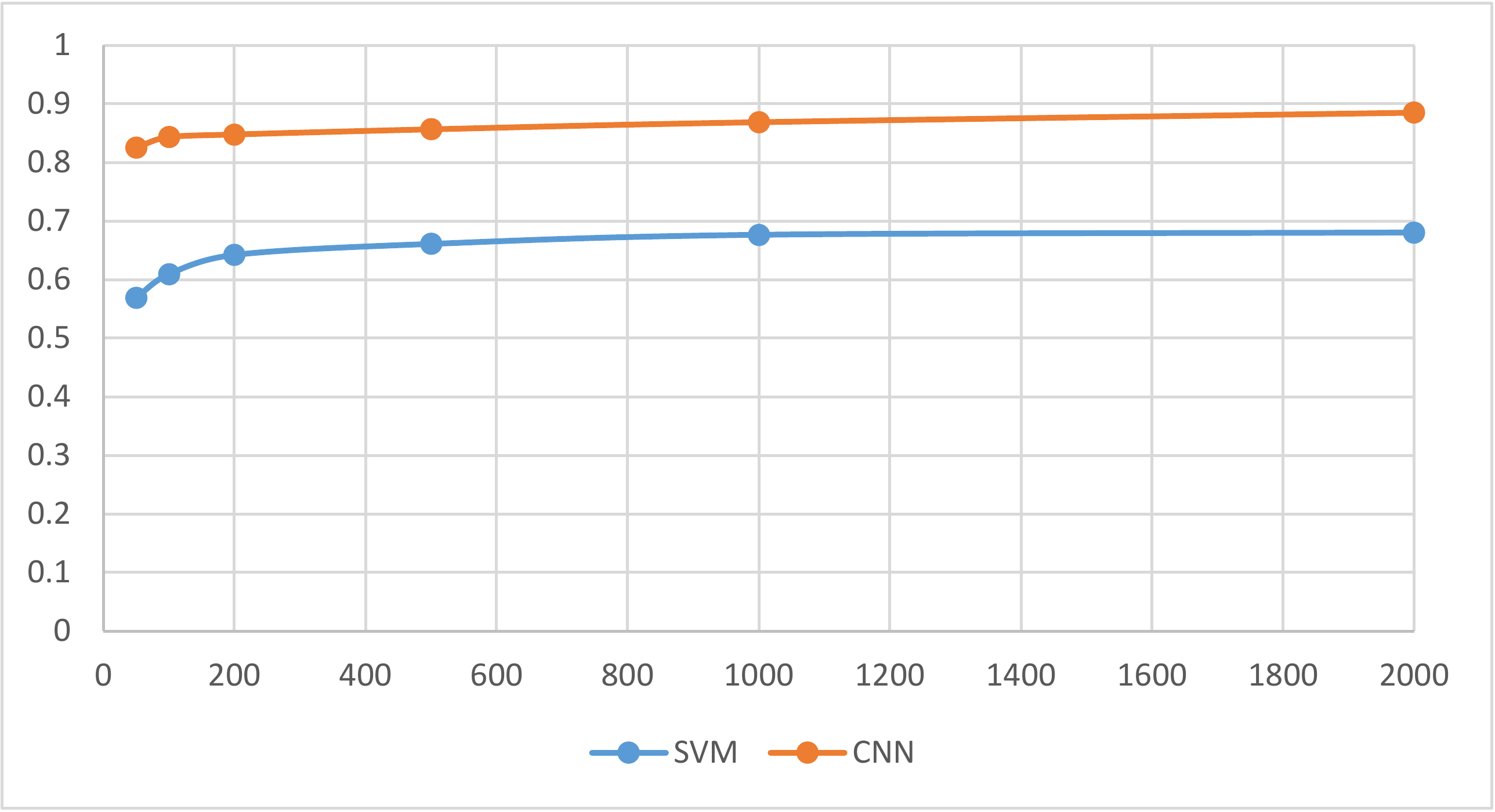}{Performance dependent on number of evaluated tokens}
Another question is how many words are needed in each document to reliably detect the correct category. The results show that as few as 50 words are needed to detect the correct category in more than 80\% of cases. The \acr{svm} classifier also still shows good performance but seems to be more sensitive to the total sequence length.

It has to be noted that for this experiment the batch size was reduced to 200 documents as the \acr{gpu} memory requirement would have been to high for a sequence length of 2000. For the base line of 500 words per document this reduction resulted in a slight increase of overall performance to 85.67\%, probably caused by the increased number of total updates to the classifier.

Also it is worth remembering that with the base line configuration of the \acr{svm} classifier, there is no restriction on the number of evaluated words. The \acr{svm} classifier will always evaluate the whole document.

\subsection{Performance Over Time and Overfitting}
\figgraph{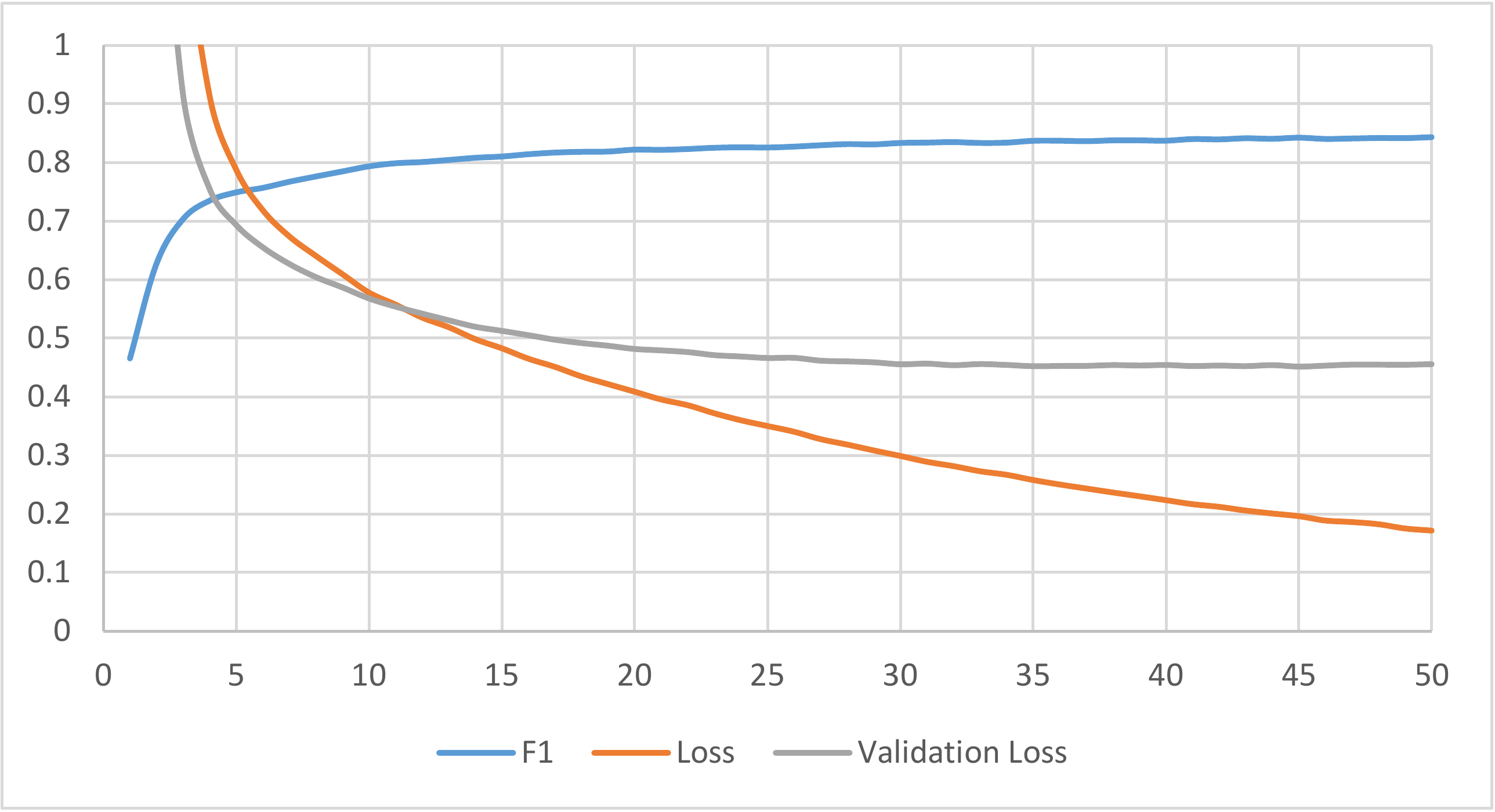}{Performance over time}
Lastly it's interesting to see how the classifier performs during the course of the training and how many epochs are really needed for optimal performance.

The graph shows the F1 score (on validation samples) as well as the loss function for the training samples and the validation samples. As can be seen, the performance as measured by the F1 score does not significantly improve anymore after about 30 epochs. Similarly the validation loss function, where lower values are better values, does no longer improve after this time. The general loss function on the training data does still improve but this does no longer translate into measurable effects on the validation data. A good sign is that there is no indication of overfitting where a decrease in training loss would actually cause an increase in validation loss when the classifier learns to memorize the training samples instead of generalizing.

Some possible explanation for this could be the extensive use of dropout in all layers, the use of dense vectors instead of sparse vectors as input (reducing the number of variables/weights to train) and the use of convolutional filters (further reducing the number of weights to be trained).

\subsection{Training Time}
The training time for a single epoch on a Core i5 6600 and a NVIDIA GeForce GTX 980 Ti graphics card with data read from a Samsung Solid State Disk was about 54 seconds or just under 1 minute on overage. In comparison, the total time needed by the \acr{svm} classifier for training was only about 5 minutes. This means that for a complete training cycle, the \acr{cnn} classifier needs almost 10 times as long.

For a fair comparison which includes also the training time, one would have to check the performance of the \acr{cnn} classifier after 5 training epochs, which is about 74.9\% and only slightly better than the \acr{svm} performance of 70.7\% described in the first chapter. It should also be noted that most of the time in the \acr{svm} classifier is spend on computing the tf-idf values for all documents, which is a task that can easily be optimized by preprocessing without requiring a lot of additional storage capacity. On the other hand, preprocessing with the \acr{cnn} classifier is more difficult since precomputing and storing the word2vec vectors for the documents requires more space than the original documents.

\section{Comparison of Word Vector Models}
\figgraph{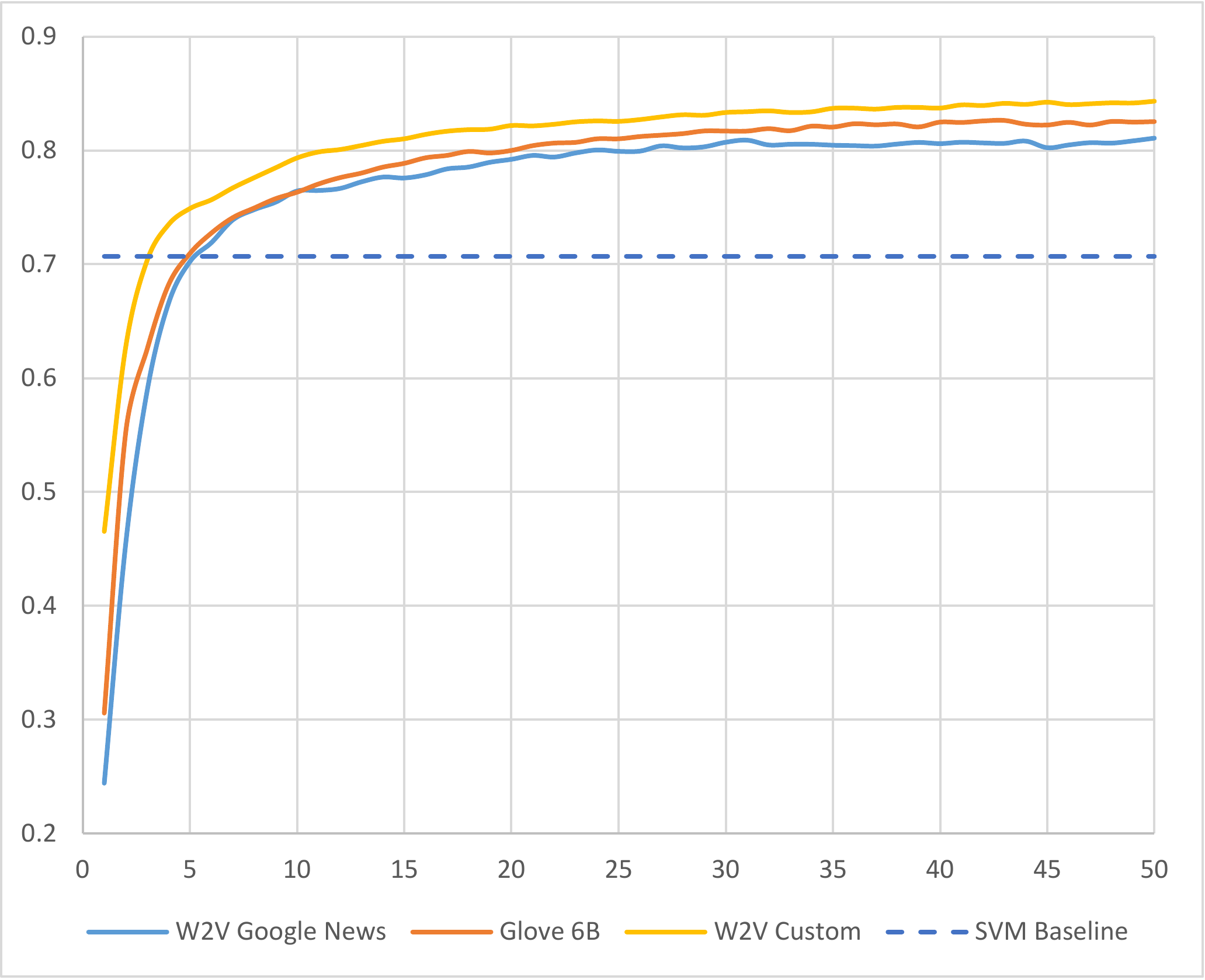}{Comparison of word vector models}
In the experiments so far, the word vector model used was a custom word2vec model trained using the skip-gram technique on 444430 documents taken from the arXiv library with about 2.5 billion words in total. The vocabulary was not trimmed and contains about 8 million different words. The total space required for this large model is just over 11 Gigabytes.

There are some pretrained word vector models available online that can be used as an alternative, when training a custom model is not possible or not desired.

One example is the Google News model\footnote{Downloadable at \url{https://code.google.com/archive/p/word2vec/}}, so called because it was trained on Google News articles with a total of 100 billion words. The vocabulary is trimmed to contain 3 million words and phrases. The model is also trained with the word2vec skip-gram technique.

Another example are the Glove\parencite{pennington14} models. Glove uses another technique than word2vec to obtain word vectors. On the project website\footnote{\url{http://nlp.stanford.edu/projects/glove/}} various pretrained models are available. For this comparison a model based on a Wikipedia dump and the Gigaword corpus was used (``Glove6B''). The training consisted of 6 billion words and the final vocabulary contains 400 thousand words.

The first interesting thing to notice is that the Glove model provides better performance than the Google News model, which is remarkably considering that the Glove model was trained on a much smaller corpus (6 billion vs. 100 billion) and the vocabulary is much smaller (444 thousand vs. 8 million). This could be due to the fact that the Wikipedia articles used to train Glove result in a better model for the arXiv dataset than the news articles used in the Google News model or it could indicate that the algorithm used in Glove results in higher quality vectors as indicated by the Stanford authors \parencite{pennington14}.

In any case the customized word2vec model seems to provide the best performance even though it was trained on only 2.5 billion words. This is a further indication that the domain in which a model is trained is more important than the corpus size.

Since the models not only use different corpora but also use different training parameters, algorithms and vocabulary sizes, it is difficult to do a fair comparison between them. Further research would be helpful to clarify the optimal word vector strategy.

%% file: chapters/06.prospect.tex
\chapter{Conclusions}
\label{chap:Conclusions}
\section{Summary}
In \autoref{chap:Algorithm} a generic model was proposed to allow classification of documents using a service based \acr{api}. This was accompanied by a classifier based on deep learning principles, which was designed for the purpose of scientific document categorization.

Both the \acr{api} and the classifier were successfully implemented in \autoref{chap:Implementation} in a proof of concept demonstration, showing that it is indeed possible to encapsulate the classification into a reusable \acr{api}.

Evaluation using the F1 metric (\autoref{chap:Evaluation}) showed that the classifier performs very well and provides superior performance to a linear \acr{svm} classifier that was implemented as a baseline. The performance was stable over a wide range of parameters. The training time was identified as a potential problem, where the \acr{svm} still provides a good alternative.

It can be concluded that deep learning based classifiers are a good choice for scientific document categorization and that they can be abstracted well using a generic \acr{api}.

\section{Prospects}
There are many aspects which could be improved in this solution and which might be topics of future research. The following will list just a few of these.

\subsection{Comparison with Other Classifiers on Various Datasets}
One of the first important steps to further prove the performance of the classifier is to compare it with additional classifiers and to do these comparisons on additional datasets. The \acr{svm} baseline showed that the \acr{cnn} classifier performs well under a specific usage scenario, but it is unclear how it would perform in comparison to other state of the art classifiers and using different work loads.

\subsection{Parameter Tuning and Model Refinement}
While many experiments have been done in order to select the current model for categorization, more research is needed to find out how the different hyperparameters effect the classifier performance and to further improve the performance. One example is the selection of the optimal word embedding model as was mentioned in the previous chapter.

\subsection{Automatic Parameter Selection}
The network hyperparameters presented showed good performance over a wide range of different test settings. It is however reasonable to assume that these parameters are not optimal for all use cases. It would be a nice addition if the network can adapt some of the parameters automatically based on the specific work load.

For example, it could make sense to increase the number of hidden neurons if the number of categories increases.

One way to do this would be to determine the optimal set of parameters for a wide range of different work loads and then trying to approximate a function for the optimal set of hyperparameters. Even more sophisticated systems could use machine learning techniques such as a separate neural network to determine the best set of hyperparameters.

\subsection{Early Stopping}
Currently the number of epochs is fixed for a single training run. There are several methods known for neural networks to allow early stopping if it becomes evident that the performance of the network does no longer increase. One simple way for example is to check if the loss function for the validation set did not decrease for several epochs. This early stopping would make typical training times faster and make the system more adaptive to the needs of the concrete work load.

\subsection{Distributed and Cloud Computing}
The solution was designed from the beginning to be easily distributed among several computers. By separating the front-end (web service) clearly from the back-end (processing) using separate processes and using Celery as a distributed task queue, it should already be possible to run worker processes on distributed computers. This however has not been tested so far and would be an interesting area of research. One step further would be enabling the system to run on cloud services so that it can scale up dynamically when needed. As was mentioned before, there is already related work done at the University of Hagen to research such possibilities in general (see \autoref{sec:related_work}).

Another interesting step would be to distribute a single training session among several machines. This is currently not possible, mainly because the libraries used internally do not support this in an easy way. However once this is technically supported it should be possible to integrate as well.

\subsection{Improved Service Functionality}
There are several improvements that could be made to the service itself to allow easier management of all resources in the system and to allow additional functionality. For example it would be nice if attributes were supported which are not category like (e.g. author or publication date) and being able to train classifiers for these attributes as well.

\subsection{Improved Robustness and Fault Tolerance}
All in all a functional prototype was built. In order to move this prototype into production, the code needs to be tested extensively in order to find and correct possible bugs. Also the system needs to be more robust in case of error conditions.

%% file: chapters/A.lessons.tex
\chapter{Lessons Learned}
This chapter is intended to give the reader some practical tips on how to train deep neural networks based on experience gathered during the development of this project.

\section{Keeping Statistics and Result Plotting}
Designing a neural network in general and maybe even more so designing a deep neural network requires a lot of experimentation with various models and hyper parameters. In order to compare the performance of these models it is very helpful to not only automatically evaluate the models and log various metrics at every training epoch but also to plot these results to an image file. It is a lot faster to make quick compares using such a graph than evaluating numbers in a spreadsheet. This is even more important during early development where bigger changes are more likely to occur than during the final evaluation stages where one might only need to do a few runs.

Independent of this the raw numbers should also be persisted automatically into a file for later use. In this project CSV files were used for this as they are easy to write and consume. Metrics which are useful for both plotting and recording include (for every epoch):

\begin{enumerate}
\item Training loss function (result of the loss function on the training samples)
\item Validation loss function (result of the loss function on the validation samples)
\item Validation F1 score or accuracy
\item Time spent
\end{enumerate}

These will help to get a general impression of the performance and also to notice problems such as overfitting.

\section{Detecting and Correcting Overfitting}
It is very important to notice when a model is overfitting and to adjust the model to compensate for this.

To notice overfitting it is very helpful to compare the training loss and the validation loss. The neural network is always trying to optimize the training loss and is often quite good at it. Overfitting occurs when the network simply learns to repeat the training samples without generalizing. This means that when using another set of samples not used during training(!) the network will perform worse or in the worst case not at all. The graph in \autoref{fig:overfitting} shows an example of overfitting.

\figgraph{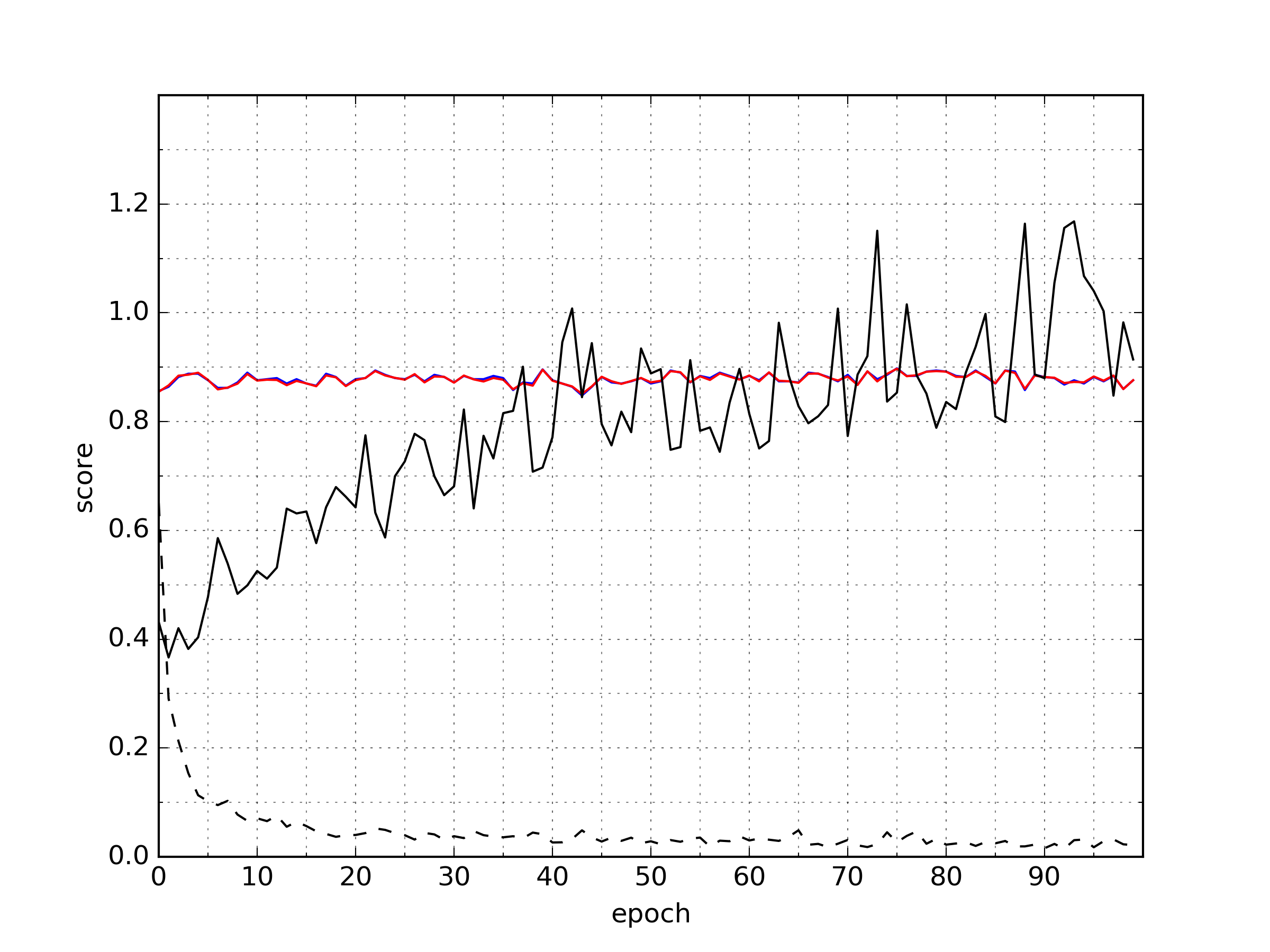}{Overfitting}

The dotted line in this graph shows the loss function for the training set after each epoch. As can be seen the value is almost steadily decreasing which in itself is a good sign and shows that the network is actually adapting to the training samples. The solid black line on the other hand is the same loss function but evaluated on the validation set after each epoch. In the good case, the function should look similar as the training loss, steadily decreasing. If the function starts to increase after some epoch (like here already after the first few epochs) it indicates that overfitting is occurring. Notice that the red and blue line in the graph which displays the F1 score of the classifier remains largely constant during this particular experiment. Overfitting doesn't always directly affect the F1 or similar scores since the classifier outputs probabilities for each class, but before the F1 score is calculated the probabilities are binarized so that a sample is either in a class or not. It doesn't matter if the classifier was 60\% sure or 80\% sure of that.

If overfitting is detected there are fortunately a number of ways to reduce or eliminate overfitting:
\begin{enumerate}
\item The recommended way is always to increase the number of training samples. With every epoch the classifier sees the same training data, so instead of using many epochs with the same training data, it is more effective to use fewer epoch with a higher number of training samples. Of course in practice the number of training samples is often limited so this is not always an option. Depending on the use case it might be possible to automatically modify existing training samples to create new training samples. This technique was not used within this project though.
\item Another option can be to reduce the complexity of the network by removing layers or inner nodes. In general the more variables are in the network the easier it is for the network to overfit. 
\item However, based on the experience of the author, the easiest and best option (if increasing the training samples is not possible) is extensive use of dropout layers. This means simply adding a dropout layer or increasing the drop out rate of the layer (e.g. from 30\% to 50\% to 80\%, sometimes even going as high as 90\%). This forces the network to adapt even if a high number of inputs is not available so that it can no longer depend on certain inputs. In a way it is similar to training various networks simultaneously and then combining their results. With a total drop out rate of 80\% only 20\% of nodes are active at any given time, effectively forming a small subnet which is trained separately.
\end{enumerate}

\section{Using Established Libraries}
During development it was very helpful to have state-of-the-art neural network libraries such as Keras and Theano available which remove a lot of overhead and let one concentrate on the actual architecture of the networks instead of dealing with low level details or having to implement all components individually. Of course this does not mean it is not important to know what the components actually do, the libraries just make it easier to use them and often provide highly performance tuned versions (e.g. executing on \acr{gpu}).

\section{Investing in Hardware}
Investing in Hardware can really speed up the time spent on experiments. Currently the most important investment is a good GPU if a library is used that supports it. During the development of this project, upgrading from a (decent, 3-4 years old) notebook that already had a dedicated graphics card to a current desktop computer with high end \acr{gpu} ran some experiments up to 10 times faster. When working with a lot of data a solid state disk is also a good investment since I/O can be a limiting factor sometimes. It also allows using the hard disk as a cache without incurring much performance penalty if the working data does not fit into main memory.

Generally it is very helpful to have a dedicated desktop computer or server which is solely used for running experiments since some of these experiments can run up to several days -- especially if multiple repetitions of the same base experiment need to be run. When working on a single computer it is a good idea to develop during the day and run experiments during the night in order to use resources optimally. Developing and testing at the same time is often not possible as testing can use a lot of computer resources slowing a PC down considerably. Having multiple computers available provides the most options of course. It is a good idea to use version control systems like Git to quickly push code changes to other computers.

%% file: chapters/B.installation.tex
\chapter{Installation}
\label{chap:installation}
The following will demonstrate a basic installation of the system using Debian Linux. Similar instructions can be used to install on other Linux systems or on Windows.

It will also be shown how to configure and execute the various components of the system.

\section{Base System}
For a base installation, a Python (2.7) development environment is required. In addition, for improved performance, some other libraries will be installed as well, such as the openblas library.

The project itself has additional dependencies, which are installed using the \coderef{pip} package manager for Python.

\begin{lstlisting}[language=bash,caption={Installing the Python environment}]
$ apt-get install python-numpy python-scipy python-dev python-pip python-nose python-matplotlib g++ libopenblas-dev git libfreetype6-dev libxft-dev
$ pip install theano keras gensim sklearn
$ pip install flask flask-cors flask-basicauth celery sqlalchemy jsonpickle concurrent-iterator unittest2 sortedcontainers matplotlib
\end{lstlisting}

All commands must be executed as a user with root privileges (using sudo is recommended).

\section{Configuration}
\label{sec:configuration}

Before starting the system, the configuration file \coderef{config.py} in the root directory of the installation must be configured. This can be done using any standard text editor.
\pagebreak[3]

The following settings can be configured:

\begin{lstlisting}[language=python,caption=Configuration of the classification service]
# Directory where service data is stored
DATA_ROOT =  r'/var/lib/classifysvc'

# Connection string for the database
DATABASE = 'sqlite:///var/lib/classifysvc/repo.db'

# Set to true to log all database statements
DATABASE_ECHO = True

# Set to true to enable authentication
SVC_AUTH = False

# Users and passwords accepted by the service
SVC_USERS = {"fuhagen": "pwd", "test": "test" }
\end{lstlisting}

The most important and mandatory setting is the \coderef{DATA\_ROOT} setting, which determines where the service stores uploaded documents and other files.

By default this is also the location for the database, but this can be changed using the \coderef{DATABASE} setting. Using the \coderef{DATABASE} setting it is possible switching to other databases than SQLite, by changing the SQLAlchemy connection string.

To enable basic authentication on the REST service, the \coderef{SVC\_AUTH} setting must be set to \coderef{True}. After this only user/password combinations listed in \coderef{SVC\_AUTH} are allowed access. This should only be seen as a very basic protection mechanism, not as a full fledged solution for user authentication. All configured users have access to all methods of the service and can create, update or delete any of the resources defined by the service.

Celery specific configuration data can be set in the \coderef{celeryconfig.py} file. This should allow running the celery workers on separate machines or even multiple machines at the same time. This has not been tested however.

\section{Running}
To start the system, both the service process and the celery workers have to be started.

The service is started using:

\begin{lstlisting}[language=bash,caption=Starting the service]
$ PYTHONPATH=. python service
 * Restarting with stat
 * Running on http://127.0.0.1:5000/ (CTRL+C to quit)
\end{lstlisting}

The worker processes are run by executing the following commands:

\begin{lstlisting}[language=bash,label=Starting the worker processes]
$ PYTHONPATH=. python -m celery -A worker worker --app=worker.tasks:app --loglevel=info -Q training_queue
$ PYTHONPATH=. python -m celery -A worker worker --app=worker.tasks:app --loglevel=info -Q classify_queue
\end{lstlisting}

The commands are exactly the same in Windows, however to set the environment variable \coderef{PYTHONPATH}, the \coderef{SET} command should be used instead:

\begin{lstlisting}[language=command.com,caption=Execution on Windows]
C:\>SET PYTHONPATH=.
C:\>python ...
\end{lstlisting}

%% file: chapters/C.usage.tex
\chapter{Using the Classification Service}
\label{chap:usage}
This chapter will show a quick overview of training a new classifier using the \acr{rest} \acr{api}. It is intended as a quick overview. For details on the individual service calls, the programming manual should be consulted.

Before a classifier can be trained, three things are needed: A collection of documents, a schema that defines attributes that a document can take and a classification set that contains labels for some of the documents in the collections.

\section{Creating a Collection}
To create a collection, a simple \coderef{POST} request has to be made to the \coderef{/collections/} endpoint of the service. The body of the request must be a \acr{json} \parencite{rfc7159} encoded dictionary containing the name for the collection and a unique code that can be used to identify the collection.

\begin{lstlisting}[language=rest,caption=Creating a collection - Request]
POST /collections/
{
	"name": "My collection",
	"code": "Collection1"
}
\end{lstlisting}

The response of this request will return a 201 status code indicating that the resource was created \parencite{rfc2616}. The location header of the response will point to the new resource and the body of the response contains a representation of the created resource.

\begin{lstlisting}[language=rest,caption=Creating a collection - Response]
HTTP/1.1 201 Created
Content-Type: application/json
Location: /collections/1/
{
		"href": "/collections/1/",
		"id": 1,
		"code": "arxiv",
		"name": "My collection"
}
\end{lstlisting}

As can be seen, the response has an additional \coderef{id} field and an \coderef{href} field. The \coderef{id} field returns the internal identifier for the collection, which is also needed when the collection is referenced elsewhere. The caller either needs to save this identifier or alternatively it can use the \coderef{code} assigned earlier to look up the \coderef{id} later on (e.g. \lstinline{GET /collections?code=Collection1}).

The \coderef{href} field contains the URL of the resource, just like the location header. This supports the concept of hypermedia, where navigation is done using links even in the context of services.

The pattern of \coderef{id}, \coderef{code} and \coderef{href} is present in most endpoints and will therefore not be explained again in the upcoming explanations.

\subsection{Adding Documents to a Collection}
After the collection was created, documents can be added to it, using:

\begin{lstlisting}[language=rest,caption=Adding documents to a collection]
POST /collections/<colid>/documents/
{
	"name": "My first document",
	"code": "doc123"
}
\end{lstlisting}

In the URL, \coderef{<colid>} refers to the id of the collection that was created earlier.

There are additional fields that can be set for a document, which are explained in the manual accompanying this work.

The response uses the same mechanics as seen previously for the collection. It also contains a new \coderef{id} that identifies the new document within the system and which is required in various service calls.

It should be noted that the request above does not include the actual content of the document. To upload the text of a document, a separate POST request has to be made:

\begin{lstlisting}[language=rest,caption=Uploading document data]
POST /collections/<colid>/documents/<docid>/content
This is the content of the document.
\end{lstlisting}

The body of the request is simply the content of the document.

\section{Creating a Schema}
Creating a schema requires the definition of one or more attributes:

\begin{lstlisting}[language=rest,caption=Creating a schema]
POST /schemas/
{
	"name": "My schema",
	"code": "Schema1",
	"attributes": [
		{
			"name": "Category",
			"code": "category",
			"values": ["physics", "math", "biology"]
		},
		{
			"name": "Type",
			"code": "type",
			"values": ["book", "article", "thesis"]
		}
	]
}
\end{lstlisting}

As can be seen the possible values for each attribute is a flat list. If the attribute represents an hierarchy/taxonomy, it is the responsibility of the caller to convert the hierarchical structure into a flat list understood by the service.

The response will not only contain an \coderef{id} for the schema itself, but also for every attribute and for every attribute value within the \coderef{values} array. These are required later to create classification sets.

\section{Creating a Classification Set}
Creating a classification set consists of first creating the set itself and then adding labels to it.

\begin{lstlisting}[language=rest,caption=Creating a classification set]
POST /classificationsets/
{
	"name": "My classification set",
	"code": "Set1",
	"collectionId": <colid>,
	"schemaId": <sid>
}
\end{lstlisting}

\coderef{<coldid>} and \coderef{<sid>} refer to the collection and schema created earlier in the process.

\subsection{Adding Labels}
After a classification set was created, labels can be created that assign one or more attribute values to a document.

\begin{lstlisting}[language=rest,caption=Adding labels to a classification set]
POST /classificationsets/<setid/labels/
[
	{
		"documentId": <docid>,
		"attributeId": <attrid>
		"valueIds": [<valueid>],
	}
]
\end{lstlisting}

The various \coderef{<*id>} fields are placeholders for the various ids created previously, namely for documents, attributes and attribute values.

The process can be repeated several time to add additional labels. Also since the root element is an array, multiple labels can be added at the same time with a single request.

\section{Creating a Classifier}
To create a classifier, the \coderef{/classifiers/} endpoint is used.

\begin{lstlisting}[language=rest,caption=Creating a classifier]
POST /classifiers/
{
	"name": "My classifier",
	"code": "Classifier1",
	"attributeId": <attrid>
}
\end{lstlisting}

A classifier only works on a single attribute of a schema, which must be specified during creation of the classifier. Of course other classifiers can be created that work on other attributes.

\subsection{Starting a Training Session}
Before a classifier can be used, it needs to be trained.

\begin{lstlisting}[language=rest,caption=Starting a training session]
POST /classifiers/<clsid>/trainings/
{
	"trainerId": <trnid>,
	"classificationSetId": <setid>,
	"settings": null
}
\end{lstlisting}

The \coderef{<trnid>} refers to the id of one of the registered trainers (e.g. \acr{cnn} or \acr{svm}). These identifiers can be retrieved using the \coderef{/trainers/} endpoint. The classification set is used for the training and validation data of the classifier. There are several more options, described in the programming manual. 

\subsection{Querying Training Progress}
Training can be a long process. It is possible to retrieve the current progress of the training by querying:

\begin{lstlisting}[language=rest,caption=Querying training progress]
GET /classifiers/<clsid>/trainings/<trnid>/
{
	"id": 150
	"state": "PROGRESS",
	...
	"progress": {
		"current_action": {
			"progress": 0.15
			...
		}
	}
	"checkpoints": [
		{
			"name": "Checkpoint 0",
			"created": "2016-05-31 12:09:23.590000",
			"statistics": {"loss": 0.426, "f1_macro": 0.7, "f1_micro": 0.7, "val_loss": 0.562},
			"score": 0.7,
			"id": 459,
		},...
	],
}
\end{lstlisting}

Many properties had to be omitted from the output for brevity. But what can be seen is that there are detailed statistics about the current progress and the checkpoints created so far.

One the \coderef{state} changes to ``SUCCESS'', the classifier training is finished and the classifier can be used.

\section{Using a Classifier}
Using a classifier is done by creating a classification request.

\begin{lstlisting}[language=rest,caption=Using a classifier]
POST /classification_requests/
{
	"classifier_id": <clsid>,
	"document_ids": [<docid>, ...]
}
\end{lstlisting}

The response will be immediately and returns the labels that should be associated with each document.